\documentclass[12pt]{article}
\usepackage{epsfig,latexsym,amsbsy,amssymb,cite,eqsection}

\def\spose#1{\hbox to 0pt{#1\hss}}
\def\ltapprox{\mathrel{\spose{\lower 3pt\hbox{$\mathchar"218$}}
 \raise 2.0pt\hbox{$\mathchar"13C$}}}
\def\gtapprox{\mathrel{\spose{\lower 3pt\hbox{$\mathchar"218$}}
 \raise 2.0pt\hbox{$\mathchar"13E$}}}

\begin{document}
\setlength{\unitlength}{0.2cm}

\title{
Operator Product Expansion on the Lattice: a Numerical Test in the 
Two-Dimensional Non-Linear $\sigma$-Model
}
\author{
  \\
  {\rm Sergio Caracciolo$^{\rm a}$, Andrea Montanari$^{\rm a}$, and
          Andrea Pelissetto$^{\rm b}$}             \\[0.2cm]
  {\small\it ${}^{\rm a}$ 
      Scuola Normale Superiore and INFN -- Sez. di Pisa, I-56100 Pisa, ITALY} 
        \\[-0.2cm]
  {\small\it ${}^{\rm b}$ 
      Dip. di Fisica and INFN -- Sez. di Roma I, 
      Universit\`a di Roma I, I-00185 Roma, ITALY} 
        \\[-0.2cm]
{\small E-mail: {\tt Sergio.Caracciolo@sns.it}, {\tt Montanari@sns.it},
                  {\tt Andrea.Pelissetto@roma1.infn.it} }\\
}
\vspace{0.5cm}

\maketitle
\thispagestyle{empty}   

\vspace{0.2cm}

\begin{abstract}
We consider the short-distance behaviour of the product of the 
Noether $O(N)$ currents in the lattice nonlinear $\sigma$-model. 
We compare the numerical results with the predictions of the 
operator product expansion, using one-loop perturbative 
renorma\-li\-za\-tion-group
improved Wilson coefficients. We find that, even on quite small
lattices ($m a \approx 1/6$), the perturbative operator product expansion
describes that data with an error of 5-10\% in a large window 
$2a \ltapprox x \ltapprox m^{-1}$. We present a detailed discussion of the 
possible systematic errors.
\end{abstract}

\clearpage

\newcommand{\be}{\begin{equation}}
\newcommand{\ee}{\end{equation}}
\newcommand{\bea}{\begin{eqnarray}}
\newcommand{\eea}{\end{eqnarray}}
\newcommand{\<}{\langle}
\renewcommand{\>}{\rangle}
\def\bsigma{\mbox{\protect\boldmath $\sigma$}}
\def\bpi{\mbox{\protect\boldmath $\pi$}}
\def\MS{$\overline{\rm MS}$ }

\def\sg{\mbox{\boldmath $\sigma$}}
\def\pg{\mbox{\boldmath $\pi$}}
\def\tg{\mbox{\boldmath $\tau$}}
\def\jg{\mbox{\boldmath $j$}}
\def\phg{\mbox{\boldmath $\phi$}}
\def\psg{\mbox{\boldmath $\psi$}}
\def\dip{\frac{d^2 p}{(2\pi)^2}}
\def\diq{\frac{d^2 q}{(2\pi)^2}}
\def\dirac{(2\pi)^2\delta^{(2)}}
\def\intBZ{\int_{BZ}}
\def\f{\hat{f}}
\def\opl{\left[}
\def\opr{ \right]_{{\overline{MS}}}}
\def\mbar{\mu }
\newcommand{\beq}{\begin{eqnarray}}
\newcommand{\eeq}{\end{eqnarray}}
\def\f{\hat{f}}
\def\pb{\overline{p}}
\def\qb{\overline{q}}
\def\MMS{{\overline{\rm MS}}}

\newcommand{\R}{\hbox{{\rm I}\kern-.2em\hbox{\rm R}}}
\newcommand{\reff}[1]{(\ref{#1})}
\def\smfrac#1#2{{\textstyle\frac{#1}{#2}}}

\section{Introduction} \label{Introduction}

The lattice regularization is nowadays the best theoretical tool for the study 
of nonperturbative phenomena in field theories.
It has been extensively applied to QCD, providing quantitative predictions for
many non-perturbative quantities, 
and it has also been
extensively used in the study of strong-interaction effects in 
weak processes. One of the basic problems one has to deal with in these
investigations is the determination of renormalized operators. Referring to
the simple case of a purely multiplicative renormalization, given a 
``bare" operator ${\cal O}_{LAT}$ on a lattice of size $L$, we want to find a 
renormalized operator related to ${\cal O}_{LAT}$ by
\begin{equation}
{\cal O}_R(\mu) = Z_{\cal O} (\Lambda/\mu,\Lambda a, a L^{-1})\ {\cal O}_{LAT}.
\label{MultiplicativeRenormalization}
\end{equation}
In Eq. (\ref{MultiplicativeRenormalization}) we have emphasized the 
dependence of the renormalization constant $Z_{\cal O}$ 
upon the various scales of the 
problem: the lattice spacing $a$, which encodes the dependence on the 
bare coupling constant $g_0$, the renormalization scale $\mu$,
the physical scale $\Lambda$ (the so called ``$\Lambda$-parameter''), 
which breaks the scale invariance of the continuum theory, 
and the size of the lattice $L$.  
The constant $Z_{\cal O}$ should be defined so that 
${\cal O}_R(\mu)$ has  a finite continuum limit. In other words, the limit 
$\Lambda a\to 0$ (i.e. $g_L\to\infty$), $L/a\to \infty$, 
at fixed $\Lambda/\mu$, 
\begin{equation}
{\cal O}_R^{\rm cont} (\Lambda/\mu) = 
   \, \lim_{\Lambda a\to 0} \lim_{L/a\to\infty}
       {\cal O}_R (\mu)
\label{calOcont}
\end{equation}
should be finite.\footnote{Eq. (\ref{calOcont}) is a symbolic notation. 
Explicitly it indicates that correlations of ${\cal O}_R (\mu)$ with 
{\em renormalized} fields have a finite continuum limit.} It 
is important to notice that the two limits in Eq. (\ref{calOcont}) do not 
commute, the infinite-volume limit should be taken
{\em before} the continuum extrapolation $\Lambda a \to 0$.
The renormalization constant admits a smooth infinite-volume limit 
at fixed $a$. The behaviour in the continuum limit is instead singular and 
it is predicted by the renormalization group (RG). 
An exception is represented by operators
that are conserved, at least in the continuum limit. In this case, the 
renormalization constant has a smooth continuum limit and does not depend on
the scale $\mu$.  Examples of such operators are 
the vector and axial flavour 
currents in QCD and the energy-momentum tensor. In this case, 
the renormalization constants can be found by requiring the validity of the 
Ward identities related to the conservation of the operator 
\cite{Bochicchio:1985xa,Curci:1987sm,Caracciolo:1988hc,Caracciolo:1992cp}.

More difficult is the computation of the renormalization constants 
when the operator is not conserved. In this case, one can use 
lattice perturbation theory.
However, these calculations are very tedious and long and one 
is usually restricted to one-loop order. 
A general non-perturbative strategy has been proposed in 
Ref. \cite{Martinelli:1995ty} and 
extensively studied since then, see, for instance, 
Ref. \cite{Gockeler:1998ye}. It provides 
a direct method for the computation of the renormalization
constant appearing in \reff{MultiplicativeRenormalization}. The main difficulty 
is the continuum limit, Eq. \reff{calOcont}, which requires 
to keep under control the lattice artifacts and the finite-size effects.
In practice, the method works only if there exists a window 
\begin{equation}
L/a\gg \frac{1}{\mu a}\gg 1, \qquad\qquad
L/a\gg \frac{1}{\Lambda a}\gg 1,
\label{ScalingWindow}
\end{equation}
in which lattice artifacts $O(\mu a,\Lambda a)$ and finite-size
effects $O(a/L)$ are sufficiently small. Conditions \reff{ScalingWindow}
are not enough since we also want to relate ${\cal O}^{\rm cont}$ 
to renormalized operators defined in continuum, say $\overline{\rm MS}$, 
perturbation
theory. Since we want to apply continuum perturbation theory, we need
additionally 
\be
\Lambda \ll \mu\, .
\label{Bound}
\ee

A different renormalization method exploits finite-size
effects and it is based on the
observation due to Symanzik \cite{Symanzik:1981wd} that renormalizability is 
not spoiled when a field theory is considered on a finite space-time manifold.
We can define a renormalization scheme by using the lattice size $L$ 
(instead of the external momentum employed in the preceding 
scheme) as the relevant energy scale. Finite-size scaling schemes have been
much studied in the last years 
\cite{Luscher:1992an,Luscher:1993zx,Luscher:1994gh,Jansen:1996ck,
Capitani:1998mq,Luscher:1998pe,Luscher:1997jn}. 
They have been successfully applied to the computation of the 
$\Lambda$-parameter, of the light-quark masses and of the renormalized
axial current in quenched QCD.
With respect to the previous method, there 
is the advantage that the requirements \reff{ScalingWindow} reduce to
the weaker ones
\begin{eqnarray}
L\gg a, \qquad \Lambda a \ll 1, \qquad \mu a \ll 1.
\end{eqnarray} 

Recently, a new method for the determination of renormalized composite 
operators in asymptotically free theories has been proposed in 
\cite{Dawson:1997ic,Rossi:1998kc,Testa:1997ne,Martinelli:1998hz}. 
It is based on the Operator Product Expansion (OPE), which 
has been postulated for the first time by Wilson \cite{Wilson:1969zs}
thirty years ago. It  has been proved in perturbation theory by Zimmermann 
\cite{Zimmerman:70}, 
and it is widely thought to hold beyond perturbation theory.
This new procedure shares many features of the infinite-volume 
schemes and in fact it is 
defined in infinite volume, but it has the advantage of being more direct.
It provides the matrix elements directly in continuum schemes and therefore,
it could be the method of choice in those cases in which 
the renormalized operators are obtained as combinations of many different
lattice operators.
Moreover, it avoids the evaluation of products of local operators at 
coincident points. 
This fact reduces lattice artifacts and allows a simpler
implementation of operator improvement. 

We proceed now to describe the general context to which this method applies.
Let us consider, for instance, the following simple example of OPE:
\begin{eqnarray}
{\cal A}(x){\cal B}(-x) \sim C_{\cal O}(x){\cal O}(0)+\dots
\label{OPEesempio}
\end{eqnarray}
where the dots $\dots$ indicate terms of higher order in $x^2$, 
corresponding to operators of higher canonical dimension.
Let us suppose that we know how to define the renormalized operators
${\cal A}$ and ${\cal B}$ nonperturbatively. For instance, this is the case
if they are conserved currents: If the lattice
does not break the corresponding symmetry, one can choose a discretization 
in which they are  exactly conserved and therefore, do not 
renormalize. 

Let us suppose that we are interested in computing some hadronic
matrix element of ${\cal O}$ which we shall denote by
$\<h_1|{\cal O}|h_2\>$.
The proposed procedure works as follows:
\begin{enumerate}
\item The Wilson coefficient is calculated using RG-improved 
perturbation theory. Any renormalization scheme can be used in 
this step, but we will usually consider the  \MS scheme. Moreover,
to avoid spurious scale dependences, we consider 
RG-invariant (RGI) Wilson coefficients and 
correspondingly RGI operators. We indicate by $C^{(l)}_{\cal O}(x)$ 
the resulting $l$-loop approximation for the Wilson coefficient.
\item The matrix element $\<h_1|{\cal A}(x){\cal B}(-x)|h_2\>$ is computed in 
a numerical simulation for a properly chosen range of $x$,
say for $r^2\le x^2\le R^2$. This step gives
a function $G_{\cal AB}(x)$.
\item Finally, $G_{\cal AB}(x)$ is fitted using the form 
$C^{(l)}_{\cal O}(x)\cdot\widehat{\cal O}$ and keeping $\widehat{\cal O}$
as the parameter of the fit.
\end{enumerate}
The result $\widehat{\cal O}$ will be a function of $\Lambda a$, 
$L/a$, and of the considered range $[r^2,R^2]$. The physically 
interesting quantity\footnote{One could also consider the OPE in 
a finite physical volume. In this case, one would take the infinite-volume
limit and the continuum limit together, keeping the physical volume
fixed.} is then
\begin{equation}
\widehat{\cal O}^{\rm cont}_{RGI} =\, 
 \lim_{R \Lambda,r\Lambda\to0}\, 
 \lim_{\Lambda a, a/r, a/R\to 0}\,
 \lim_{L/a\to \infty}   \widehat{\cal O}.
\end{equation}
Note that the limits cannot be interchanged. First, we should 
consider the infinite-volume limit, then the scaling (continuum) limit
at fixed $R \Lambda$, and $r\Lambda$, and finally
take the short-distance limit. The outcome of this step,\footnote{
Notice the close resemblance with the so-called ``point-splitting''
renormalization procedure proposed a long time ago by Dirac
\cite{Dirac:1934}.}
$\widehat{\cal O}^{\rm cont}$, is identified with the matrix element we are 
looking for,
$\<h_1|{\cal O}|h_2\>$, renormalized in the same scheme 
in which we computed the Wilson coefficient $C^{(l)}_{\cal O}(x)$.

In practice, one works at finite values of $a$ and $L$. However, one may
still hope that there exists
a window of values of $|x|$ such that $a\ll |x| \ll \Lambda^{-1} \ll L$, 
so that $|x|$ is small enough to allow the application of 
perturbative OPE and still large enough to make lattice artifacts small.
The volume must of course be large enough to avoid size effects.
Notice that the condition $a\ll |x| \ll \Lambda^{-1}$ replaces here 
the condition $a\ll \mu^{-1}\ll \Lambda^{-1}$, $\mu$ being the momentum
at which renormalization is performed, present in standard 
infinite-volume schemes. In this respect the OPE method is not different
from the standard renormalization methods. 

The OPE has already been used in lattice simulations. For instance, in
Refs. \cite{D'Elia:1998ji,D'Elia:1998wm,D'Elia:1997en,D'Elia:1997ne,
Campostrini:1984xr}
the authors consider quark and gluon two-point functions in QCD.
Their aim is to compute the (chiral and gluon) condensates. 
They use an ``OPE-inspired'' phenomenological fitting form, considering 
the condensates as fitting parameters.
In all these and similar studies vacuum expectation values of short-distance 
products are considered. 
The possibility of varying the external states, investigated
in this paper, provides a stronger check of the OPE and could be
a fruitful resource.

An approach which allows to bypass the perturbative computation of
the Wilson coefficient (with the related limitations) has been proposed in
Refs. \cite{Capitani:1998fe,Capitani:1999fm}. 
The idea has been tentatively applied to the evaluation of 
the momenta of the nucleon structure functions. 
This method allows to study the structure functions at energy scales 
$Q^2 \ll 1/a^2$. 

In this paper we perform a feasibility study of the procedure outlined
above in the 
two-dimensional nonlinear $\sigma$-model. This model is quite 
interesting since it shares with QCD the property of being 
asymptotically free with a nonperturbatively generated mass gap
\cite{Polyakov:1975rr,Brezin:1976qa,Bardeen:1976zh}. It has also 
been extensively studied in perturbation theory, both in the 
continuum \cite{Brezin:1976ap,Hikami:1977vr,Bernreuther:1986js}
and on the lattice \cite{Falcioni:1986xb,Caracciolo:1994sk,Caracciolo:1995bc,
Alles:1997rr,Shin:1998bh,Alles:1999ca,Alles:1999fh,Alles:1999gm}. 
Extensive simulations have been done to check whether the theory shows the 
correct behaviour predicted by the perturbative RG
\cite{Wolff:1990dm,Wolff:1990hv,Edwards:1992eg,Caracciolo:1995ud,%
Caracciolo:1995ah,Alles:1997rr,Alles:1998ms}  
and to test the exact predictions for the mass gap 
obtained using the thermodynamic Bethe ansatz 
\cite{Hasenfratz:1990zz,Hasenfratz:1990ab}. Another advantage of this 
model is the availability of a very efficient algorithm
\cite{Wolff:1989uh,Wolff:1989kw,Wolff:1990hv}, 
which shows no critical slowing down 
(for a general discussion see Ref. \cite{Caracciolo:1993nh}). 
Therefore, we can work
on very large lattices keeping the size effects completely under control.

We consider the OPE of products of Noether currents, a particularly simple case
since the current is exactly conserved and therefore does not need 
any renormalization. We discuss in detail all possible sources of systematic 
errors, by considering several different cases.

The paper is organized as follows. In Sec. \ref{sec2} we define the 
model and the basic notations, and in Sec. \ref{sec.OPE} we give the 
OPE for the product of the Noether currents that will be studied numerically.
Sec. \ref{sec4} presents a detailed discussion of our numerical results.
Our conclusions are presented
in Sec. \ref{sec5}. In App. \ref{secA} and \ref{secB}
we give our perturbative results, while in App. \ref{RunningCoupling}
we critically discuss different methods for the determination of the 
running coupling constant.

Preliminary results of this work have been presented
at the Lattice conferences in 1998 and 1999
\cite{Caracciolo:1998gf,Caracciolo:1999yv}.

\section{The lattice model} \label{sec2}

In this paper we consider the nearest-neighbor lattice nonlinear 
$\sigma$-model in two dimensions. The fields
are unit-length spins $\bsigma_x\in R^N$ and the action is given by
\be
{S}^{\rm latt} \, =\,  {a^2\over 2 g_L} 
  \sum_{x\in {\mathbb Z}^2_a}\sum_{\mu=1}^2 \partial_\mu\bsigma_x \cdot 
   \partial_\mu\bsigma_{x},
\label{action-lattice}
\ee
where $\partial_\mu f_x = (f_{x + a\mu} - f_x)/a$, and $g_L$ is the bare lattice
coupling constant (often one introduces $\beta \equiv 1/g_L$). 
The partition function is given by
\be
Z\, =\, \int \exp (-{S}^{\rm latt})\, \prod_x d^N\bsigma_x  \,
                \delta (\bsigma_x^2 - 1)\; .
\ee
As usual, we introduce the lattice $\Lambda$-parameter
\begin{eqnarray}
\Lambda^{\rm latt} (a,g_L) &=& {1\over a}\
   \left( \beta_0 g_L\right)^{-\beta_1/\beta_0^2}
       \exp\left[ - {1\over \beta_0 g_L} - \int^{g_L}_0 dt
       \left( {1\over \beta^{\rm latt}(t)} + {1\over \beta_0 t^2} -
              {\beta_1\over \beta_0^2 t}\right)\right] \nonumber \\
&=& {1\over a}\
\left({(N-2) g_L\over 2\pi}\right)^{- 1/(N-2)}
\exp\left( - {2\pi\over (N-2) g_L}\right) \left(1 +
\sum_{n=1}^\infty{a_n g_L^n} \right),
\label{Lambdalatt}
\end{eqnarray}
where $\beta^{\rm latt}(g_L)$ is the lattice $\beta$-function which can be 
expanded as
\be 
\beta^{\rm latt}(g_L) = - g_L^2 \sum_{k=0} \beta_k^{\rm latt} g_L^k.
\ee
The coefficient $\beta_k^{\rm latt}$ are known up to four loops, i.e. for 
$k\le 3$, see \cite{Falcioni:1986xb,Caracciolo:1995bc,Alles:1999fh}.
The first two coefficients are universal, and therefore, we have not added
the superscript.

The mass gap of the model is related to the $\Lambda$-parameter by
\be
m = C_N \Lambda^{\rm latt} (a,g_L).
\label{MassGapLattice}
\ee
The constant $C_N$ is a nonperturbative quantity, and for generic models,
it is usually unknown. For the nonlinear $\sigma$-model, an explicit expression
has been obtained using the thermodynamic Bethe ansatz 
\cite{Hasenfratz:1990zz,Hasenfratz:1990ab}. Explicitly
\be
C_N = \left( {8\over e} \right)^{1/(N-2)}
{1\over \Gamma\left(1+{1\over N-2}\right)} 2^{5/2} 
  \exp\left[{\pi\over 2(N-2)}\right].
\label{Hasenfratzconst}
\ee
In this paper we will consider the OPE of products of the Noether 
$O(N)$ current, which is defined by
\be
j_{\mu,x}^{L,ab} = {1\over g_L} 
  \left(\sigma_x^a \partial_\mu \sigma_x^b - \sigma_x^b \partial_\mu \sigma_x^a
  \right).
\label{jlattice}
\ee
Because of the $O(N)$ invariance of the theory, this current satisfies 
the Ward identity 
\be
\partial^-_\mu \< j_{\mu,x}^{L,ab} {\cal O}\>\ =\, 
 \left\< {\delta{\cal O}\over \delta \sigma_x^a} \sigma_x^b -
         {\delta{\cal O}\over \delta \sigma_x^b} \sigma_x^a \right\>,
\label{Ward-ON}
\ee
where ${\cal O}$ is a generic operator and 
$(\partial^-_\mu f)_x = (f_x - f_{x-a\mu})/a$. 
Since (\ref{Ward-ON}) holds exactly
on the lattice, $j_{\mu,x}^{L,ab}$ does not need to be renormalized.

A second operator we will be interested in is the energy-momentum tensor.
In the continuum it is given by
\be
T_{\mu\nu} = {1\over g} \left[ 
    \partial_\mu\bsigma\cdot \partial_\nu\bsigma - {1\over2}
    \delta_{\mu\nu} \left( \partial \bsigma \right)^2 \right].
\label{Tmunucontinuum}
\ee
The energy-momentum tensor is the Noether current of the translations,
and therefore it is exactly conserved. On the lattice translation invariance
is lost, and thus, there is no lattice operator whose naive continuum limit
corresponds to (\ref{Tmunucontinuum}) and that is exactly conserved.
However, as shown in Ref. \cite{Buonanno:1995us}, it is possible to 
define an operator which is conserved in the continuum limit.\footnote{
The general ideas have been presented in \cite{Caracciolo:1988hc},
following closely the method used in \cite{Bochicchio:1985xa} 
for the definition of the axial current.}
Explicitly one defines 
\be
T_{\mu\nu,x}^{\rm latt} =\,
  Z_{TT}^{L,(2,0)} T^L_{\mu\nu,x} + 
  Z_{T2}^{L,(2,0)} {1\over g_L} \delta_{\mu\nu} 
      (\overline{\partial}_\mu \bsigma)^2_x + 
  \widetilde{Z}_{T3}^{L,(2,0)} {1\over g_L} \delta_{\mu\nu}
      (\overline{\partial} \bsigma)^2_x,
\label{Tmunulattice}
\end{equation}
where $(\overline{\partial}_\mu f)_x = (f_{x+a\mu} - f_{x-a\mu})/(2a)$,
$T^L_{\mu\nu,x}$ is the naive lattice energy-momentum tensor,
\be
T^L_{\mu\nu,x} = \, {1\over g_L} \left[
    (\overline{\partial}_\mu \bsigma)_x \cdot 
    (\overline{\partial}_\nu \bsigma)_x
  - {1\over2} \delta_{\mu\nu} (\overline{\partial} \bsigma)^2_x \right],
\label{Tmunulatticenaive}
\ee
and $Z_{TT}^{L,(2,0)}$, $Z_{T2}^{L,(2,0)}$, $\widetilde{Z}_{T3}^{L,(2,0)}$
are renormalization constants.
In Eq. (\ref{Tmunulattice}) we have not considered terms that 
are not $O(N)$ invariant and that vanish because of the equations
of motion.\footnote{Similar terms appear also in QCD. Indeed, in this case,
one should also consider, beside gauge-invariant operators, 
operators which are BRS variations and operators that are proportional
to the equations of motion,
see \cite{Caracciolo:1992cp}.} We have also discarded the contribution
proportional to the identity operator (diverging as $1/a^2$), since it
does not contribute to connected correlation functions. 
The renormalization constants have been computed to one-loop
order in \cite{Buonanno:1995us}. We have found a small error in one of the 
expressions. The correct results are reported in App. \ref{secA.2}.

\section{OPE for the $O(N)$ Noether currents} \label{sec.OPE}

In this Section we report the general form of the OPE for 
the products of the $O(N)$ currents.
The coefficients are computed in perturbation theory, the explicit 
expressions to one-loop order being reported in App. \ref{secB}.
As is well known, a perturbative expansion cannot be defined 
directly for the model (\ref{action-lattice}) because of infrared 
singularities. A standard way out consists in adding a magnetic field
that breaks the $O(N)$ invariance of the theory and 
plays the role of infrared regulator. One considers therefore
\be
{S}^{\rm latt} \, =\,  {a^2\over 2 g_L} 
  \sum_{x\in {\mathbb Z}^2_a}\sum_{\mu=1}^2 \partial_\mu\bsigma_x \cdot 
   \partial_\mu\bsigma_x -\, 
   {h a^2\over g_L} \sum_{x\in {\mathbb Z}^2_a} \sigma^N_x.
\label{action-lattice-conmassa}
\ee
A perturbative expansion is obtained 
setting $\bsigma_x = (\bpi_x, \sqrt{1 - \bpi_x^2})$,
and expanding in powers of $\bpi_n$. 
When the integration measure is kept into account, the total action reads:
\be
{S}^{\rm latt}_{\rm TOT} = {S}^{\rm latt} +\sum_{x\in {\mathbb Z}_a^2}
\log \sigma^N_x \; .
\ee
The calculation of the OPE
is fairly standard and we report below the results in the continuum 
\MS scheme and on the lattice for the cases that 
will be studied numerically.
\subsection{The continuum $\overline{\rm MS}$ scheme.} \label{sec3.1}
We begin by considering the OPE of the scalar product of two currents.
The general form of the OPE is dictated by $O(N)$ and rotational
invariance and by power counting. The result can be written as follows:
\begin{eqnarray}
\frac{1}{2}\jg_\mu   (x)\cdot   \jg_{\rho}(-x)&\equiv&
\frac{1}{2}\sum_{a,b}j^{ab}_\mu   (x)
j^{ab}_{\rho}(-x)= 
\nonumber \\
& = & \left[\frac{\delta_{\mu   \rho}x_{\nu}x_{\sigma}}{x^2}W_1(x)+
\frac{x_\mu   x_{\rho}x_{\nu}x_{\sigma}}{(x^2)^2}W_2(x)+
\frac{x_\mu   x_{\nu}\delta_{\rho\sigma}+
x_{\rho}x_{\sigma}\delta_{\mu   \nu}}{x^2}W_3(x)+\right.\nonumber\\
&&\left.+\frac{\delta_{\mu   \nu}\delta_{\rho\sigma}+\delta_{\mu   \sigma}
\delta_{\rho\nu}}{2}
W_4(x)\right]\frac{1}{g} \opl T_{\nu\sigma}\opr (0)+\nonumber\\
&&+\left[\frac{x_\mu   x_{\rho}}{x^2}W_5(x)+\delta_{\mu   \rho}W_6(x)\right]
\frac{1}{g^2}\opl(\partial\sg)^2\opr(0)+\nonumber\\
&&+\left[\frac{x_\mu   x_{\rho}}{x^2}W_7(x)+\delta_{\mu   \rho}W_8(x)\right]
\frac{1}{g^2}\opl\alpha\opr(0)+\nonumber\\
&&+\frac{1}{x^2}W_{0,\mu   \rho}(x)\,\frac{1}{g}\mbox{\boldmath $1$}\, ,
\label{scalcorrcont}
\end{eqnarray}
where $W_i(x)$ and $W_{0,\mu   \rho}(x)$ are functions of $x$, of the \MS 
coupling $g$, and of the renormalization scale $\mu$. Explicit one-loop
expressions are reported in App. \ref{secB.1}.
With $\opl \hphantom{\cdot} \cdot \hphantom{\cdot}\opr$ we indicate the \MS 
renormalized operator, while $T_{\nu\sigma}(x)$ is the energy-momentum 
tensor \reff{Tmunucontinuum}. Notice the appearance of the operator 
$\alpha(x)$,
\be
\alpha(x) \equiv {1\over \sigma_B^N(x)} 
  \left[h_B + \partial^2 \sigma_B^N(x)\right],
\label{defalphacont}
\ee
where the subscript $B$ indicates bare quantities.
This operator is not $O(N)$ invariant and its presence is due to the 
breaking of the $O(N)$ invariance by the magnetic term. However, 
using the equations of motion, $\alpha(x)$ can be rewritten as 
\be
\opl \alpha \opr (x) = h \sigma^N(x) - \opl (\partial\sg)^2 \opr + 
 g \bpi \cdot {\delta S\over \delta \bpi}.
\ee
Therefore, on-shell and in the limit $h\to0$, we recover an $O(N)$-invariant
expansion. Explicitly:
\begin{eqnarray}
\frac{1}{2}\jg_\mu   (x)\cdot   \jg_{\rho}(-x)
& = & \left[\frac{\delta_{\mu   \rho}x_{\nu}x_{\sigma}}{x^2}W_1(x)+
\frac{x_\mu   x_{\rho}x_{\nu}x_{\sigma}}{(x^2)^2}W_2(x)+
\frac{x_\mu   x_{\nu}\delta_{\rho\sigma}+
x_{\rho}x_{\sigma}\delta_{\mu   \nu}}{x^2}W_3(x)+\right.\nonumber\\
&&\left.+\frac{\delta_{\mu   \nu}\delta_{\rho\sigma}+\delta_{\mu   \sigma}
\delta_{\rho\nu}}{2}
W_4(x)\right]\frac{1}{g}\opl T_{\nu\sigma}\opr(0)+\nonumber\\
&&+\left[\frac{x_\mu   x_{\rho}}{x^2}W'_5(x)+\delta_{\mu   \rho}W'_6(x)\right]
\frac{1}{g^2}\opl(\partial\sg)^2\opr(0)+\nonumber\\
&&+\frac{1}{x^2}W_{0,\mu   \rho}(x)\, \frac{1}{g}\mbox{\boldmath $1$} ,
\end{eqnarray}
with
\be
W'_5(x)\equiv W_5(x)-W_7(x), \qquad\qquad
W'_6(x)\equiv W_6(x)-W_8(x).
\ee
The Wilson coefficients satisfy the following RG equations:
\begin{eqnarray}
\left[\mu   \frac{\partial}{\partial\mu   }+\beta(g)\frac{\partial}{\partial g}
-\frac{\beta(g)}{g}\right]W_{0,\mu   \rho}(x;g,\mu   )&=&0, \\
\left[\mu   \frac{\partial}{\partial\mu   }+\beta(g)\frac{\partial}{\partial g}
-\frac{\beta(g)}{g}\right]W_i(x;g,\mu   )&=&0,\;\;\;\; i = 1,\dots,4,\\
\left[\mu   \frac{\partial}{\partial\mu   }+\beta(g)\frac{\partial}{\partial g}
-\frac{\beta(g)}{g}-g\frac{\partial}{\partial g}
\left(\frac{\beta(g)}{g}\right)\right]W'_i(x;g,\mu   )&=&0,\;\;\;\; i = 5,6,
\end{eqnarray}
where $\beta(g)$ is the \MS $\beta$-function. 

We also consider the OPE of the antisymmetric product of the currents. 
Neglecting terms of order $O(x \log^p x)$, we have\footnote{
Note that one could also add a contribution proportional to 
$(x_\mu   x_{\alpha}\delta_{\nu\beta}-
  x_\nu   x_{\alpha}\delta_{\mu\beta})/x^2 
\left[\partial_\alpha j_\beta^{ab}(0) -
                 \partial_\beta j_\alpha^{ab}(0)\right]$. 
However, in two dimensions, 
$(x_\mu   x_{\alpha}\delta_{\nu\beta}-
  x_\nu   x_{\alpha}\delta_{\mu\beta} - (\alpha \leftrightarrow\beta) ) =
  x^2 (\delta_{\mu\alpha} \delta_{\nu\beta} - 
       \delta_{\nu\alpha} \delta_{\mu\beta})$, and thus 
this term is equivalent to that proportional to $U_1(x)$.} 
\begin{eqnarray}
&& \hskip -1truecm
\sum_c\left[j^{ac}_\mu   (x)j^{bc}_{\nu}(-x)-
j^{bc}_\mu   (x)j^{ac}_{\nu}(-x)\right]  =  
\nonumber \\
&& \left[{x_\mu x_\nu x_\alpha\over (x^2)^2} U_{00}(x) + 
      {\delta_{\mu\nu} x_\alpha\over x^2} U_{01}(x) + 
      {\delta_{\mu\alpha}x_\nu + \delta_{\nu\alpha} x_\mu\over x^2} U_{02}(x)
\right]
\frac{1}{g}j^{ab}_{\alpha}(0)
\nonumber \\
&&+ (\delta_{\mu   \alpha}\delta_{\nu\beta}-
\delta_{\mu   \beta}\delta_{\nu\alpha})U_1(x)
\frac{1}{4 g}\left[\partial_\alpha j_\beta^{ab}(0) - 
                 \partial_\beta j_\alpha^{ab}(0)\right]
\nonumber\\
&&+\frac{x_\mu   x_{\alpha}\delta_{\nu\beta}-
x_{\nu}x_{\alpha}\delta_{\mu   \beta}}{x^2}U_2(x)\frac{1}{2 g}
\left[\partial_\alpha j_\beta^{ab}(0) +
                 \partial_\beta j_\alpha^{ab}(0)\right].
\label{anticorrcont}
\end{eqnarray}
The coefficients $U_i(x)$ and $U_{0i}(x)$ satisfy the RG equations:
\begin{eqnarray}
\left[\mu   \frac{\partial}{\partial\mu   }+\beta(g)\frac{\partial}{\partial g}
-\frac{\beta(g)}{g}\right]U(x;g,\mu   )&=&0.
\label{RG-Wilson-antisymmetric}
\end{eqnarray}
%
%
%
\subsection{Lattice} \label{sec3.2}
On the lattice one can write expansions completely analogous to those 
holding in the continuum. The only difference is the appearance of terms that 
are cubic but not rotationally invariant. For the scalar product of two
currents we obtain
\begin{eqnarray}
\frac{1}{2}\jg^L_{\mu   ,x}\cdot   \jg^L_{\rho,-x}&=&
\left[\frac{\delta_{\mu   \rho}x_{\nu}x_{\sigma}}{x^2}W^L_1(x)+
\frac{x_\mu   x_{\rho}x_{\nu}x_{\sigma}}{(x^2)^2}W^L_2(x)+
\frac{x_\mu   x_{\nu}\delta_{\rho\sigma}+
x_{\rho}x_{\sigma}\delta_{\mu   \nu}}{x^2}W^L_3(x)+\right.\nonumber\\
&&\left.+\frac{\delta_{\mu   \nu}\delta_{\rho\sigma}+\delta_{\mu   \sigma}
\delta_{\rho\nu}}{2}
W^L_4(x)\right]\frac{1}{g_L}T^L_{\nu\sigma,0}+\nonumber\\
&&\left[\frac{\delta_{\mu   \rho}x_{\nu}x_{\sigma}}{x^2}{\widehat W^L_1}(x)+
\frac{x_\mu   x_{\rho}x_{\nu}x_{\sigma}}{(x^2)^2}{\widehat W^L_2}(x)+
\frac{x_\mu   x_{\nu}\delta_{\rho\sigma}+
x_{\rho}x_{\sigma}\delta_{\mu   \nu}}{x^2}{\widehat W^L_3}(x)+\right.\nonumber\\
&&\left.+\frac{\delta_{\mu   \nu}\delta_{\rho\sigma}+\delta_{\mu   \sigma}
\delta_{\rho\nu}}{2}
{\widehat W^L_4}(x)\right]\frac{1}{g_L^2}\delta_{\nu\sigma}
(\overline{\partial_{\nu}}\sg)^2_0+\nonumber\\
&&+\left[\frac{x_\mu   x_{\rho}}{x^2}W^L_5(x)+\delta_{\mu   \rho}W^L_6(x)\right]
\frac{1}{g_L^2}(\overline{\partial}\sg)_0^2+\nonumber\\
&&+\left[\frac{x_\mu   x_{\rho}}{x^2}W^L_7(x)+\delta_{\mu   \rho}W^L_8(x)\right]
\frac{1}{g_L^2}\alpha^L_0+\nonumber\\
&&+\frac{1}{x^2}W^L_{0,\mu   \rho}(x)\, \frac{1}{g_L}\mbox{\boldmath $1$} \, .
\label{scalcorrlatt}
\end{eqnarray}
Here $T^L_{\nu\sigma,x}$ is the lattice energy-momentum tensor defined in 
Eq. \reff{Tmunulatticenaive}, and $\alpha^L_x$ is the lattice 
analogue of the quantity defined in Eq. \reff{defalphacont},
\be
\alpha_x^L\equiv {1\over \sigma_x^N} \left(h + \partial^2 \sigma_x^N - 
  {g_L\over a^2 \sigma_x^N}\right),
\label{defalphalatt}
\ee
where $\partial^2=\sum_\mu\partial_\mu^-\partial_\mu$. Note the last term
in Eq. \reff{defalphalatt} that is not present in the continuum and which is 
due to the measure term in the functional integral. Using the lattice 
equations of motion, one can show that
\be
\alpha_x^L = h \sigma_x^N  +\sg_x\cdot\partial^2 \sg_x + 
   {1\over a^2}
   g_L \bpi_x\cdot {\delta S^{\rm latt}_{\rm TOT}\over \delta \bpi_x}-
   {g_L\over a^2}.
\ee
Therefore, on-shell and in the limit $h\to0$, it is possible to get rid 
of the terms proportional to $\alpha^L_x$ as we did before.

In the antisymmetric sector we can write 
\begin{eqnarray}
&& \hskip -1truecm
\sum_c\left[j^{L,ac}_{\mu   ,x}j^{L,bc}_{\nu,-x}-
j^{L,bc}_{\mu   ,x}j^{L,ac}_{\nu,-x}\right]  =  
\nonumber \\ 
&&
\left[{x_\mu x_\nu x_\alpha\over (x^2)^2} U_{00}^L(x) +
      {\delta_{\mu\nu} x_\alpha\over x^2} U_{01}^L(x) +
      {\delta_{\mu\alpha}x_\nu + \delta_{\nu\alpha} x_\mu\over x^2} 
      U_{02}^L(x)
\right]
\frac{1}{g_L}j^{L,ab}_{\alpha,0}
\nonumber \\[1mm]
&&+ \left(\delta_{\mu   \alpha}\delta_{\nu\beta}-
\delta_{\mu   \beta}\delta_{\nu\alpha}\right) U_1^L(x)
\frac{1}{4 g_L}\left[(\partial_\alpha^- j_\beta^{L,ab})_0 -
                   (\partial_\beta^- j_\alpha^{L,ab})_0\right]
\nonumber\\[1mm]
&&+\frac{x_\mu   x_{\alpha}\delta_{\nu\beta}-
x_{\nu}x_{\alpha}\delta_{\mu   \beta}}{x^2}U_2^L(x)\frac{1}{2 g_L}
\left[(\partial_\alpha^- j_\beta^{L,ab})_0 +
      (\partial_\beta^- j_\alpha^{L,ab})_0\right].
\label{anticorrlatt}
\end{eqnarray}
\subsection{RG-improved Wilson coefficients} \label{sec3.3}

We want now to resum the Wilson coefficients using the RG.
We assume that the Wilson coefficient $W(\mu x; g)$ satisfies the 
equation
\be
\left[\mu {\partial\over\partial \mu} + 
      \beta(g){\partial\over\partial g} + \gamma^W(g)\right] 
 W(\mu x; g) = 0,
\label{RGeqW}
\ee
where $\beta(g)$ is the (scheme-dependent) $\beta$-function and 
$\gamma^W(g)$ is related to the anomalous dimensions of the operators 
appearing in the OPE. Such an equation is not valid in general, 
since for generic operators $\gamma^W(g)$ is a matrix. However, in the 
cases we are interested in, cf. Sec. \ref{sec3.1} and \ref{sec3.2},
we can consider $\gamma_W(g)$ as a scalar 
function of $g$ and thus apply Eq. \reff{RGeqW}.

The general solution of Eq. \reff{RGeqW} is easily found. 
If we define a running 
coupling $\overline{g}(g,t)$ by
\be
\exp\left[ - \int_g^{\overline{g}(g,t)} {dx\over\beta(x)}\right] =\, t,
\label{defgbar}
\ee
then
\be
W(\mu x; g) = W(1;\overline{g}(g,\mu x)) 
   \exp\left[ \int_g^{\overline{g}(g,\mu x)} dx 
                 {\gamma^W(x)\over\beta(x)}\right]\; .
\label{Wilson-RGimproved}
\ee
It is useful to rewrite this equation as 
\be
W(\mu x; g) = U(g) W_{RGI}(\overline{g}(g,\mu x)),
\ee
where
\begin{eqnarray}
W_{RGI}(\overline{g}(g,\mu x)) &=& W(1;\overline{g}(g,\mu x))\
   \left[\overline{g}(g,\mu x)\right]^{\gamma_0^W/\beta_0}\
    \exp\left[ \int_0^{\overline{g}(g,\mu x)} dx \left(
                 {\gamma^W(x)\over\beta(x)} + {\gamma_0^W\over\beta_0}\right)
       \right]   , \nonumber 
\\
U(g) &=& g^{\gamma_0^W/\beta_0} 
   \exp\left[ - \int_0^{g} dx \left(
                 {\gamma^W(x)\over\beta(x)} + {\gamma_0^W\over\beta_0}\right)
       \right] ,
\end{eqnarray}
where we have expanded $\gamma^W(g)=\gamma_0^W g + O(g^2)$. The factor 
$U(g)$ may be included in the definition of the operator, introducing
\be
{\cal O}_{RGI} \equiv\, g^{\gamma_0^{\cal O}/\beta_0}
   \exp\left[ - \int_0^{g} dx \left(
    {\gamma^{\cal O}(x)\over\beta(x)} + {\gamma_0^{\cal O}\over\beta_0}\right)
   \right]
 \opl {\cal O}\opr,
\ee
where $\gamma^{\cal O}(g) = \gamma^{\cal O}_0 g + O(g^2)$ is 
the anomalous dimension of ${\cal O}$.
The new operator ${\cal O}_{RGI}$ is called RG
invariant, in the sense that it satisfies a RG equation of the form
\begin{eqnarray}
\left[\mu   \frac{\partial}{\partial \mu   }+\beta(g)
\frac{\partial}{\partial g}+\gamma_h(g)h\frac{\partial}{\partial h}-
\frac{n}{2}\gamma(g)\right]
\Gamma_{{\cal O}_{RGI}}^{(n)} = 0,
\end{eqnarray}
where $\Gamma_{{\cal O}_{RGI}}^{(n)}$ is the $n$-point irreducible 
correlation function with one insertion of $\cal O$. In other words, 
with this new definition, we have eliminated the running due to the 
anomalous dimension of the operator. Note, however, that, in spite of the 
name, this definition is not scheme independent. 

Eq. \reff{defgbar} can be written as
\be
\Lambda(1,\overline{g}(g,\mu x)) = \Lambda(\mu,g) x,
\ee
which shows that $\overline{g}$ is a function of $\Lambda(\mu,g) x$ only,
i.e. $\overline{g}(g,\mu x) = \overline{g}(\Lambda x)$.
Moreover, defining $z = - \log \Lambda x$, we can expand 
$\overline{g}(\Lambda x)$ in inverse powers of $z$. Explicitly, we obtain
\be
\overline{g}(\Lambda x) = {1\over \beta_0 z} -
  \frac{\beta_1\log z}{\beta_0^3 z^2} + O(\log^2 z/z^3).
\label{overlinegdef}
\ee
Since the $\beta$ function is known to four-loop order, both on the lattice 
and in the \MS scheme, two additional terms can be trivially 
added to this expansion, neglecting terms of order $\log^4 z/z^5$.

Finally, we wish to point out a further interpretation of $\overline{g}$. 
Consider for instance the \MS scheme and correspondingly the 
\MS coupling
$g_{\overline{MS}}$ which is a function of $m/\mu$, where $m$ is the mass gap.
Then, it is easy to show using the RG equations
that $\overline{g}(g,\mu x) = g_{\overline{MS}} (mx)$. In other words, 
$\overline{g}(g,\mu x)$ is nothing but the coupling constant computed 
at the scale $\mu = 1/x$.


Finally, notice that the results presented in this Section
can be applied to any renormalization scheme. In the following,
we will use them to resum continuum \MS predictions as well as lattice 
perturbative expressions.
%
%
\section{Numerical results} \label{sec4}

In this Section we present our numerical computations meant 
to verify the feasibility of the nonperturbative
renormalization method presented in the Introduction.

We consider matrix elements of the product of two Noether currents between
one-particle states. This case is particularly simple, because the 
currents are exactly conserved and the matrix 
elements of the operators that appear in the OPE expansion 
can be computed exactly in most of the cases.

The typical procedure we adopt is the following:
\begin{enumerate}
\item We compute a matrix element
$\<1|{\cal A}(x){\cal B}(y)|2\>$ measuring a suitable lattice 
correlation function.
\item We compute the matrix element
$\<1|{\cal O}|2\>$ appearing in the OPE expansion either exactly 
(as is possible in most of the cases) or numerically 
by means of some different numerical nonperturbative technique.
\item We divide $\<1|{\cal A}(x){\cal B}(y)|2\>$ by the OPE prediction
$\sum_{\cal O}C_{\cal O}(x-y)\<1|{\cal O}|2\>$.
\end{enumerate}
The goal is to see if there is a window of values of $|x-y|$ in which the 
OPE works, i.e. the result of step 3 is 1 independently of $|x-y|$.
For the cases we will consider here, the OPE gives an accurate
description (at the level of 5-10\%) of correlation functions for distances 
$2\lesssim |x-y|\lesssim\xi=m^{-1}$ (from now on we set $a=1$).
This result is quite encouraging for future applications of this method.

\subsection{The observables} \label{sec4.1}

\begin{figure}
\hspace{2cm}
\epsfig{figure=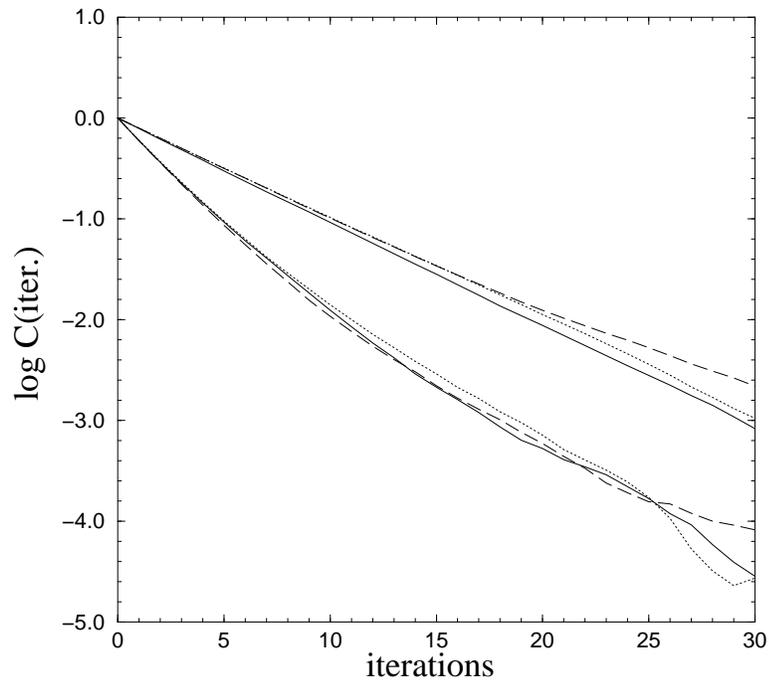,angle=-90,
width=0.7\linewidth}
\caption{Logarithm of the autocorrelation function $\log A(i)$
for the three lattices employed. Continuous lines refer to lattice
\ref{Lattice64x128}, dotted lines to lattice \ref{Lattice128x256}
and dashed lines to lattice \ref{Lattice256x512}. The upper curves
refer to the short-distance observable ${\cal O}_d$, $d=1$.
The lower curves refer to the long-distance observable
${\cal O}_d$ for $d\approx \xi^{\rm exp}$:
$d=8$ for lattice \ref{Lattice64x128}, 
$d=15$ for lattice \ref{Lattice128x256} and
$d=30$ for lattice \ref{Lattice256x512}.}
\label{Autocorrelation}
\end{figure}

We have simulated the 
$O(3)$ $\sigma$-model with action \reff{action-lattice}
using a Swendsen-Wang cluster algorithm 
with Wolff embedding 
\cite{Wolff:1989uh,Wolff:1989kw,Wolff:1990dm,Caracciolo:1993nh}. 
We did not try to optimize the updating procedure:
Most of the CPU time was employed in evaluating the relevant observables
(four-point functions) 
on the spin configurations of the ensemble. 
In order to estimate the scaling corrections,
we simulated three different lattices, of size $T\times L$, using 
in all cases periodic boundary conditions:
\renewcommand{\theenumi}{(\Alph{enumi})}
\begin{enumerate}
\item \label{Lattice64x128}
Lattice of size $128\times 64$ with $g_L = 1/1.40$. 
\item \label{Lattice128x256}
Lattice of size $256\times 128$ with $g_L = 1/1.54$. 
\item \label{Lattice256x512}
Lattice of size $512\times 256$ with  $g_L = 1/1.66$.
\end{enumerate}
The algorithm is extremely efficient---the dynamic critical exponent 
$z$ is approximately 0---and the autocorrelation time is very small. 
We performed a preliminary study in order to determine how many iterations 
are needed to obtain independent configurations. For this purpose we measured 
the normalized autocorrelation function
\be
A(j) = {{\cal N}\over {\cal N} -j} {\sum_{i=1}^{{\cal N}-j} 
     ({\cal O}_i - \overline{{\cal O}})
     ({\cal O}_{i+j} - \overline{{\cal O}}) \over 
           \sum_{i=1}^{{\cal N}} ({\cal O}_i - \overline{{\cal O}})^2 }
\ee
for different observables $\cal O$. Here 
${\cal N}$ is the number of Monte Carlo iterations, ${\cal O}_i$ is the 
value of $\cal O$ at the $i$-th iteration, and 
$\overline{{\cal O}}$ the sample mean of $\cal O$. 
In Fig. \ref{Autocorrelation} we report $A(\tau)$ for the observable
\begin{eqnarray}
{\cal O}_{d} &\equiv& {1\over 2 L T} \sum_{x} 
( \bsigma_x\cdot\bsigma_{x+v}+ \bsigma_x\cdot\bsigma_{x+w}),
\end{eqnarray}
where $v = (d,0)$ and $w = (0,d)$. 
We have considered $d=1$ ---a short-distance 
observable---and $d\approx \xi^{\rm exp}=m^{-1}$. 
As expected, local observables have a slower dynamics---this is due to the 
fact that the dynamics is nonlocal---than long-distance ones. 
In any case, for all observables the autocorrelation function shows a fast 
decay: indeed, for $d=1$ we have $\tau^{\rm exp}\approx 10$, while for 
$d\approx \xi^{\rm exp}$ we have $\tau^{\rm exp}\approx 5$. Moreover,
$A(j)$ is independent of the lattice used, confirming the fact that 
$z\approx 0$ (as we shall report below, all lattices have the same 
$L/\xi^{\rm exp}$). Since the measurement of the observables is quite CPU-time
consuming, we computed the relevant correlation functions only every 15 
iterations, which should be enough to make the measurements independent. 
Nonetheless, most of the CPU time is employed in
evaluating the observables on each spin configuration of the ensemble.
We computed the product $j(0)j(x)$ of two Noether currents
for distances $x$ smaller than some fixed fraction of
the correlation length: $|x|\lesssim k\xi$.
If the physical size $L/\xi$ of the lattice is kept constant (as we did)
we expect the CPU time to scale as $L^4$.
The CPU time per iteration turns out to be roughly independent
of the particular product considered.
As an example we give the CPU time per measurement for the
simulation in which we compute the antisymmetric product of two currents 
between states with opposite momentum. For the three different lattices,
on an SGI Origin2000, we have:
$\tau_{128\times 64}\approx 5.4\, \rm{sec}$,
$\tau_{256\times 128}\approx 71\, \rm{sec}$,
$\tau_{512\times 256}\approx 1100\, \rm{sec}$.

We measure several different observables. First, we measure the two-point 
function $C(\pb;t)$ (here and in the following the ``temporal" direction
is the first one, of extent $T$)
\begin{eqnarray}
C(\pb;t)\equiv \frac{1}{LT}\sum_{t_1=1}^T\sum_{x_1,x_2=1}^L
   e^{i\pb(x_1-x_2)} \<\sg_{t_1,x_1}\cdot\sg_{t+t_1,x_2}\>.
\label{TwoPointCorrelationFunction}
\end{eqnarray}
We computed the correlation function $C(\pb;t)$ on the lattices
\ref{Lattice128x256} and \ref{Lattice256x512} for momenta
$\pb=2\pi n/L,\, n=0\dots 3$ and times separations
$0\le t\le 100$; on lattice \ref{Lattice64x128} we considered the same set of
momenta and time separations $0\le t\le 40$. 
The number of independent configurations we generated is: 
$N_{\rm conf}\simeq 6\cdot 10^6$
for lattice \ref{Lattice64x128}; $N_{\rm conf}=590000$ for
lattice \ref{Lattice128x256}; $N_{\rm conf}= 180000$ for
lattice \ref{Lattice256x512} and $\pb=0$; finally
$N_{\rm conf}= 139000$ for lattice \ref{Lattice256x512} and $\pb\neq 0$. 

A  check of our simulation is provided by the results of Ref. 
\cite{Luscher:1990ck}, who computed, among other things, the mass gap for 
lattices \ref{Lattice64x128} and \ref{Lattice128x256}.
For the exponential correlation length $\xi^{\rm exp} = m^{-1}$ 
we obtain 
\be
\xi^{\rm exp} = 6.878(2), \, 13.638(10), \, 27.054(25) ,
\ee
for lattices  \ref{Lattice64x128}, \ref{Lattice128x256}, \ref{Lattice256x512}
respectively. They are in good agreement with the 
results of Ref. \cite{Luscher:1990ck}: they obtain 
$\xi^{\rm exp} = 6.883(3)$ and $\xi^{\rm exp}=13.632(6)$ 
for the first two lattices.
The three lattices we simulate 
have approximately the same physical size,
$mL\sim 9$, which is large enough to make finite-size effects 
much smaller than our statistical errors. This is confirmed by the analysis
of Ref. \cite{Luscher:1990ck}. 

\begin{table}
\begin{center}
\begin{tabular}{|c|c|c|c|}
\hline 
$p$ 
& ${\cal O} = (\overline{\partial}_0\sg)^2$
& ${\cal O} = (\overline{\partial}_0\sg\cdot\overline{\partial}_1\sg)$
& ${\cal O} = (\overline{\partial}_1\sg)^2$ \\
\hline
$0$ &$34.619(25)$&$0.00053(48)$ & $34.663(25)$\\
$2\pi/L$ & $34.707(18)$ &$0.03065(26)$ & $34.776(18)$\\
$4\pi/L$ & $34.735(18)$ &$0.06080(39)$ & $34.857(18)$\\
$6\pi/L$ & $34.741(26)$ &$0.09026(64)$ & $34.923(26)$\\
\hline
\end{tabular}
\end{center}
\caption{Estimates of $\sum_a\widehat{C}^{aa}_{\cal O}(p,p;10)$ for different 
operators measured on lattice \ref{Lattice128x256}. 
For $(\overline{\partial}_0\sg\cdot\overline{\partial}_1\sg)$ we report 
the imaginary part, the real part being zero. 
The matrix element of the other two operators is real.}
\label{DiagonalMatrixElement}
\end{table}
\begin{table}
\begin{center}
\begin{tabular}{|c|c|c|c|}
\hline 
$p$  
& ${\cal O} = (\overline{\partial}_0\sg)^2$
& ${\cal O} = (\overline{\partial}_0\sg\cdot\overline{\partial}_1\sg)$
& ${\cal O} = (\overline{\partial}_1\sg)^2$ \\
\hline
$2\pi/L$ & $0.25434(60)$ &
$ 0.01452(47) $&$ 0.20241(68) $\\
$4\pi/L$ & $0.25153(85)$ &
$ 0.02824(61) $&$ 0.18394(94) $\\
$6\pi/L$ & $0.25597(128)$ & 
$ 0.04024(96) $&$ 0.16643(140) $\\
\hline
\end{tabular}
\end{center}
\caption{Estimates of $\sum_a\widehat{C}^{aa}_{\cal O}(p,0;20)$ 
for different operators measured on 
lattice \ref{Lattice128x256}. We report here
the real part for $(\overline{\partial}_0\sg)^2$ and 
$(\overline{\partial}_1\sg)^2$, and the
imaginary part for $(\overline{\partial}_0\sg\cdot\overline{\partial}_1\sg)$.}
\label{OutOfDiagonalMatrixElement}
\end{table}
In order to verify the OPE, we need the values of the matrix elements
which appear in the r.h.s. of Eq. \reff{OPEesempio}.
Matrix elements of lattice operators can be computed from properly
defined three-point correlation functions. If
${\cal O}_{t,x}$ is a lattice operator,
we define the correlation function 
\begin{eqnarray}
C^{ab}_{\cal O} (\pb,\qb;2t) \equiv 
{1\over L T} \sum_{t_0=1}^T \sum_{x_0=1}^L \sum_{x_1,x_2=1}^L
e^{i\pb x_1-i\qb x_2}\<\sigma^a_{t_0-t,x_0+x_1}{\cal O}_{t_0,x_0}
\sigma^b_{t_0+t,x_0+x_2}\>,
\end{eqnarray}
and the corresponding normalized correlation
\be
\widehat{C}^{ab}_{\cal O} (\pb,\qb;2t) \equiv 
   {C^{ab}_{\cal O} (\pb,\qb;2t)\over 
    \sqrt{C(\pb;2t) C(\qb;2t)} }.
\label{NormalizedCorrelation}
\ee
The function $\widehat{C}^{ab}_{\cal O} (\pb,\qb;2t)$ has a finite limit
for $t\to\infty$. In this work we will only need the matrix elements of the 
naive lattice energy-momentum tensor \reff{Tmunulatticenaive}. For this reason,
we have computed $C^{aa}_{\cal O} (\pb,\pb;2t)$ with 
${\cal O} = \overline{\partial}_{\mu}\sg\cdot\overline{\partial}_{\rho}\sg$.
Such a correlation function has been computed on lattice \ref{Lattice128x256},
using $N_{\rm conf} = 320000$ configurations, for
$t=5,\dots,20$ and $\pb = 2n\pi/L$, $n=0,\dots,3$. 
For these observables $\widehat{C}^{ab}_{\cal O} (\pb,\pb;2t)$ 
is independent of $t$, within the statistical errors, already at $t=5$.
The results  obtained for $t=5$ are reported in Table 
\ref{DiagonalMatrixElement}. For 
${\cal O} = \overline{\partial}_0\sg\cdot\overline{\partial}_1\sg$, 
statistical errors are dominated by the error on the evaluation of 
$C^{ab}_{\cal O} (\pb,\pb;2t)$. On the other hand,
for ${\cal O} = (\overline{\partial}_\mu\sg)^2$, 
the statistical error of the numerator
in Eq. (\ref{NormalizedCorrelation}) is roughly equal to that of 
the denominator. The reason is clear: since 
in the continuum limit $(\overline{\partial}_\mu\sg)^2$ is proportional
to the identity operator, we are computing
essentially (up to $O(a^2)$ corrections) the same quantity in the numerator
and in the denominator with approximately the same statistics. 
The reported errors
on the ratios are obtained using the independent error formula. 
For $(\overline{\partial}_\mu\sg)^2$ smaller error bars could have been 
obtained by taking into account the statistical correlations between
numerator and denominator.

We also measured $C^{aa}_{\cal O}(0,\pb;2t)$ for the same operators 
on lattice \ref{Lattice128x256}, using $N_{\rm conf} = 62000$ independent
configurations, for $\pb = 2n\pi/L$ with $n=1,\dots,3$ and 
$t=5,\dots,10$. The normalized three-point function shows a plateau for
$t\gtrsim 10$ when $\mu = \nu $ and for $t\gtrsim 6$ when $\mu\neq\nu$.
The results obtained for $t=10$ 
are reported in Tab. \ref{OutOfDiagonalMatrixElement}. 
In this case the statistical errors are dominated by
the uncertainty on $C^{aa}_{\cal O}(0,\pb;2t)$.

In this paper we study the OPE of the product of two currents. 
Therefore, we have computed the following correlation functions:
\begin{eqnarray}
G^{(s)}(t,x;\pb,\qb;2t_s)&\equiv& \frac{1}{2}
\sum_{x_1,x_2}\<(\jg_{0,(0,0)}^L\cdot
\jg_{1,(t,x)}^L)(\sg_{-t_s,x_1}\cdot \sg_{t_s,x_2})\>e^{i\pb x_1-i\qb x_2},
\\
G^{(a)}_{\mu\nu}(t,x;\pb,\qb;2t_s) &\equiv&
\sum_{x_1,x_2}\sum_{abc}\<(j^{L,ac}_{\mu,(0,0)}j^{L,bc}_{\nu,(t,x)}-
j^{L,bc}_{\mu,(0,0)}j^{L,ac}_{\nu,(t,x)})
\sigma^a_{-t_s,x_1} \sigma^b_{t_s,x_2}\>\,e^{i\pb x_1-i\qb x_2},
\end{eqnarray}
where $j_{\mu,x}^{L,ab}$ is the lattice Noether current defined in
Eq. \reff{jlattice}. Of course, we averaged over lattice translations.

In particular we have measured:
\begin{itemize}
\item[a)] $G^{(s)}(t,x;\pb,\pb;2t_s)$ using 
 $N_{\rm conf} \simeq 1.3\cdot 10^6$
 configurations on lattice \ref{Lattice64x128} and using 
 $N_{\rm conf} = 58350$ independent configurations on lattice
 \ref{Lattice128x256}.
\item[b)] ${\rm Im}\, G^{(s)}(t,x;\pb,0;2t_s)$ using $N_{\rm conf} = 112000$
independent configurations on lattice \ref{Lattice128x256}.
\item[c)] 
${\rm Re}\, G^{(a)}_{11}(t,x;\pb ,-\pb ;2t_s)$
using $N_{\rm conf}\simeq 2.4\cdot 10^6$
configurations on lattice \ref{Lattice64x128},
$N_{\rm conf}= 69500$ independent configurations on
lattice \ref{Lattice128x256}, and $N_{\rm conf}= 31550$ configurations
on lattice \ref{Lattice256x512}.
\item[d)]
${\rm Im}\, G^{(a)}_{01}(t,x;\pb ,0 ;2t_s)$
using $N_{\rm conf}= 41750$ independent configurations on
lattice \ref{Lattice128x256}. 
\end{itemize}
In all cases we consider $\pb=2\pi n/L$, $n=1,2,3$; 
$t_s=7,8,9$ and $|t|\le 5,|x|\le 5$ on lattice \ref{Lattice64x128},
$t_s=10,11,12$ and $|t|\le 8,|x|\le 8$ on lattice \ref{Lattice128x256}, and
$t_s=20,23, 26$ and $|t|\le 12,|x|\le 12$ on lattice \ref{Lattice256x512}.

Using the four-point correlation function determined above, we constructed
the normalized ratios
\begin{eqnarray}
\widehat{G}^{(\cdot)} (t,x;\pb,\qb;2t_s) \equiv 
   {G^{(\cdot)} (t,x;\pb,\qb;2t_s)\over 
    \sqrt{C(\pb;2t_s) C(\qb;2t_s)} },
\end{eqnarray}
which have a finite limit for $t_s\to \infty$.
We verified that $\widehat{G}^{(\cdot)} (t,x;\pb,\qb;2t_s)$ is independent
of $t_s$ in the range considered, and thus we have taken the estimate
corresponding to the lowest considered value of $t_s$ as an estimate
of $\widehat{G}^{(\cdot)} (t,x;\pb,\qb;\infty)$.

In the paper we will usually consider averages over two-dimensional 
rotations, i.e., given a function $f(z_t,z_x)$, we consider 
\begin{eqnarray}
\overline{f}(r) \equiv 
   {\sum_{z\in\mathbb{Z}^2} f(z)\, \Theta_r(z)  \over 
    \sum_{z\in\mathbb{Z}^2} \Theta_r(z) },\quad
  \Theta_r(z) \equiv  \theta\left(|z| - r+\frac{1}{2}\right) 
                      \theta\left(r+\frac{1}{2} -|z|\right) ,
\label{definizione-mediarotazionale}
\end{eqnarray}
with $|z|\equiv\sqrt{z_t^2+z_x^2}$ and $r=n+1/2$, $n$ integer.

\subsection{One-particle states} \label{sec4.2}

In the conventional picture
the lowest states of the model are one-particle
states transforming as $O(N)$ vectors. On a lattice of finite spatial 
extent $L$, we normalize the states and the fields as follows:
\begin{eqnarray}
\<\pb,a|\qb,b\> &=& 2\omega(\pb)L\,\delta^{ab}\delta_{\pb,\qb}, \\
\<\pb,a|\hat{\sigma}^b_x|0\> &=& 
 \sqrt{\frac{Z(\pb)}{N}}\,\delta^{ab}e^{ i\pb x}.
\end{eqnarray}
The function $\omega(\pb)$, which is the energy of the state $|\pb,a\>$, and 
the field renormalization $Z(\pb)$ can be determined from the 
large-$|t|$ behavior of the two-point function $C(\pb;t)$:
\begin{eqnarray}
C(\pb;t)\sim \frac{Z(\pb)}{2\omega(\pb)}e^{-\omega(\pb)t}
\qquad
\mbox{for}
\quad
t\gg 1.
\label{AsymptoticCorrelation}
\end{eqnarray}
In the continuum (scaling) limit we have $Z(\pb)=Z$ independent of 
$\pb$ and $\omega(\pb) = \sqrt{\pb^2 + m^2}$. In Table \ref{LatticeSpectrum}
we report our results for the three lattices. 
In order to evaluate $\omega(\pb)$ and $Z(\pb)$ 
we determined effective values at time $t$ by solving the equations
\begin{eqnarray}
\frac{C(\pb;t+1)}{C(\pb;t)}&\equiv&
\frac{\cosh[\omega_{\rm eff}(\pb,t)(t+1-T/2)]}
{\cosh[\omega_{\rm eff}(\pb,t)(t-T/2)]}, 
\\
C(\pb;t)&\equiv &\frac{Z_{\rm eff}(\pb,t)}{2\omega_{\rm eff}(\pb,t)}
\left\{ e^{-\omega_{\rm eff}(\pb,t)t}
+e^{-\omega_{\rm eff}(\pb,t)(T-t)}\right\}.
\end{eqnarray}
Then we looked for a plateau in the plot of 
$\omega_{\rm eff}(\pb,t)$ and $Z_{\rm eff}(\pb,t)$ versus $t$. 
Both functions become independent of $t$ for $t\gtrsim \xi$.
The values reported in Table \ref{LatticeSpectrum} correspond to 
one particular value of $t$ of order $\xi^{\rm exp}$: 
$t = 8, 16, 20$ respectively for
lattice \ref{Lattice64x128}, \ref{Lattice128x256}, \ref{Lattice256x512}.
For the two largest lattices there is evidence of scaling at the 
error-bar level. 
Instead, for lattice \ref{Lattice64x128} 
there are tiny scaling violations, that are however so small 
(at most 1\%) that we can neglect them in the following discussion. 

One can also investigate asymptotic 
scaling, i.e. the dependence of $\omega(\pb)$ and $Z$ on $g_L$. 
The dependence of $\omega(\pb)$ can be determined from 
Eq.~\reff{MassGapLattice}. There exists also an exact prediction for $Z$,
including the nonperturbative constant 
\cite{Balog:NPB97,Campostrini:PLB97}. However, as is well known, 
lattice perturbation theory is not predictive at these values of the 
correlation length, and indeed, the perturbative four-loop predictions 
show large discrepancies compared to the numerical data.
The agreement is instead quite good \cite{Caracciolo:1995ah,Campostrini:PLB97}
if one uses the improved coupling $g_E$ defined in 
Eq. \reff{DressedCouplingEnergy}.
\begin{table}
\footnotesize
\begin{center}
\begin{tabular}{|c|c|c|c|c|c|c|}
\hline
 &\multicolumn{2}{|c|}{lattice \ref{Lattice64x128}}&
\multicolumn{2}{|c|}{lattice \ref{Lattice128x256}}&
\multicolumn{2}{|c|}{lattice \ref{Lattice256x512}}\\
\hline
$\pb$ & $\omega(\pb)$ & $Z(\pb)$ & $\omega(\pb)$ & $Z(\pb)$
& $\omega(\pb)$ & $Z(\pb)$\\
\hline
$0$      & $  0.145393(40)$ & $1.6593(8)$
         & $  0.073327(55)$ & $1.3563(18)$       
         & $  0.036963(34)$ & $1.1295(14)$\\
$2\pi/L$ & $  0.175380(40)$ & $1.6582(8)$
         & $  0.088244(67)$ & $1.3557(21)$
         & $  0.044348(35)$ & $1.1284(13)$\\
$4\pi/L$ & $  0.243657(74)$ & $1.6510(13)$
         & $  0.12263(15)$  & $1.3591(42)$
         & $  0.061456(66)$ & $1.1292(21)$\\
$6\pi/L$ & $  0.326327(192)$& $1.6378(32)$
         & $  0.16415(32) $ & $1.3499(86)$ 
         & $  0.082494(109)$& $1.1336(35)$\\
 \hline
\end{tabular}
\end{center}
\caption{The one-particle spectrum and the field normalization
for lattices \ref{Lattice64x128}, \ref{Lattice128x256}, \ref{Lattice256x512}.}
\label{LatticeSpectrum}
\end{table}

\subsection{Nonperturbative renormalization constant for the lattice 
energy-momentum tensor} \label{sec4.3}
In this Section we want to compute nonperturbatively the renormalization
constant of the lattice energy-momentum tensor \reff{Tmunulattice}. 
In general, given an operator $\cal O$ on the lattice we define
its matrix element by
\be
\<\pb,a|\hat{\cal O}|\qb,b\> \equiv 
    N \sqrt{4\omega(\pb)\omega(\qb)} 
  \lim_{t\to\infty} \widehat{C}^{ab}_{\cal O}(\pb,\qb;2t).
\ee
For $\overline{\partial}_\mu\bsigma\cdot \overline{\partial}_\nu\bsigma$ 
the matrix elements can be obtained from the results given in Tables 
\ref{DiagonalMatrixElement} and \ref{OutOfDiagonalMatrixElement}.
The matrix elements of $(\overline{\partial}_\mu\bsigma)^2$ are dominated 
by the mixing with the identity operator and thus, in order to define
the renormalized operator for $\mu=\nu$, we should perform a 
nonperturbative subtraction of the large $1/a^2$ term, which is practically
impossible.\footnote{In general we expect 
$\<\pb,a|(\overline{\partial}_0\bsigma)^2|\pb,b\> = 
2AL\sqrt{\pb^2+m^2}+B - C(\pb^2 + m^2)$ and 
$\<\pb,a|(\overline{\partial}_1\bsigma)^2|\pb,b\> = 
2AL\sqrt{\pb^2+m^2}+B + C \pb^2$. The quantity
we are interested in is $C$. However, from Table \ref{DiagonalMatrixElement}
we immediately realize that much smaller errors are required 
to really observe the momentum dependence of the matrix elements 
and thus to determine the constant $C$. }
Therefore, we only compute the renormalized operator for $\mu\not=\nu$,
which amounts to determining the constant $Z^{L,(2,0)}_{TT}$. 
This constant is obtained by requiring
\be
\<\pb,a| T^{\rm latt}_{01,0} |\pb,b\> =\, 
2 i \,\pb \sqrt{\pb^2 + m^2}\, \delta^{ab}.
\label{EnergyMomentumCondition}
\ee
In practice, we first compute an effective (momentum-dependent) 
renormalization constant
\be
\zeta(\pb) \equiv {2\, \pb\, \omega(\pb)\,g_L \over 
     {\rm Im}\, \<\pb,a|(\overline{\partial}_0\sg\cdot
\overline{\partial}_1\sg)  |\pb,a\>},
\ee
which, in the continuum limit, becomes independent of $\pb$ and converges to
$Z^{L(2,0)}_{TT}$. Using the data of Table \ref{DiagonalMatrixElement},
on lattice \ref{Lattice128x256}, 
we obtain $\zeta(\pb) = 1.040(8)$, $1.048(6)$, $1.059(7)$ 
for $\pb = 2\pi/L$, $4\pi/L$, $6\pi/L$ respectively. 
Clearly the scaling corrections are small, and thus we can estimate 
$Z^{L(2,0)}_{TT} = 1.05(2)$ on this lattice.
This compares very well with the result 
of one-loop lattice perturbation theory given in Eq. \reff{EMZLatticePT},
which yields $Z^{L(2,0)}_{TT} \simeq 1.044357$, 
and with the result of ``improved'' (sometimes called ``boosted") 
perturbation theory in terms of the coupling $g_E$, cf. Eq. 
\reff{DressedCouplingEnergy}, 
$Z^{L(2,0)}_{TT} = 1.052471(3)$ (the error is due to the error on $g_E$).
%
%

\subsection{OPE for the scalar product of currents} \label{sec4.4}

In this Section we consider the product $j_{0,0}^{ab} j_{1,x}^{ab}$ averaged 
over rotations. Using Eq. \reff{scalcorrcont}, we have in the continuum
scheme 
\be
\frac{1}{2}\,\overline{j_{0,0}^{ab} j_{1,x}^{ab}} = 
  \left[{1\over4} W_2(r/2) + W_3(r/2) + W_4(r/2)\right]
  {1\over g} \opl  T_{01}(0) \opr .
\ee
All other contributions vanish after the angular average. Using the results
of App. \ref{secB.1} we have at one loop in the \MS scheme
\be
\frac{1}{2}\,\overline{j_{0,0}^{ab} j_{1,x}^{ab}} = \,
\left[1 - {N-2\over 2\pi} g \left(\log\left({{\mu} r\over2}\right) + 
      \gamma - {5\over 4}\right) + O(g^2)\right] 
  {1\over g} \opl  T_{01}(0) \opr,
\ee
where $\mu$ is the renormalization scale and $\gamma$ Euler's constant. 
We will not use this form of the OPE expansion,
but instead the RG-improved Wilson coefficients. 
Thus, cf. Sec. \ref{sec3.3}, we write 
\be
\frac{1}{2}\,\overline{j_{0,0}^{ab} j_{1,x}^{ab}} = 
   W_{RGI}(\overline{g}(\Lambda_\MMS r))  \opl  T_{01}(0) \opr,
\ee
where 
\be 
W_{RGI}(\overline{g}) = \frac{1}{\overline{g}}
\left[1 + {5(N-2)\over 8\pi} \overline{g}\right] ,
\label{Wilson1ms}
\ee
and $\overline{g}(\Lambda_\MMS r)$ is the running coupling constant defined 
by\footnote{This equation should be intended as follows: 
$\Lambda_{\MMS}(1,g) = \Lambda_{\MMS}(\mu,g)/\mu$ in the right-hand side
is assumed given, while for $\Lambda_{\MMS}(1,\overline{g})$ 
we intend its expression in terms of the $\beta$-function,
Eq. \reff{LambdaDefinitionMS}.}
\be
\Lambda_{\MMS}(1,\overline{g}) = \, \Lambda_{\MMS}(1,g) 
   {r e^{\gamma}\over 2}.
\label{gdix}
\ee
Using the perturbative expression 
\reff{overlinegdef}, we can also rewrite \reff{Wilson1ms} as 
\begin{eqnarray}
W_{RGI}(r) = \beta_0 z +\left[\frac{\beta_1}{\beta_0}\log z+
\frac{5(N-2)}{8\pi}\right] =\, 
{N-2\over2\pi} z + \left[{1\over 2\pi}\log z + \frac{5(N-2)}{8\pi}\right]\ ,
\label{Wilson2ms}
\end{eqnarray}
where $z = -\log(\Lambda_{\MMS} r e^{\gamma}/2)$.

We will also use the lattice Wilson coefficients defined in Sec. \ref{sec3.2}. 
Using the one-loop results 
of App. \ref{secB.3} and the general expressions reported in Sec. \ref{sec3.3},
proceeding as before, we obtain the prediction
\be
\frac{1}{2}\,\overline{j_{0,0}^{L,ab} j_{1,x}^{L,ab}} = 
U(g_L) W^L_{RGI}(\overline{g}_L(\Lambda^{\rm latt} r))   T^L_{01,0},
\ee
where, at one loop, 
\begin{eqnarray}
W^L_{RGI}(\overline{g}_L) & = & \frac{1}{\overline{g}_L}\left[ 1 + 
    \left({5N-2\over 8\pi}-\frac{1}{4}\right) \overline{g}_L\right],
\label{WRGIlatt}\\
U(g_L) & = & 1+\left(\frac{1}{\pi}-\frac{1}{4}\right)g_L,
\label{Ulatt}
\end{eqnarray}
and $\overline{g}_L(\Lambda^{\rm latt} r)$ is the running coupling 
constant defined by
\be
\Lambda^{\rm latt}(1,\overline{g}_L) =   
  \, \Lambda^{\rm latt} (1,g_L)\, r e^\gamma \sqrt{8},
\label{eq:gbarraL}
\ee
where $\Lambda^{\rm latt}$ is the lattice 
$\Lambda$-parameter \reff{Lambdalatt}.

Finally we shall test perturbation theory in the ``improved'' expansion 
parameter $g_E$ defined in Eq. \reff{DressedCouplingEnergy}. 
The OPE becomes:
\be
\frac{1}{2}\,\overline{j_{0,0}^{L,ab} j_{1,x}^{L,ab}} = 
U(g_E) W^E_{RGI}(\overline{g}_E(\Lambda_E r))   T^L_{01,0},
\ee
where
\begin{eqnarray}
W^E_{RGI}(\overline{g}_E) & = & \frac{1}{\overline{g}_E}\left[ 1 + 
    \left({5(N-2)\over 8\pi}-\frac{1}{8}\right) \overline{g}_E\right] \ ,
\label{WRGIE}
\end{eqnarray}
$U(\cdot)$ is the same as in Eq. (\ref{Ulatt}), 
and $\overline{g}_E(\Lambda_E r)$ is the running coupling constant  defined by
\be
\Lambda_E(1,\overline{g}_E) =   
  \, \Lambda_E (1,g_E)\, r e^\gamma \sqrt{8},
\label{eq:gbarraE}
\ee
where $\Lambda_E$ is the $\Lambda$-parameter \reff{LambdaEdef}. 

\begin{figure}
\begin{tabular}{c}
\hspace{2cm}
\epsfig{figure=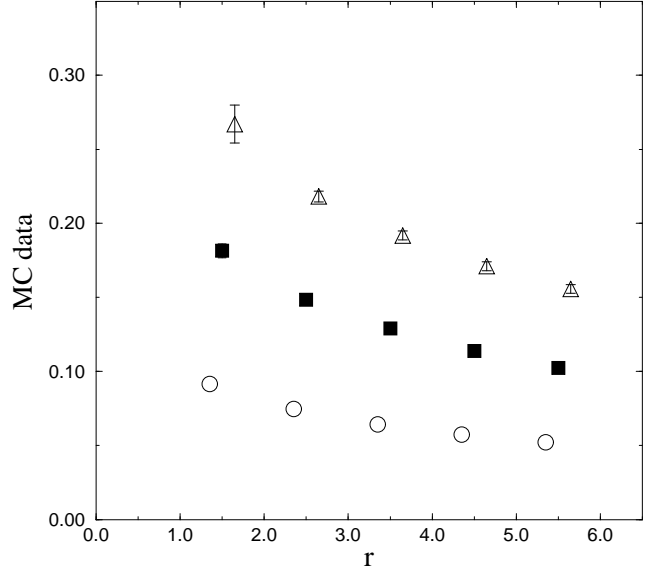,angle=-90,
width=0.6\linewidth} \\
\hspace{2cm}
\epsfig{figure=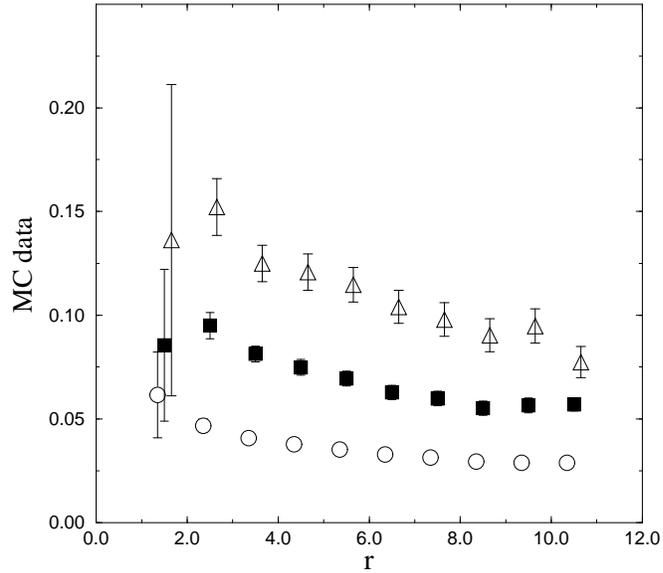,angle=-90,
width=0.6\linewidth}
\end{tabular}
\caption{Estimates of ${\rm Im}\, \widehat{G}^{(s)}(t,x;\pb,\pb;2 t_s)$
averaged over rotations on lattice \ref{Lattice64x128} (upper graph,
here $t_s = 10$) and
\ref{Lattice128x256} (lower graph, $t_s = 6$). Circles,
filled squares, and triangles correspond to $\pb=2\pi/L$, $4\pi/L$,
and $6\pi/L$ respectively. }
\label{ScalarCurrentsMCData}
\end{figure}

\begin{figure}
\begin{tabular}{cc}
\hspace{0.0cm}\vspace{-2.5cm}\\
\hspace{-0.5cm}
\epsfig{figure=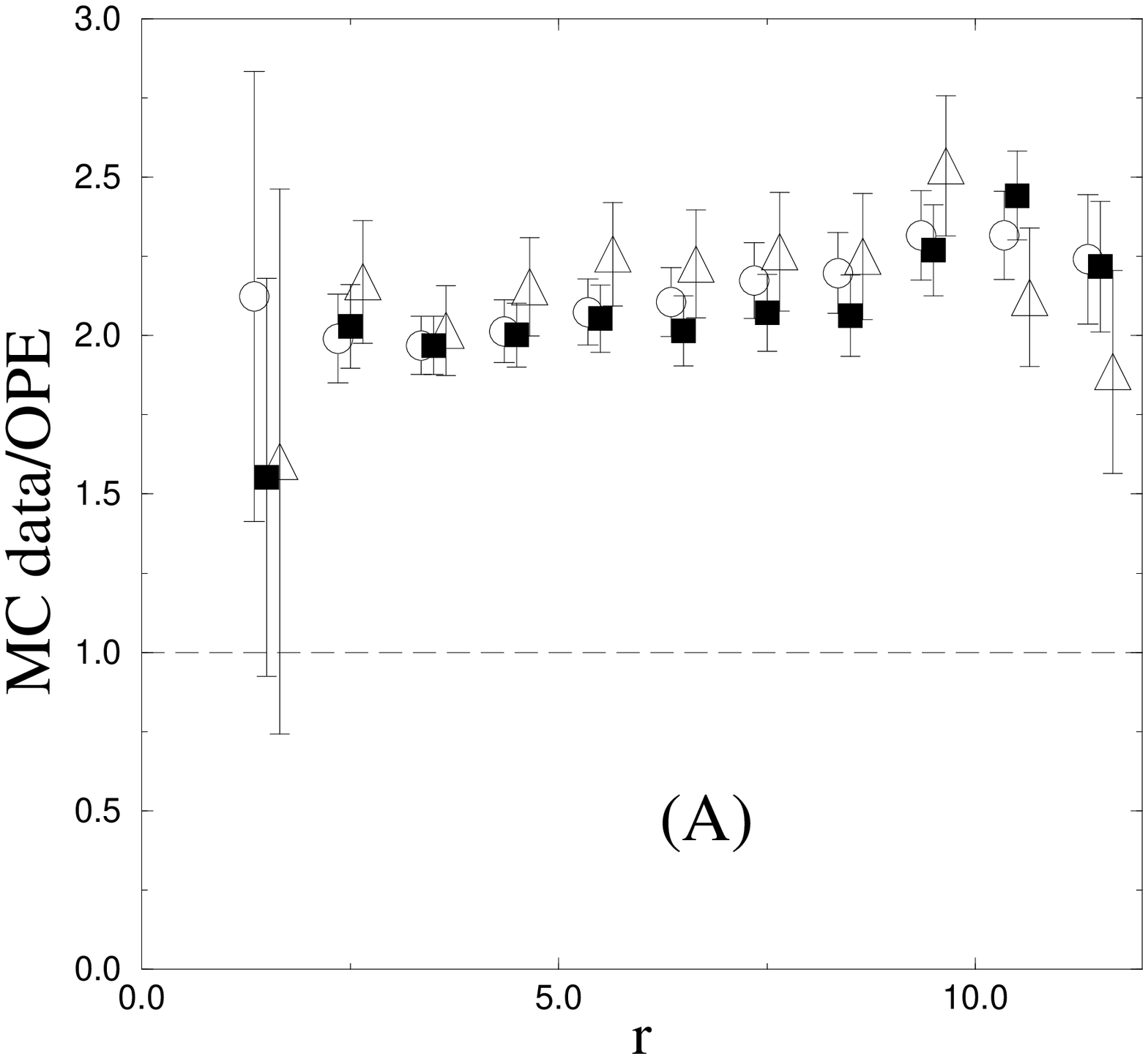,angle=-90,
width=0.5\linewidth}&\hspace{-0.5cm}
\epsfig{figure=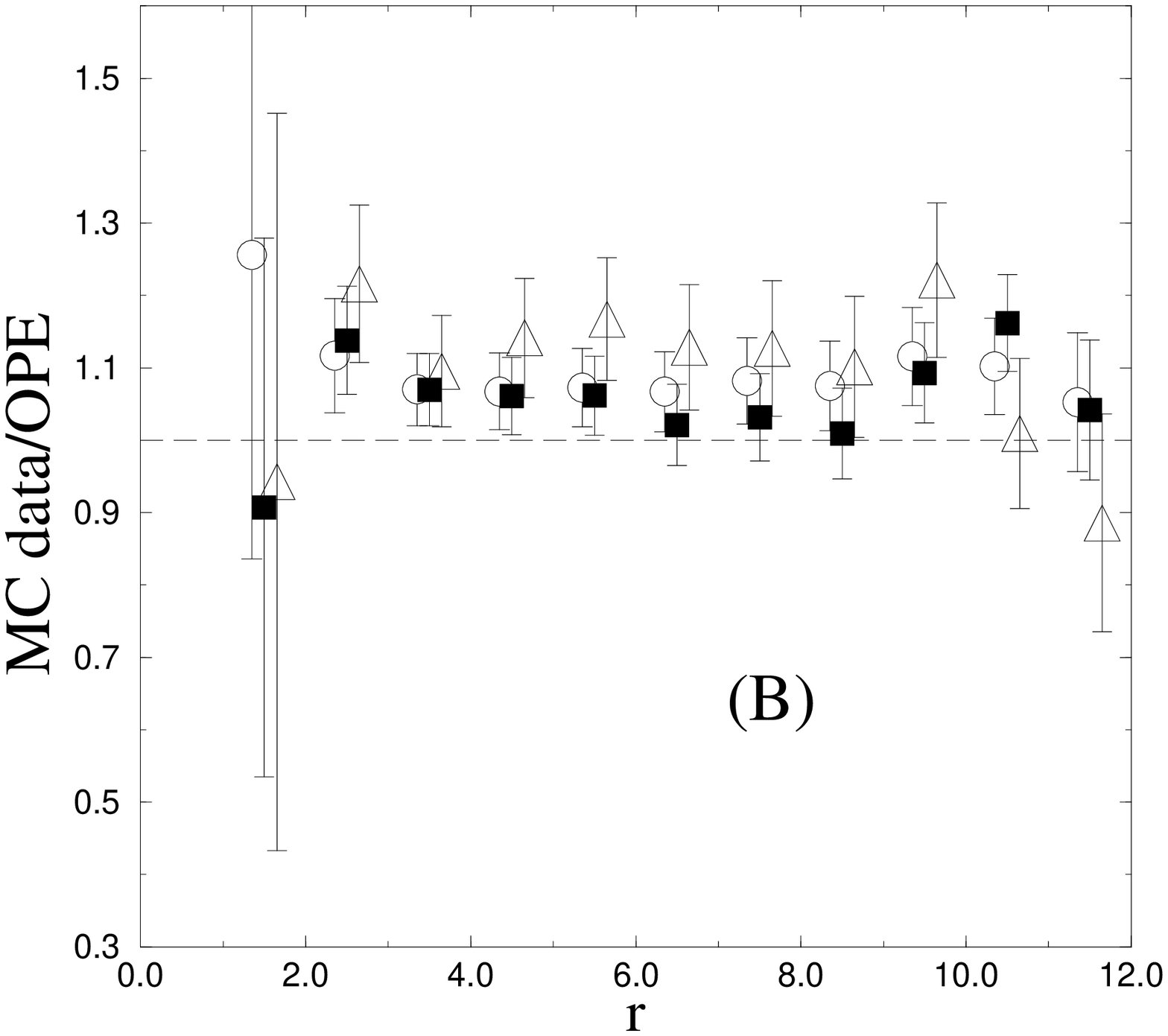,angle=-90,
width=0.5\linewidth}\vspace{-1cm}\\
\hspace{-0.5cm}
\epsfig{figure=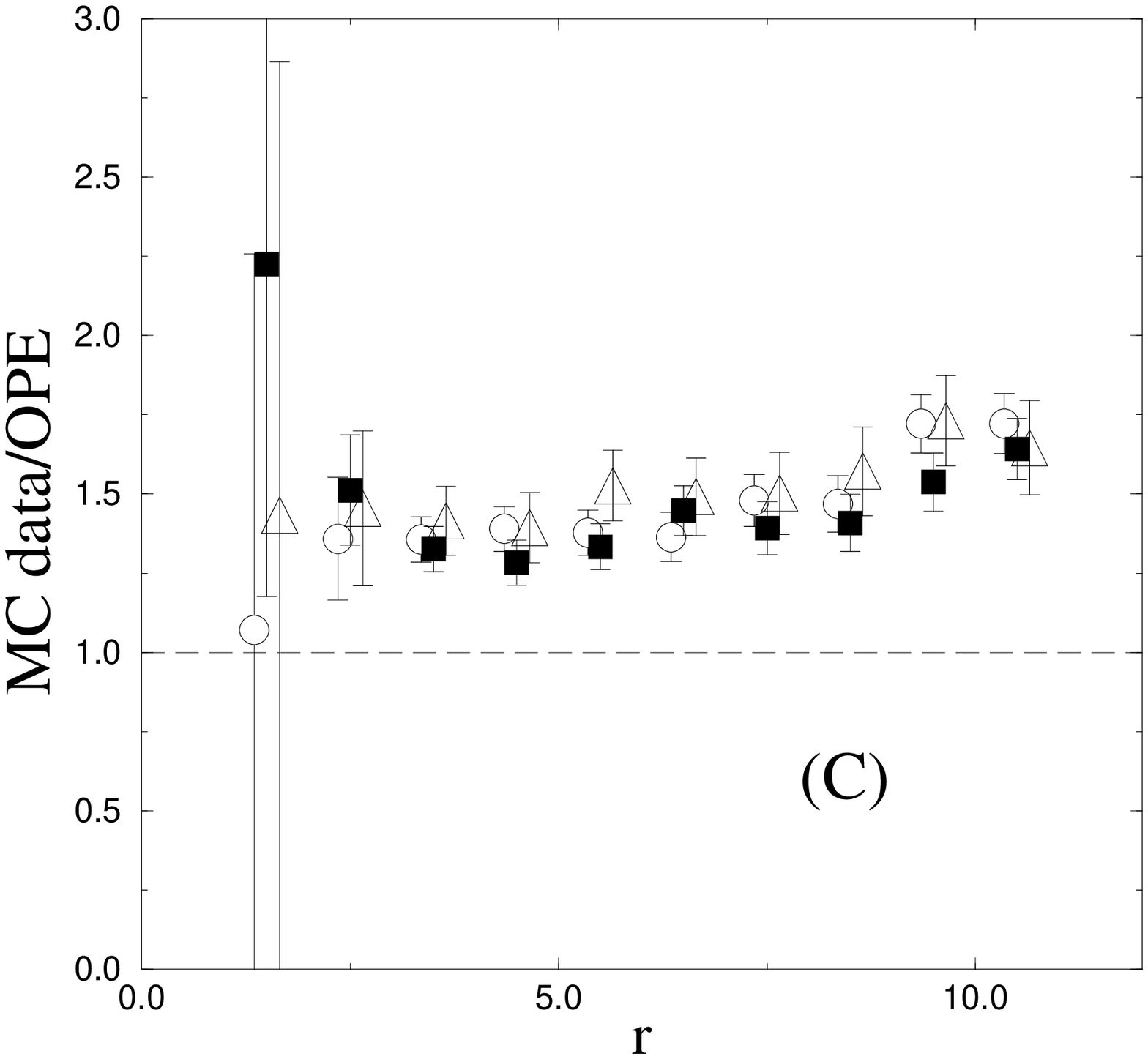,angle=-90,
width=0.5\linewidth}&\hspace{-0.5cm}
\epsfig{figure=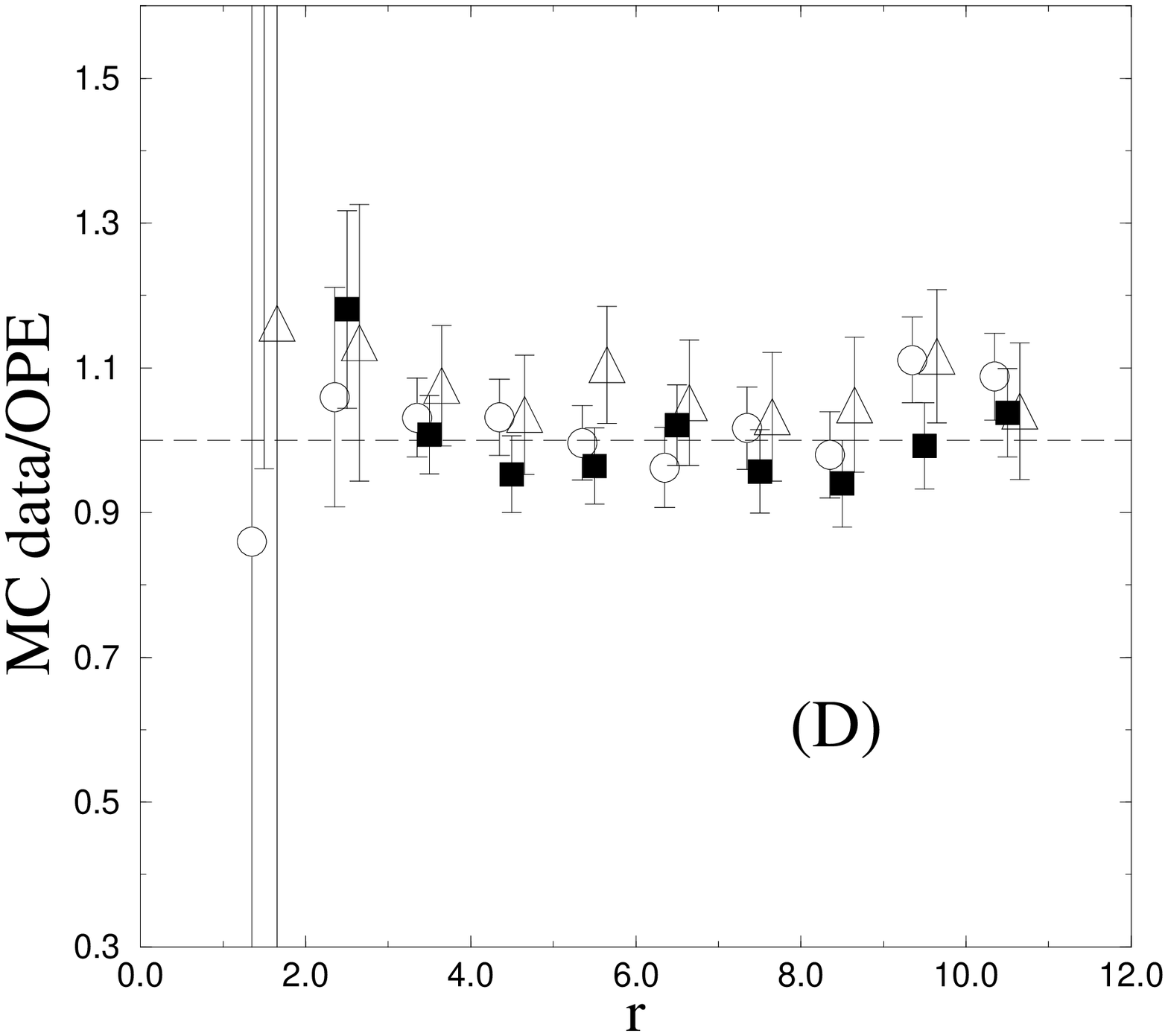,angle=-90,
width=0.5\linewidth}\vspace{-1cm}\\
\hspace{-0.5cm}
\epsfig{figure=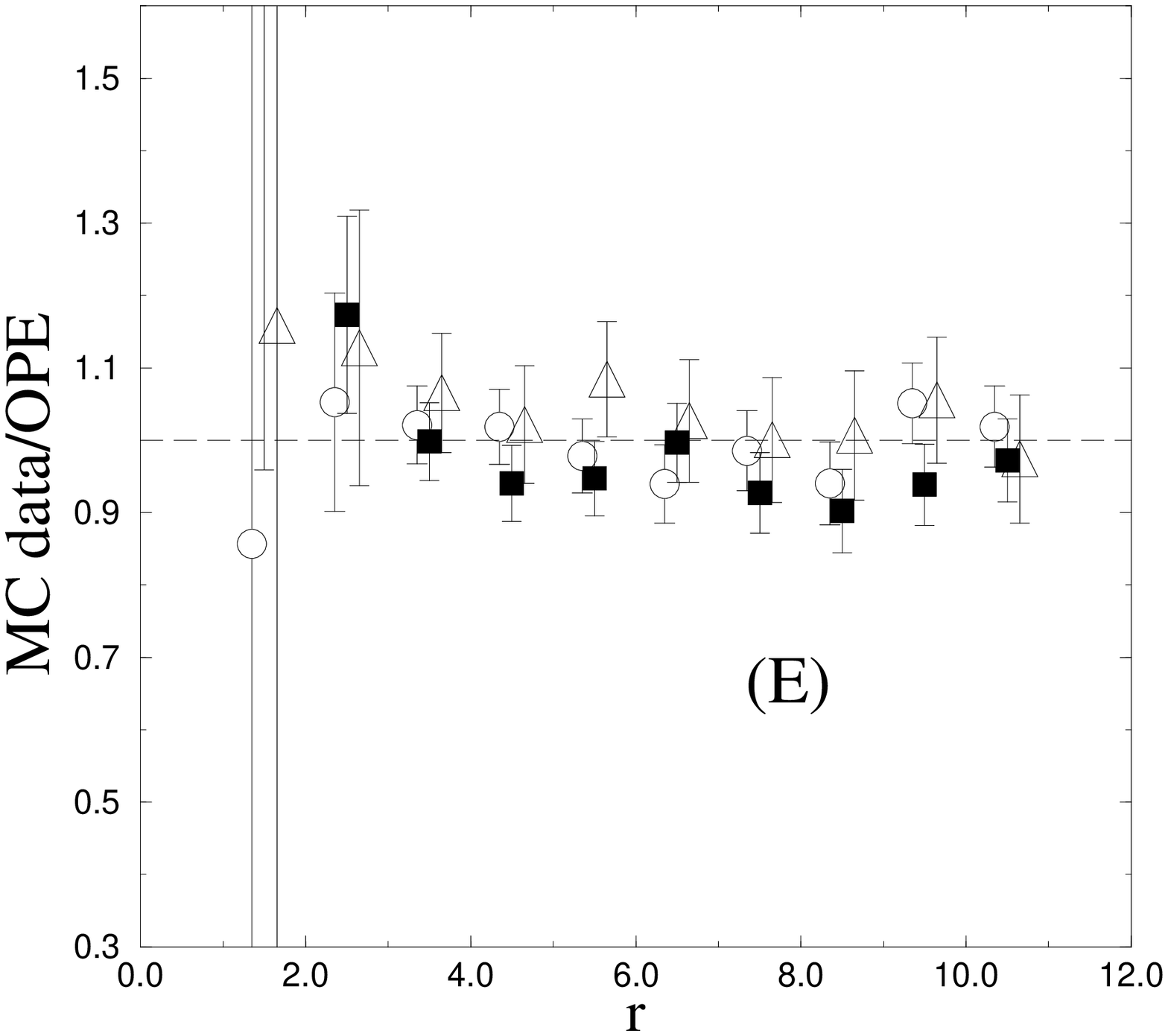,angle=-90,
width=0.5\linewidth}&\hspace{-0.5cm}
\epsfig{figure=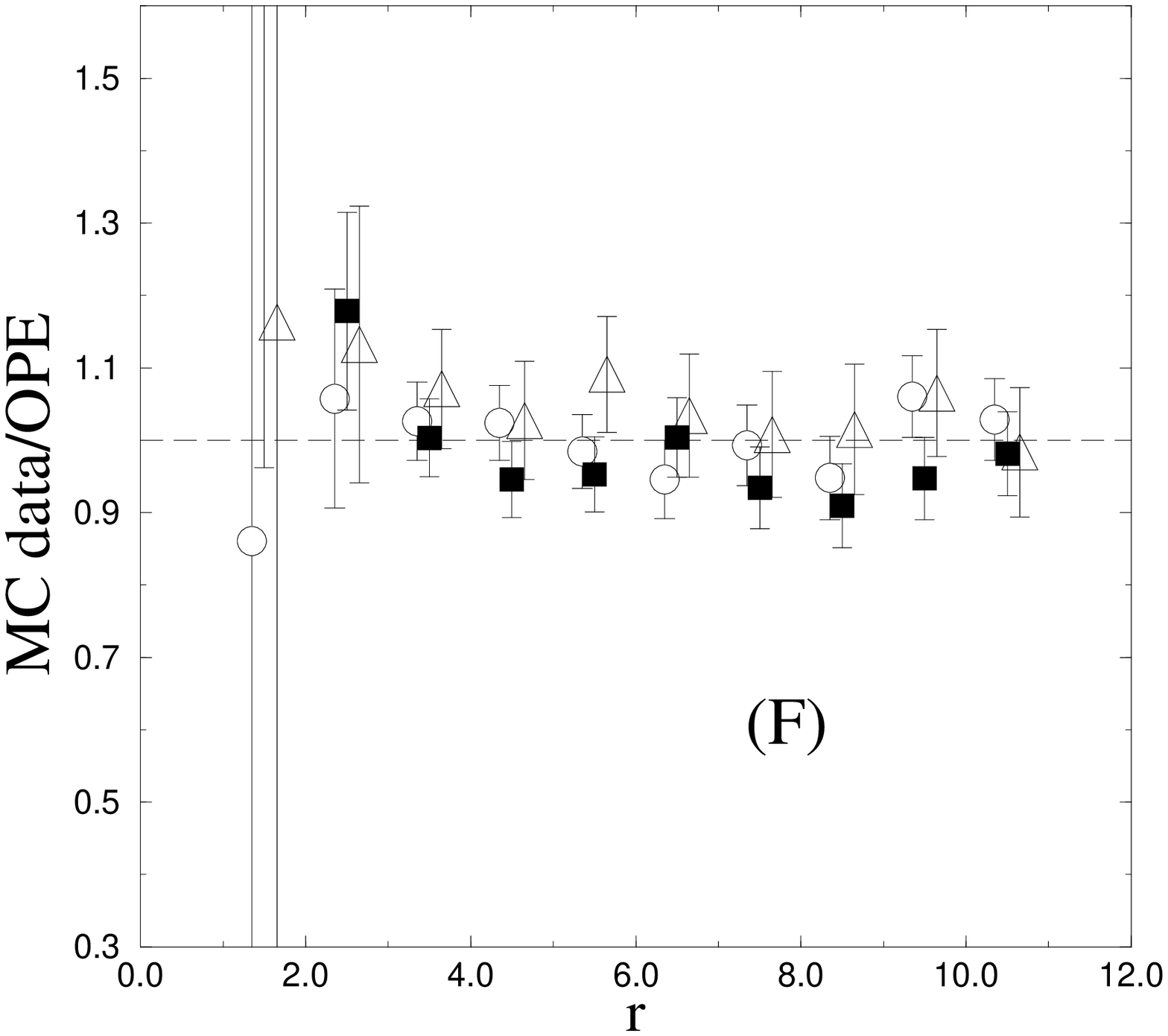,angle=-90,
width=0.5\linewidth}
\end{tabular}
\caption{The scalar product of two Noether currents compared with the OPE 
prediction: graphs of $R(r)$, cf. Eq. \protect\reff{defRatioR}, obtained
using \MS RG-improved perturbation theory. Circles,
filled squares, and triangles correspond to $\pb=2\pi/L$, $4\pi/L$,
and $6\pi/L$ respectively. The data are for lattice 
\ref{Lattice128x256}, $\xi^{\rm exp} = 13.636(10)$.}
\label{ScalarCurrentsMS}
\end{figure}

\begin{figure}
\begin{tabular}{cc}
\hspace{-2.5cm}
\epsfig{figure=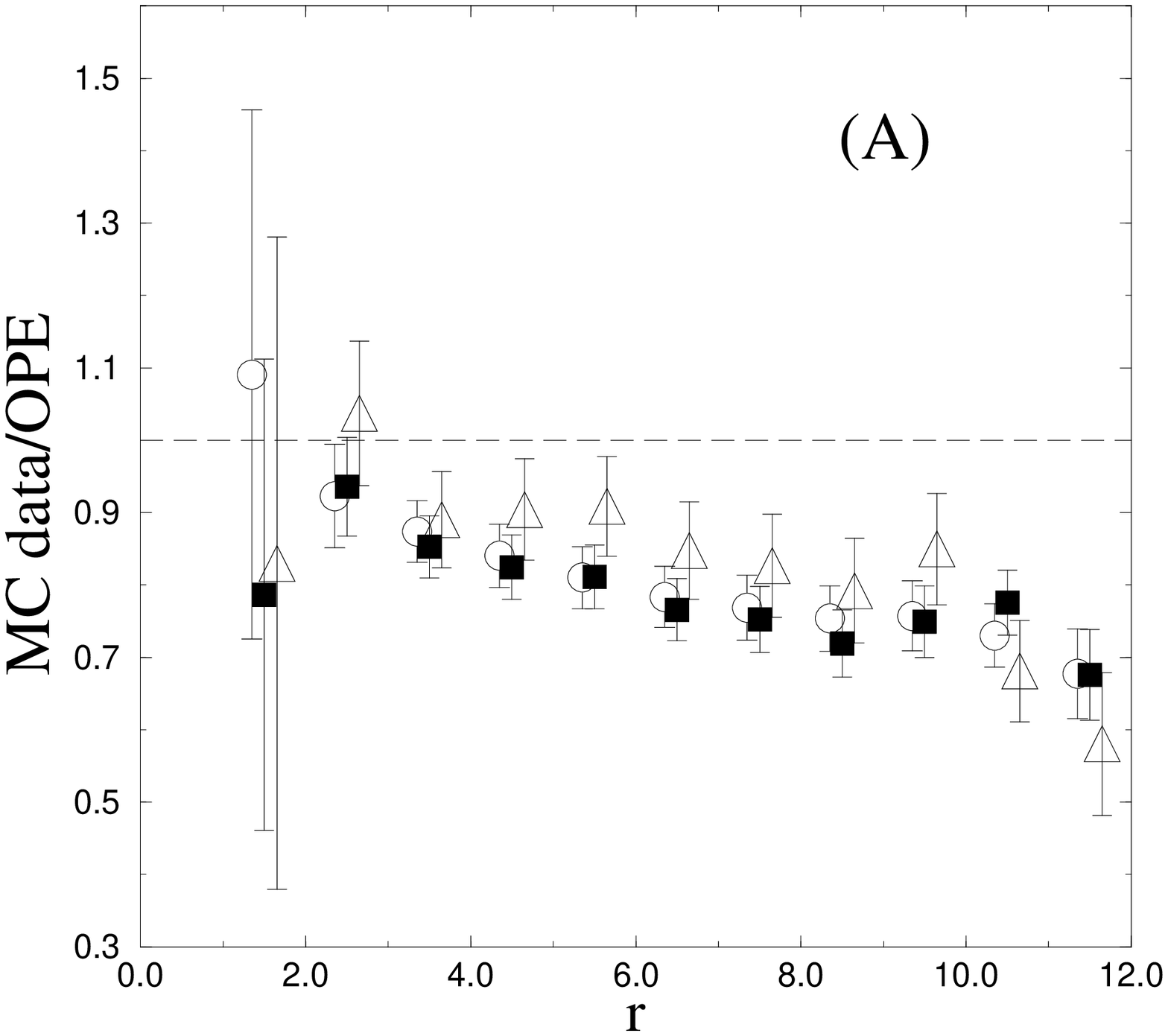,angle=-90,
width=0.6\linewidth}&
\hspace{-0.5cm}
\epsfig{figure=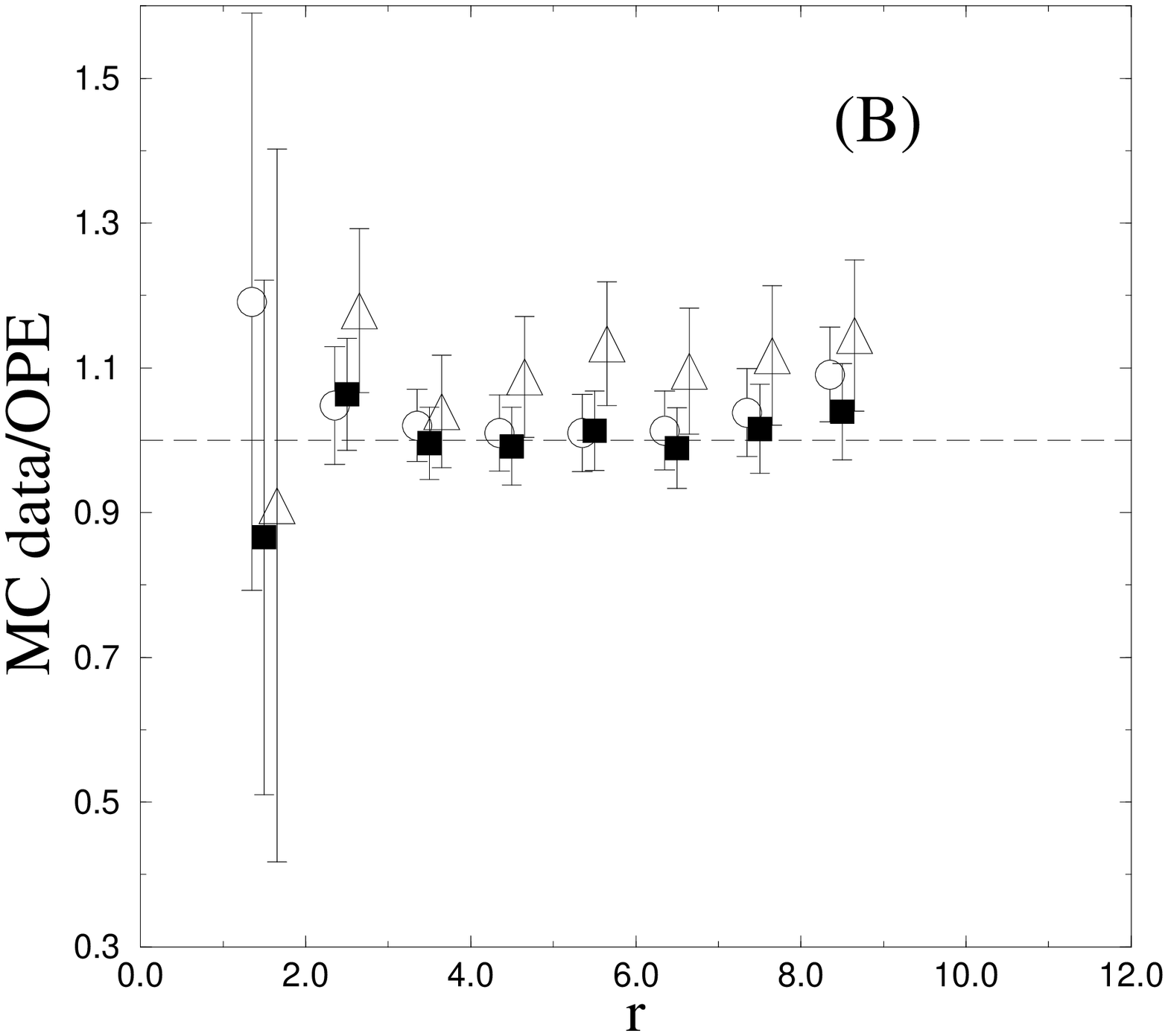,angle=-90,
width=0.6\linewidth}\vspace{-1cm}\\
\hspace{-2.5cm}
\epsfig{figure=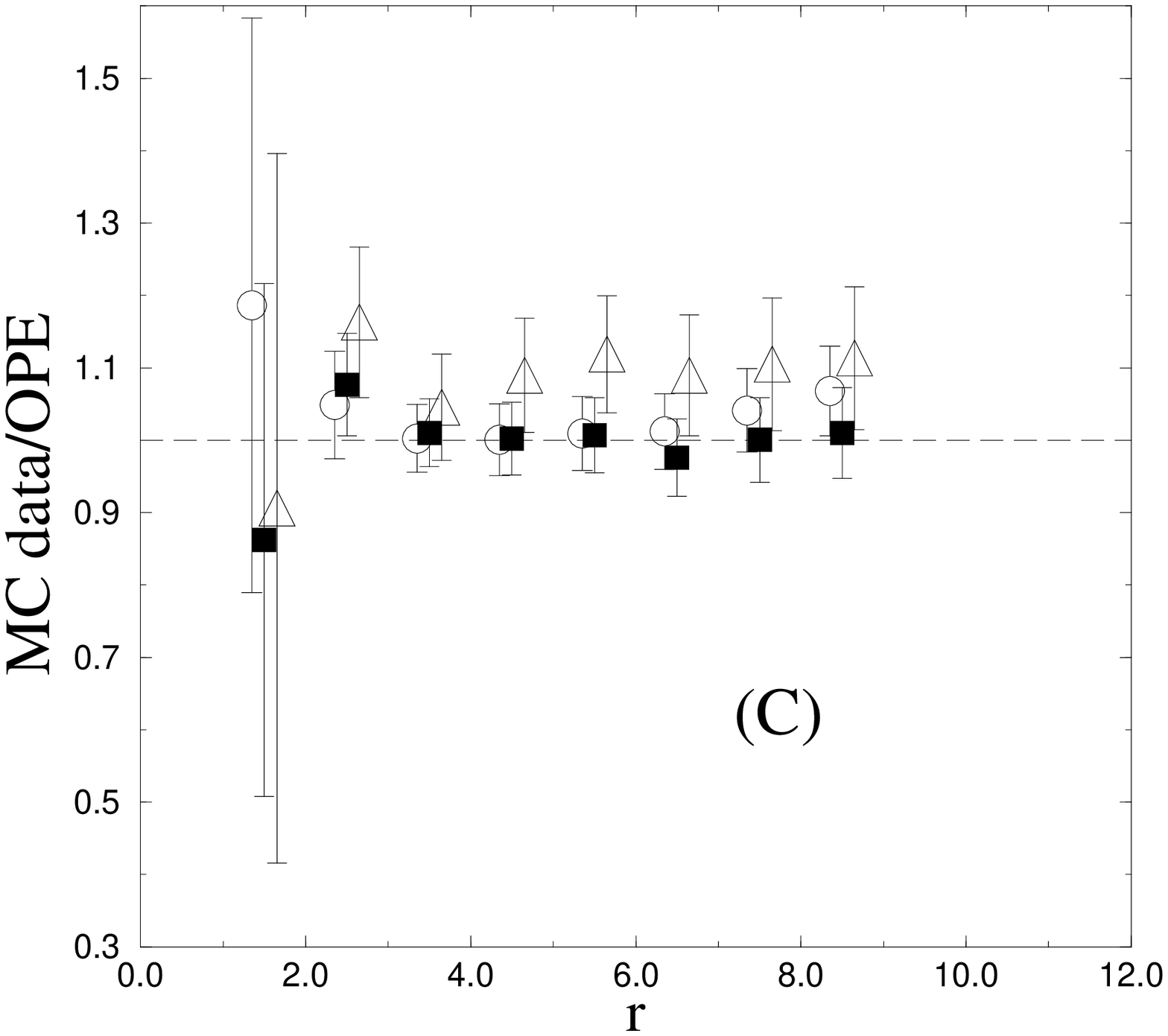,angle=-90,
width=0.6\linewidth}&
\hspace{-0.5cm}
\epsfig{figure=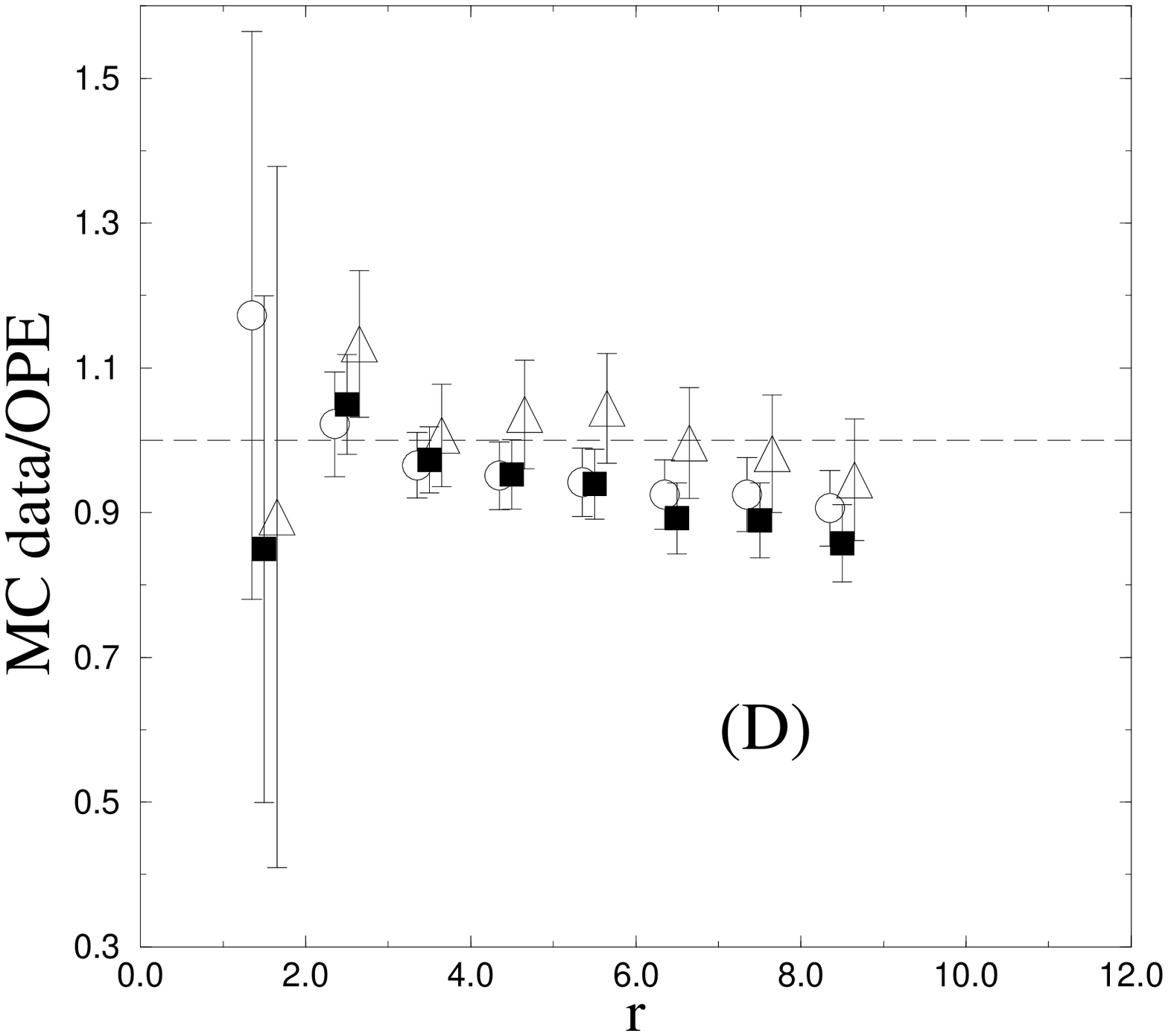,angle=-90,
width=0.6\linewidth}
\end{tabular}
\caption{The scalar product of two Noether currents compared with the OPE 
prediction: graphs of Eq. $R^{\rm latt}(r)$, cf. \protect\reff{defRatioRlatt}, 
obtained using RG-improved perturbation theory in the coupling 
$g_L$ and in the improved coupling $g_E$. Circles,
filled squares, and triangles correspond to $\pb=2\pi/L$, $4\pi/L$,
and $6\pi/L$ respectively. The data are for lattice 
\ref{Lattice128x256}, $\xi^{\rm exp} = 13.636(10)$.}
\label{ScalarCurrentLatticePT}
\end{figure}

\begin{figure}
\begin{tabular}{cc}
\hspace{-1.5cm}
\epsfig{figure=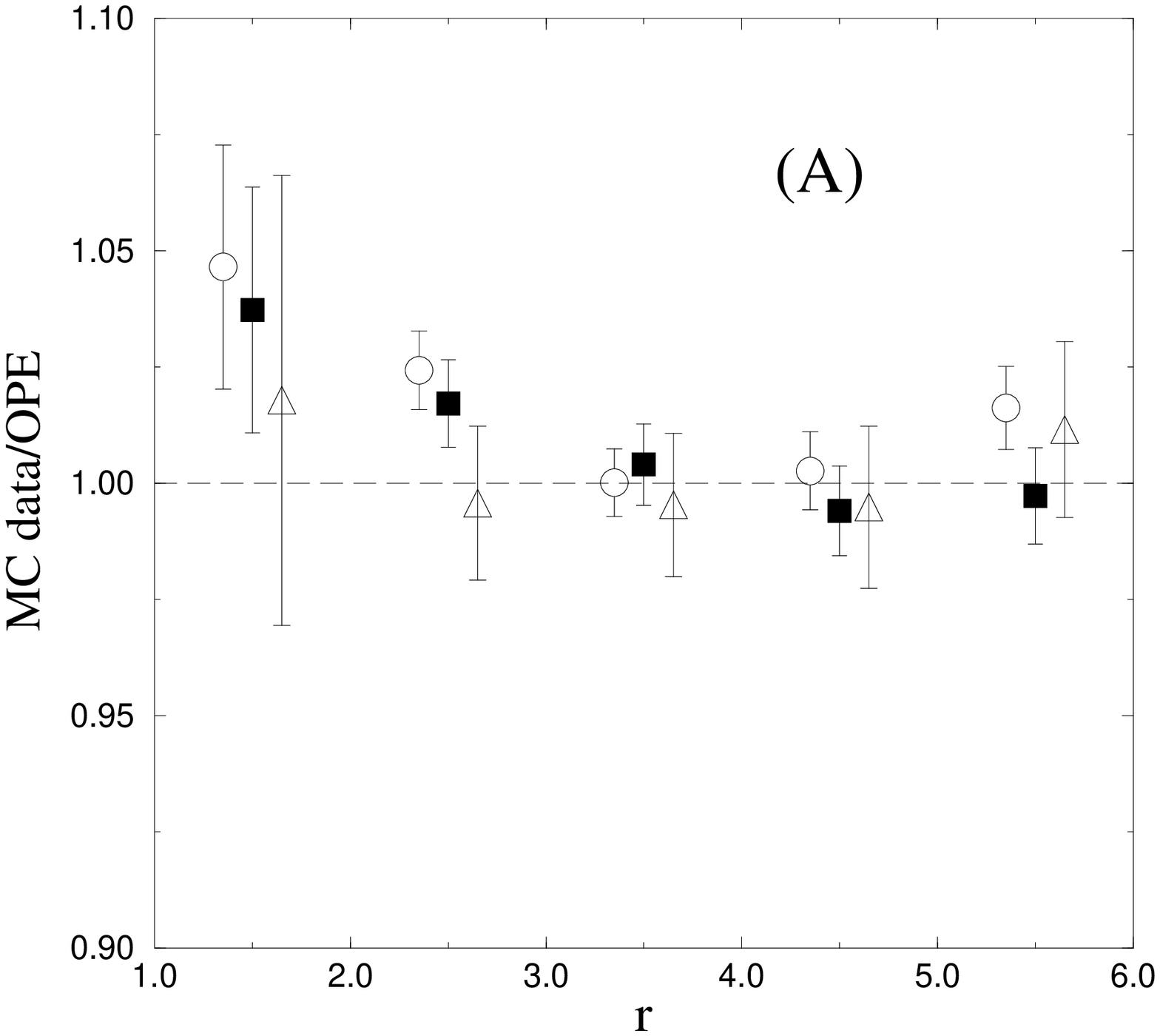,angle=-90,
width=0.55\linewidth}
&
\hspace{-0.5cm}
\epsfig{figure=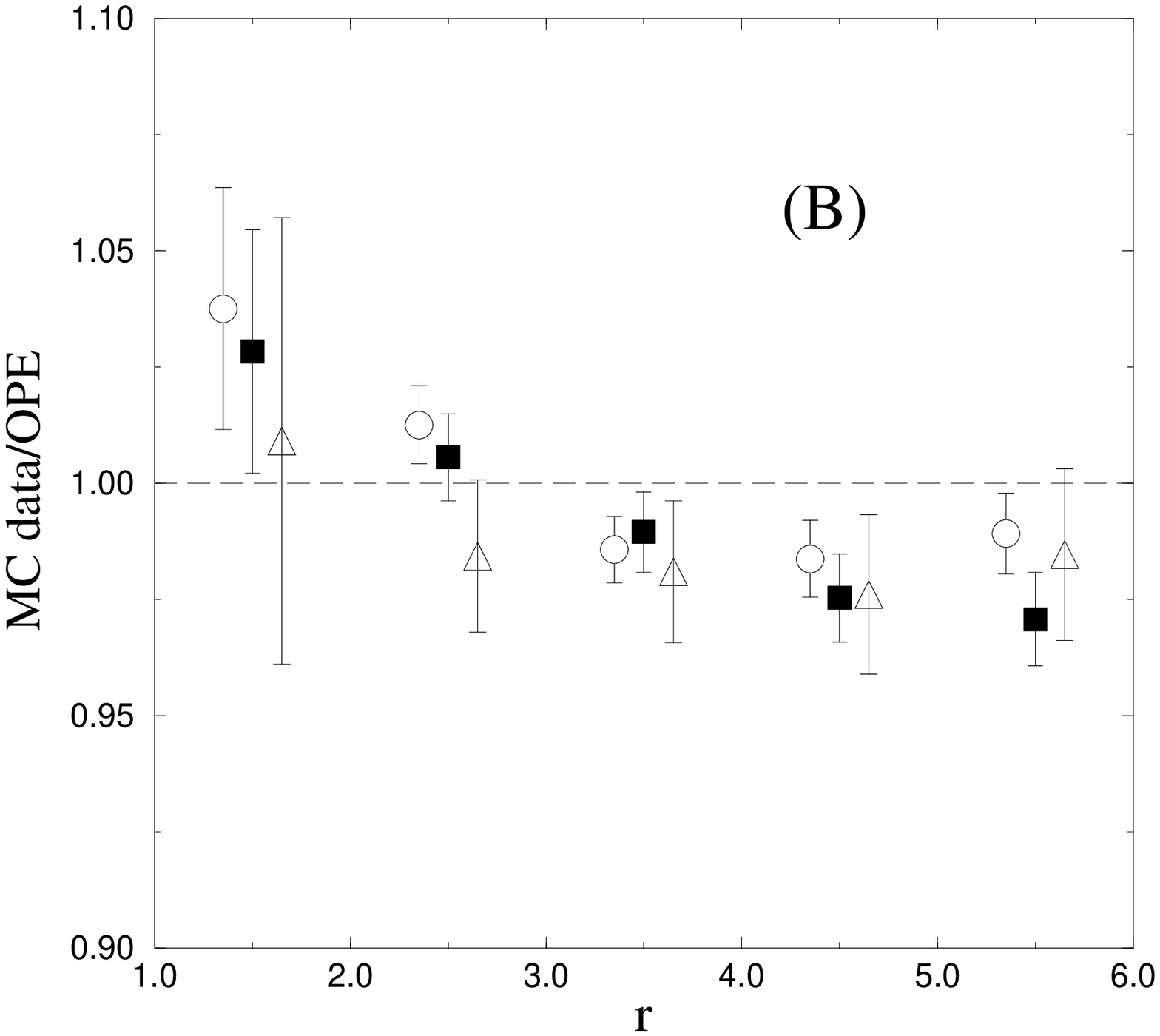,angle=-90,
width=0.55\linewidth} \\
\hspace{-1.5cm}
\epsfig{figure=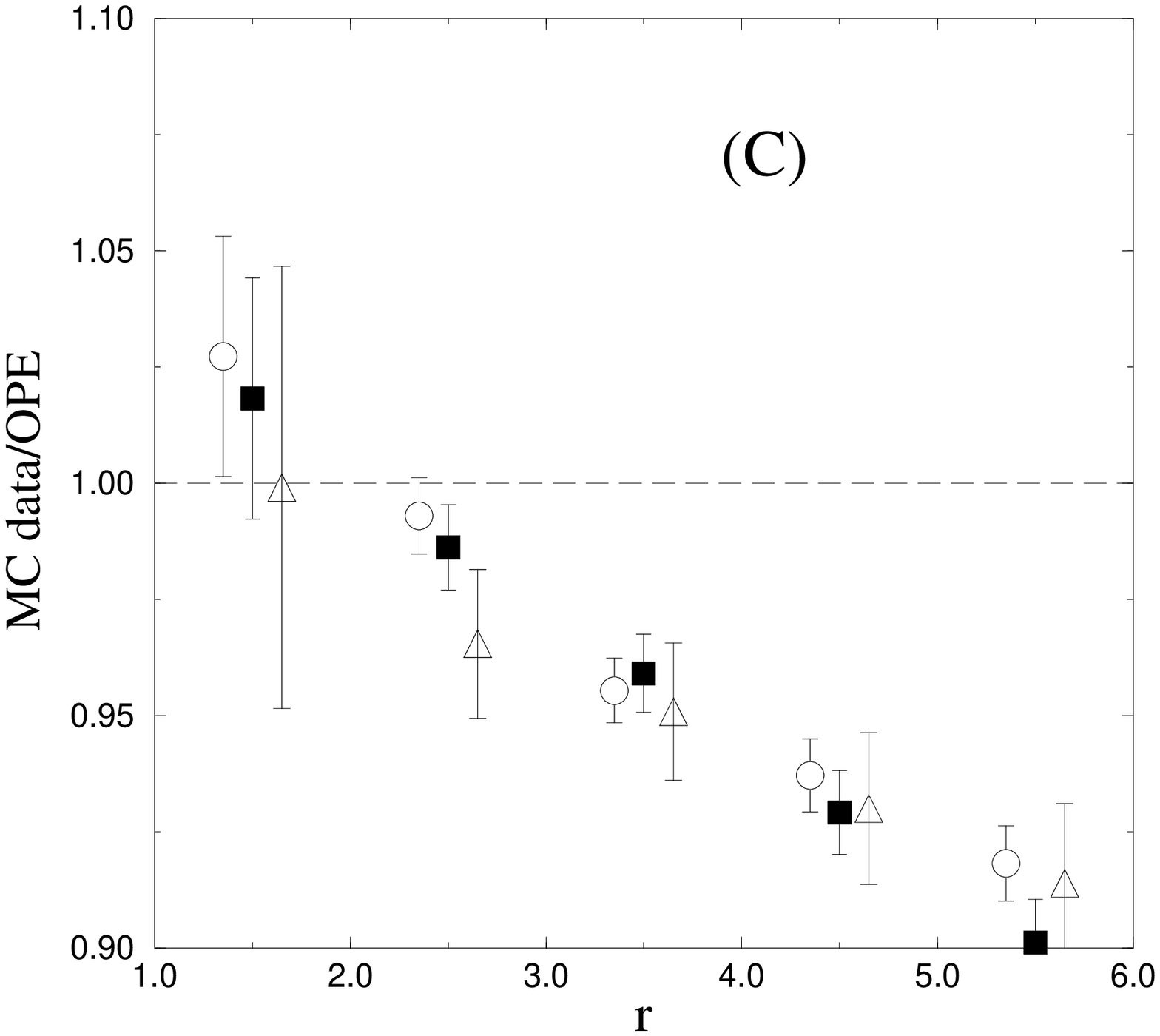,angle=-90,
width=0.55\linewidth} 
&
\hspace{-0.5cm}
\epsfig{figure=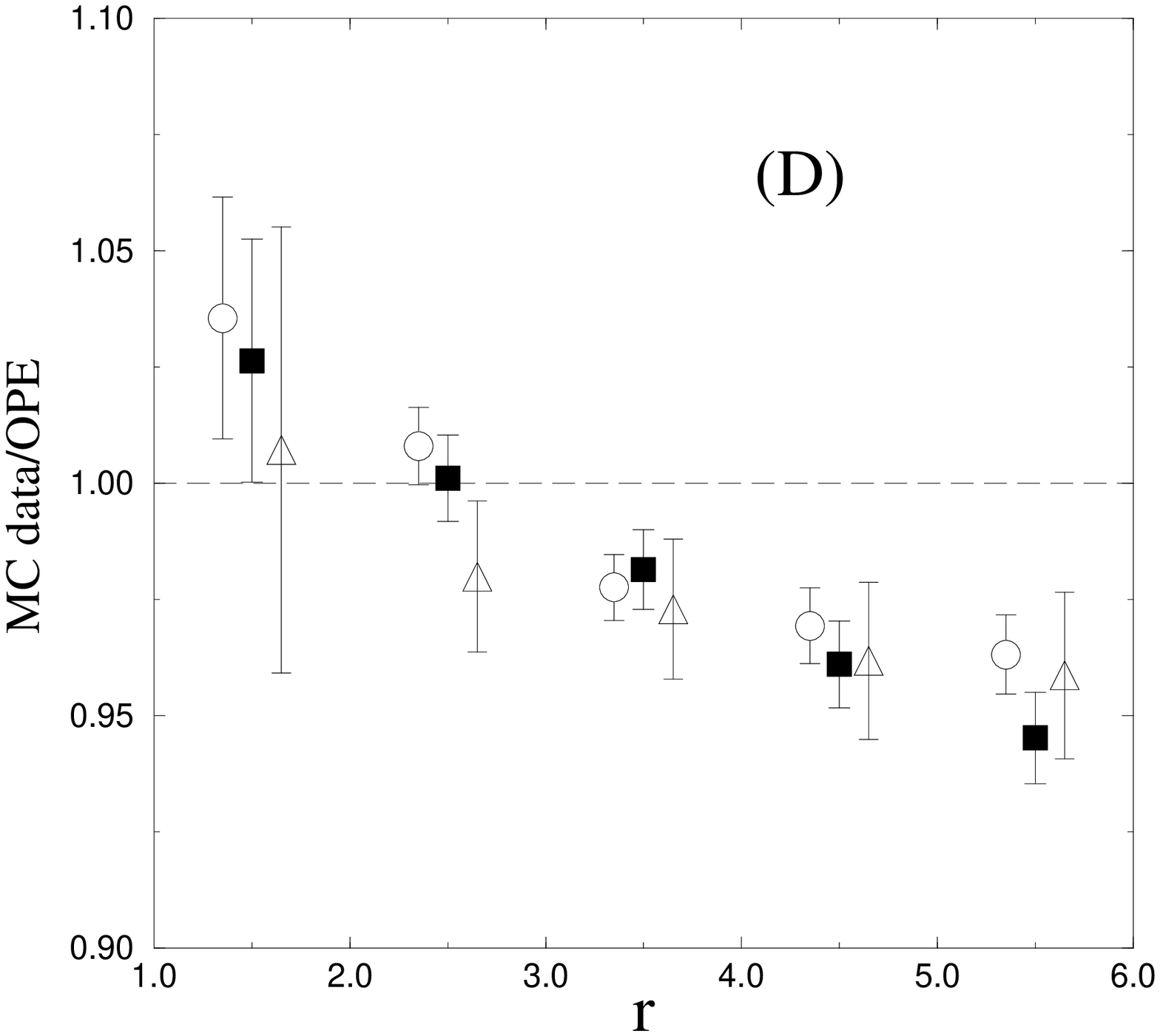,angle=-90,
width=0.55\linewidth}
\end{tabular}
\caption{The scalar product of two Noether currents compared with the OPE 
prediction: graphs of $R(r)$, cf. Eq. \protect\reff{defRatioR}, obtained
using \MS RG-improved perturbation theory. Circles,
filled squares, and triangles correspond to $\pb=2\pi/L$, $4\pi/L$,
and $6\pi/L$ respectively. The data are for lattice 
\ref{Lattice64x128}, $\xi^{\rm exp} = 6.878(3)$. Notice the change of vertical 
scale compared to Figs. \ref{ScalarCurrentsMS}, \ref{ScalarCurrentLatticePT}.}
\label{ScalarCurrentLattice64x128}
\end{figure}

We have tested the validity of the OPE by considering matrix elements
between one-particle states. The matrix elements of the product of 
the currents\footnote{
Since the currents are exactly conserved, both on the lattice and in the 
continuum, there is no need to make a distinction between lattice and 
\MS-renormalized operators.} 
can be determined in terms of $G^{(s)}(t,x;\pb,\qb;2t_s)$, since
\begin{eqnarray}
\frac{1}{2}\,\<\pb,c|\jg_{0,(0,0)}\cdot\jg_{1,(t,x)}|\qb,c\> =
N \sqrt{4\omega(\pb)\omega(\qb)}
\lim_{t_s\to\infty}
  \widehat{G}^{(s)}(t,x;\pb,\qb;2t_s).
\end{eqnarray}
In Fig. \ref{ScalarCurrentsMCData} we report\footnote{
On lattice \ref{Lattice128x256} 
$\widehat{G}^{(s)}(t,x;\pb,\pb;2t_s)$ has only been measured in 
${\cal D} = \{(t,x):|t|,|x|\le 8\}$. The points with $r > 8$ appearing in the 
figure correspond to ``partial" angular averages, i.e. they have 
been obtained using Eq. \reff{definizione-mediarotazionale} and 
restricting $z$ to $\cal D$. The 
same comment applies also to the subsequent figures.}
the angular average 
of ${\rm Im}\ \widehat{G}^{(s)}(t,x;\pb,\pb;2t_s)$ for lattices 
\ref{Lattice64x128} and \ref{Lattice128x256}: here $t_s=6$, $10$
for the two lattices respectively.
In Figs. \ref{ScalarCurrentsMS}, \ref{ScalarCurrentLatticePT}, and 
\ref{ScalarCurrentLattice64x128} we compare 
these numerical data with the predictions of perturbation theory. 

In Fig. \ref{ScalarCurrentsMS} we use continuum RG-improved
perturbation theory in the \MS scheme. 
In this case, the matrix element of the energy-momentum tensor is immediately 
computed: $\<\pb,a|T_{01}|\pb,b\> = 2 i \pb \omega(\pb) \delta^{ab}$.
Therefore, we consider the ratio
\be
R(r) = \frac{1}{2}
 {{\rm Im}\<\pb,c|\overline{\jg_{0,(0,0)}\cdot\jg_{1,(t,x)}}|\pb,c\> \over
   2 \pb \omega(\pb) W_{RGI}(\overline{g}(\Lambda_\MMS r))},
\label{defRatioR}
\ee
which should approach 1 in the short-distance limit.
In Fig. \ref{ScalarCurrentsMS} we present several determinations 
of $R(r)$ that differ in the way in which the running coupling 
constant $\overline{g}(\Lambda_\MMS r)$ and the \MS coupling $g$ are determined.

There are several different methods that can be used to compute the 
\MS coupling $g$ and the strictly related $\overline{g}(\Lambda_\MMS x)$.
They are compared in detail in App. \ref{RunningCoupling}.
It turns out that the finite-size scaling method proposed by L\"uscher 
\cite{Luscher:1982aa,Luscher:1991wu} and what we call ``the 
RG-improved perturbative method" are essentially equivalent, see, e.g.,
Table \ref{RunningComparison1} in App. \ref{RunningCoupling}. Therefore,
we can use either of them,\footnote{In QCD the RG-improved perturbative method
cannot be used since no exact prediction for the mass gap exists.
In this case the finite-size scaling method would be the method of choice.
Note that for large values of the scale also the improved-coupling
method \cite{Parisi:1980aa,Martinelli:1981tb}, which can be  
generalized to QCD
\cite{Lepage:1993xa}, works well, see App. \ref{RunningCoupling}.} 
obtaining completely equivalent results. In Fig. \ref{ScalarCurrentsMS} 
we have used the finite-size scaling method to be consistent with what
we would do in QCD. Elsewhere, we have used the RG improved perturbative method
because of its simplicity. 

The first step in the computation of $R(r)$ is the determination of 
$\overline{g}(\Lambda_\MMS r)$. This is obtained as follows: 
we fix $\mu/m$, and using the finite-size scaling results 
reported in Table \ref{RunningComparison1} in App. \ref{RunningCoupling}, 
we compute $g_\MMS(\mu)$. Then, using Eq. (\ref{LambdaTruncatedDefinition}) 
we determine $\Lambda_\MMS$ at $l$-loops. Finally, 
$\overline{g}(\Lambda_\MMS r)$ is obtained either by solving 
Eq. (\ref{gdix}), again using Eq. (\ref{LambdaTruncatedDefinition})
for $\Lambda_\MMS(1,\overline{g})$ appearing in the right-hand side, 
or by using Eq. \reff{overlinegdef}. As we discussed
in Sec. \ref{sec3.3}, the final result should be independent of the 
chosen value of $\mu/m$ and therefore we can evaluate 
the systematic error on $\overline{g}(\Lambda_\MMS x)$ by considering 
different values of $\mu/m$. If we fix $m/\mu = 0.00071$, 
cf. first row of Table \ref{RunningComparison1} in App. \ref{RunningCoupling},
on lattice \ref{Lattice128x256}, we have 
$\overline{g}(\Lambda_\MMS x) = 1.498$, $2.206$, $3.363$ 
respectively for $m x = 0.2$, $0.5$, $0.8$, while if we fix $m/\mu = 0.1033$, 
cf. 10th row of Table \ref{RunningComparison1} in App. \ref{RunningCoupling},
we have $\overline{g}(\Lambda_\MMS x) = 1.490$, $2.185$, $3.273$
at the same distances. 
The dependence is tiny and, as expected, it increases for larger values of 
$mx$. In practice, it does not play any role, the main source of 
error being instead the truncation of the OPE coefficients.
Notice that the independence on $\mu$ 
is obvious if we use the RG-improved perturbative method.

Let us now describe the various graphs appearing in 
Fig. \ref{ScalarCurrentsMS}.  In graphs (A) and (B) 
we fix $\mu = m/0.00071$ and then, using the results 
presented in Table \ref{RunningComparison1}, cf. first row, we obtain 
$g_{\MMS}(\mu) = 0.587016$. We then compute 
$\Lambda_{\MMS}(\mu,g_{\MMS})$ 
using the four-loop expression (\ref{LambdaTruncatedDefinition})
with $l=4$. Finally, the Wilson coefficient is given by \reff{Wilson2ms}:
in graph (A) we use the leading term only,
while in graph (B) we include the next-to-leading one. 

In graphs (C) and (D) we compute $\Lambda_{\MMS}$ as in (A) and (B),
choosing a different scale, $\mbar = m/0.1033$, cf. $10^{\rm th}$ row
of Table \ref{RunningComparison1}. Then we compute $\overline{g}(x)$ 
by solving numerically Eq.  (\ref{gdix}) using the four-loop expression 
\reff{LambdaTruncatedDefinition}
with $l=4$. Finally, the Wilson coefficient is obtained using
Eq. (\ref{Wilson1ms}). While in graph (C) we keep only the leading 
term $1/\overline{g}(x)$, in graph (D) we use the complete expression.

Finally, graphs (E) and (F) are identical to graph (D) except that 
$\Lambda_{\MMS}$ and $\overline{g}(x)$ are computed using the 
two-loop expression (\ref{LambdaTruncatedDefinition}) with $l=2$.
The two graphs correspond to different choices of $\mu$:
$\mbar = m/0.00071$ (graph (E)) and $\mbar = m/0.1033$ (graph (F)).

Comparing the different graphs, we immediately see that the choice of 
scale $\mu$ and the order of perturbation theory used for 
$\Lambda_\MMS$ (two loops or four loops) are not relevant with the 
present statistical errors.
Much more important is the role of the Wilson coefficients. If one 
considers only the leading term (graphs (A) and (C)) there are indeed 
large discrepancies and in the present case one would obtain estimates of the 
matrix elements with an error of 50--100\%. Inclusion of the next-to-leading
term---this amounts to considering one-loop Wilson coefficients and two-loop
anomalous dimensions---considerably improves the results, and now the 
discrepancy is of the order of the statistical errors, 
approximately 10\%. For the practical application of the method,
it is important to have criteria for estimating the error on the results. It is 
evident that the flatness of the ratio of the matrix element by the 
OPE prediction is not, in this case, a good criterion: The points in
graph (A) show a plateau---and for a quite large set of values of $r$---even
if the result is wrong by a factor of two. However, this may just be a 
peculiarity of the case we consider, in which the $r$-dependence of the 
data and of the OPE coefficients is very weak. On the other hand, 
the comparison of the results obtained using 
different methods for determining $\overline{g}(x)$ seems to provide 
reasonable estimates of the error bars. For instance, if only the leading term
of the Wilson coefficients were available, we could have obtained 
a reasonable estimate of the error by comparing graphs (A) and (C).

In Fig. \ref{ScalarCurrentLatticePT} we consider lattice
RG-improved perturbation theory, computing
\be
R^{\rm latt} (r) = \frac{1}{2} 
 {{\rm Im}\, \<\pb,c|\overline{\jg_{0,(0,0)}\cdot\jg_{1,(t,x)}}|\pb,c\> \over
   U(g_L) W_{RGI}^L(\bar{g}_L){\rm Im}\, \<\pb,c|T^L_{10}|\pb,c\>} .
\label{defRatioRlatt}
\ee

In graph (A) we use $g_L$ as an expansion parameter. We compute 
$\Lambda^{\rm latt}(1,g_L)$ using the value of the mass gap 
$m^{-1} = 13.632(6)$ and Eq.~\reff{MassGapLattice}, with the 
nonperturbative constant \reff{Hasenfratzconst}. Then, we determine 
$\overline{g}_L(\Lambda^{\rm latt} r)$ by solving numerically 
\reff{eq:gbarraL} and using 
for $\Lambda^{\rm latt}(1,\overline{g}_L)$ appearing in the left-hand side
its truncated four-loop expression \reff{Lambdalatt}. Finally,  we use 
Eq. \reff{WRGIlatt} for the Wilson coefficient. The results are quite 
poor: there is a downward trend as a function of $r$ and the data are 
far too low. Naive lattice perturbation theory 
is unable to provide a good description of the numerical data.

The results improve significantly if we use the improved coupling $g_E$:
The estimates obtained using this coupling, graphs (B), (C), (D), are 
not very different from those obtained using \MS continuum perturbation
theory.
Graph (B) has been obtained exactly as graph (A), except that in this 
case we used $g_E$ as an expansion parameter. The $\Lambda$-parameter 
is defined in \reff{LambdaEdef}, $m/\Lambda_E$ in 
\reff{costanteHasenfratz_schemaE}, and the relevant Wilson coefficient
is given in Eq. (\ref{WRGIE}). Graphs 
(C) and (D) are analogous to graph (B). The difference is in the 
determination of $\Lambda_E(1,g_E)$. We do not compute it
nonperturbatively by using the mass gap
but we determine it directly from the 
perturbative expression \reff{LambdaEdef}. In this case we use 
the perturbative expression truncated at four loops 
(C) and two loops (D).

In Figs. \ref{ScalarCurrentsMS}, \ref{ScalarCurrentLatticePT} we have checked
the validity of the OPE on lattice \ref{Lattice128x256}. If one has in mind
QCD applications this is quite a large lattice since 
$\xi^{\rm exp} \approx 14$. For this reason, we have tried to understand 
if the nice agreement we have found survives on a smaller lattice,  
by repeating the computation on lattice \ref{Lattice64x128}.
The results are reported in Fig. \ref{ScalarCurrentLattice64x128}. 
Graphs (A) and (B) should be compared with graph (D) of 
Fig. \ref{ScalarCurrentsMS}.
In (A) we compute $\Lambda_{\MMS}$ from Eq. \reff{MassGapLattice}, with the 
nonperturbative constant \reff{Hasenfratzconst} and  $m^{-1} = 6.878(3)$.
Then, we compute $\overline{g}(x)$ solving numerically
Eq.  (\ref{gdix}) using the four-loop expression 
\reff{LambdaTruncatedDefinition}
with $l=4$. The Wilson coefficient is obtained from Eq. \reff{Wilson1ms}.
In (B) we repeat the same calculation as in Fig. \ref{ScalarCurrentsMS}
graph (D) at the scale $\mbar = m/0.00071$.
In (C) and (D) we repeat the calculation of graph (B) using 
the two-loop and the three-loop $\beta$-function 
for the  determination of the coupling $\overline{g}$. 
In all graphs (B), (C), (D)  we use the result $m^{-1} = 6.878(3)$.

Graphs (A) and (B) show a nice flat behaviour and for 
$2\ltapprox r\ltapprox \xi$, the 
ratio $R(r)$ is approximately 1 with 2--3\% corrections (notice that 
the vertical scale in Figs. \ref{ScalarCurrentsMS} and 
\ref{ScalarCurrentLattice64x128} is different): The OPE works 
nicely even on this small lattice. (A) and (B) differ in the 
method used in the determination of the \MS coupling. 
As we explained in App. \ref{RunningCoupling} and it appears clearly
from the two graphs, the effect is very small. 
Graph (C)---and to a lesser extent graph (D)---shows instead significant 
deviations: Clearly, $\overline{g}(\Lambda_\MMS x)$ must be determined
using four-loop perturbative expansions in order 
to reduce the systematic error to a few percent. Notice that such 
discrepancies are probably present also for lattice 
\ref{Lattice128x256}: however, in this case, the statistical errors 
on $R(r)$ are large---approximately 5-6\% (for $\pb = 2\pi/L$
and  $\pb = 4\pi/L$) and 9\% (for $\pb = 6\pi/L$)---and thus they
do not allow to observe this effect.

\begin{figure}
\hspace{2cm}
\epsfig{figure=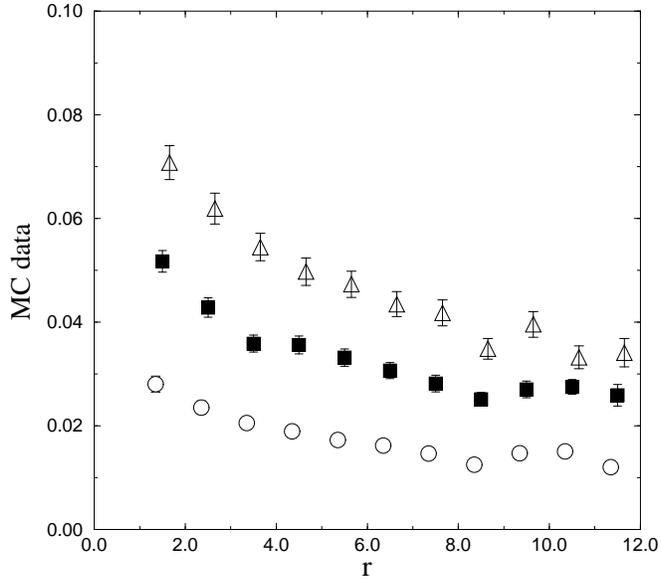,angle=-90,
width=0.6\linewidth}
\caption{Estimates of ${\rm Im}\,\widehat{G}^{(s)}(t,x;\pb,0;20)$
averaged over rotations on lattice \ref{Lattice128x256}. Circles,
filled squares, and triangles correspond to $\pb=2\pi/L$, $4\pi/L$,
and $6\pi/L$ respectively. }
\label{ScalarCurrentsOutOfDiagonalData}
\end{figure}

\begin{figure}
\hspace{2cm}
\epsfig{figure=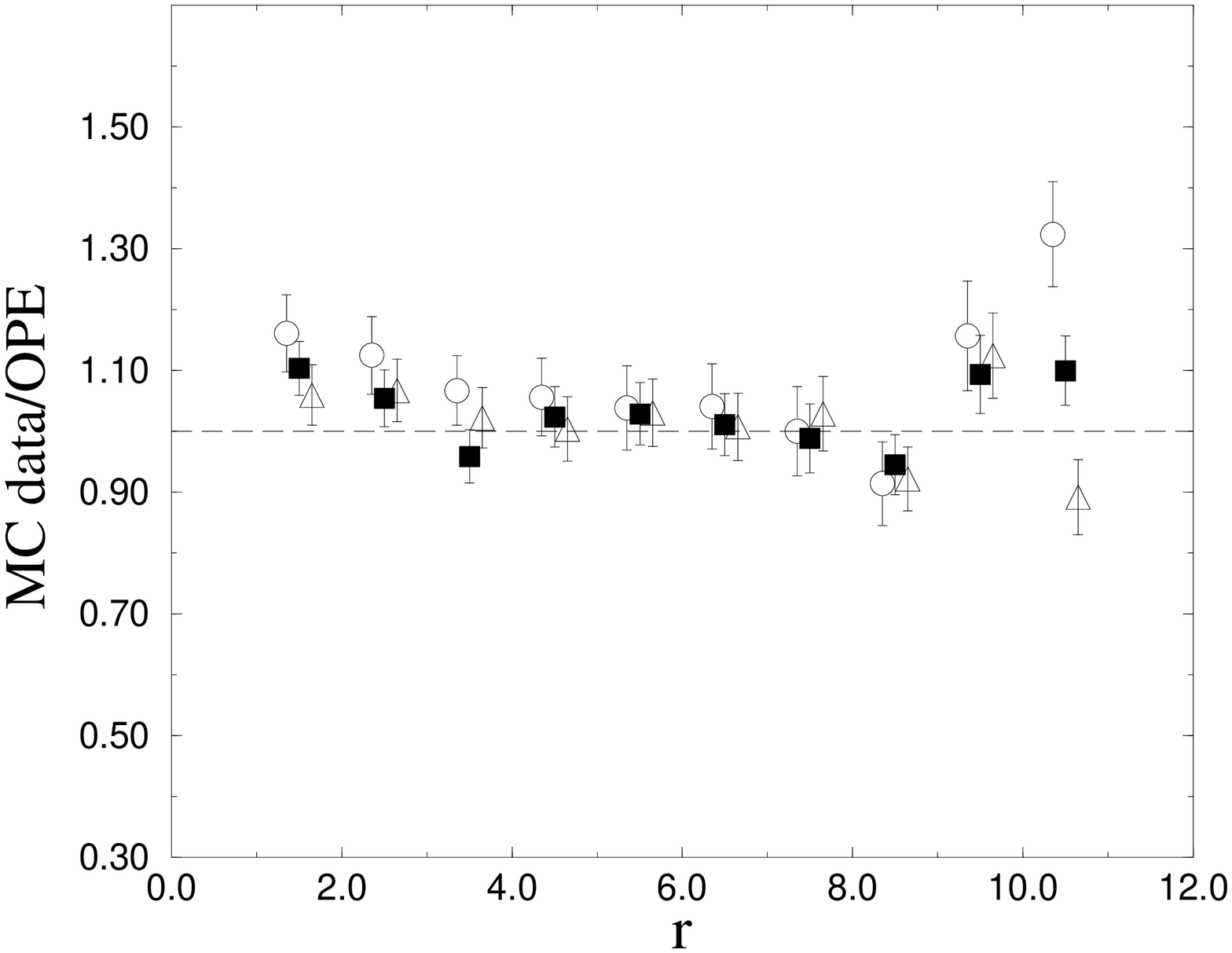,angle=-90,
width=0.6\linewidth}
\caption{The scalar product of two Noether currents compared with the OPE 
prediction: graphs of $S(r)$, cf. Eq. \protect\reff{defRatioS}, obtained
using \MS RG-improved perturbation theory. Circles,
filled squares, and triangles correspond to $\pb=2\pi/L$, $4\pi/L$,
and $6\pi/L$ respectively. The data are for lattice 
\ref{Lattice128x256}, $\xi^{\rm exp} = 13.636(10)$.}
\label{ScalarCurrentsMSOutOfDiagonal}
\end{figure}

As a further check we considered matrix elements between states of 
different momentum.
In Fig. \ref{ScalarCurrentsOutOfDiagonalData} we report the angular 
average of ${\rm Im}\ \widehat{G}^{(s)}(t,x;0,\pb;20)$ for lattice 
\ref{Lattice128x256}.
In Fig. \ref{ScalarCurrentsMSOutOfDiagonal} we compare the numerical data 
with the OPE prediction, by considering 
\be
S (r) = \frac{1}{2}
 {{\rm Im}\, \<0,c|\overline{\jg_{0,(0,0)}\cdot\jg_{1,(t,x)}}|\pb,c\> \over
   W^{RGI}(\bar{g}){\rm Im}\<0,c|T_{01}^{\rm latt}|\pb,c\>},
\label{defRatioS}
\ee
where $T_{\mu\nu}^{\rm latt}$ is defined in Eq. \reff{Tmunulattice} 
and $Z_{TT}^{L,(2,0)} $ is computed in Sec. \ref{sec4.3}.
In Fig. \ref{ScalarCurrentsMSOutOfDiagonal} we report $S(r)$ for 
lattice \ref{Lattice128x256}. The Wilson coefficients are computed as 
in graph (A) of Fig. \ref{ScalarCurrentLattice64x128}, using 
$m^{-1} = 13.636(10)$. The numerical data are again well described by the 
OPE prediction for a quite large set of values of $r$.
%

%
\subsection{OPE for the antisymmetric product of currents} \label{sec4.5}

In this Section we consider the antisymmetric product of two currents
and compare our numerical results with the perturbative predictions. 
With respect to the previous case, we have here a better knowledge 
of the perturbative Wilson coefficients---some of them are known to two 
loops---and moreover we have an exact expression for the 
one-particle matrix elements of the current $j_{\mu,x}$. Indeed,
we have \cite{Karowski:1978a,Luscher:1990ck}:
\be
\< \pb, c |j^{ab}_{\mu,(t,x)} |\qb, d \> = -
 i (p_\mu + q_\mu)\, G(k)\,
(\delta^{ac}\delta^{bd}-\delta^{ad}\delta^{bc})\, e^{i(p-q)\cdot x} ,
\label{form-factor}
\ee
where $p\cdot x\equiv p_0 t + p_1 x$, 
$p_\mu \equiv (i\sqrt{\pb^2+m^2},\pb)$, 
$k \equiv\frac{1}{2}\sqrt{(p_0-q_0)^2+(p_1-q_1)^2}$, 
and, for $N=3$, 
\be
   G(k) = \frac{\theta}{2\tanh\theta/2}\cdot\frac{\pi^2}{\pi^2+\theta^2},
\ee
where the rapidity variable $\theta$ is defined by $k = m\sinh\theta/2$.

\begin{figure}
\hspace{2cm}
\epsfig{figure=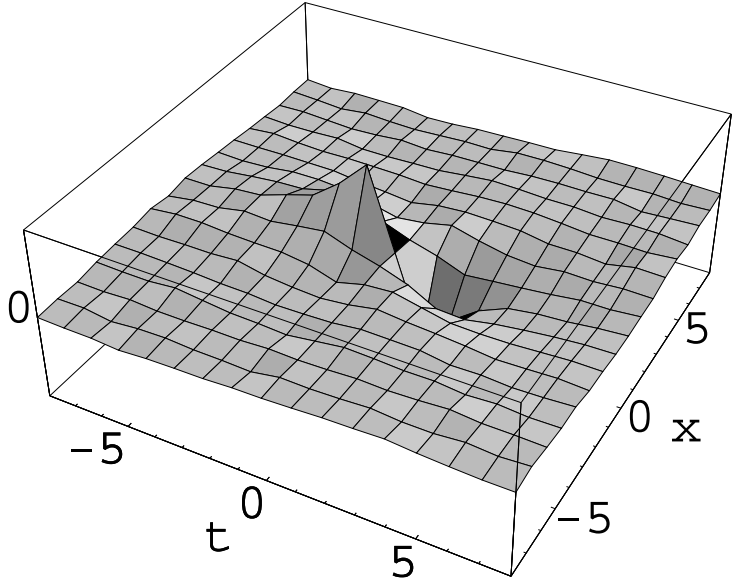,width=0.6\linewidth}
\caption{Estimate of ${\rm Re}\, \widehat{G}^{(a)}_{11}(t,x;\pb,-\pb;20) $
on lattice \ref{Lattice128x256}. 
Here $\pb =2\pi/L$.}
\label{AntisymmetricCurrentsLeadingData}
\end{figure}

\begin{figure}
\hspace{2cm}
\epsfig{figure=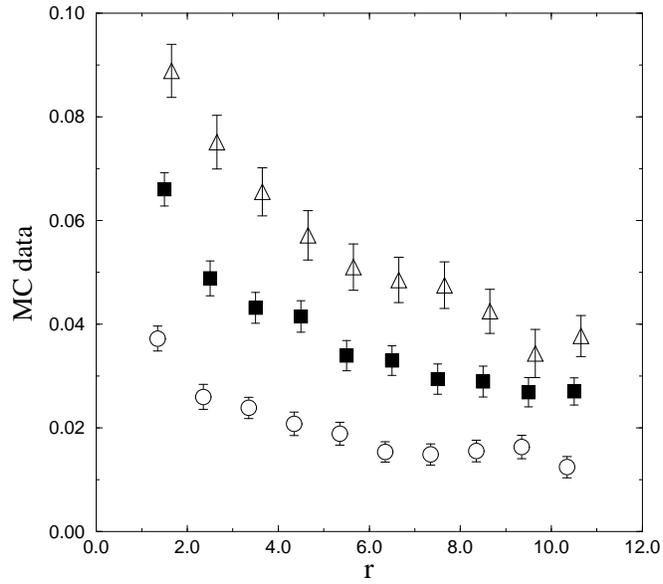,angle=-90,
width=0.6\linewidth}
\caption{Angular average of 
${\rm Im}\, \widehat{G}^{(a)}_{01}(t,x;\pb,0;20)$
on lattice \ref{Lattice128x256}.
Circles, filled squares, and triangles correspond to $\pb=2\pi/L$, $4\pi/L$,
and $6\pi/L$ respectively.}
\label{AntisymmetricCurrentsNextToLeadingData}
\end{figure}

We first consider the product $(j^{ac}_{\mu,(0,0)}j^{bc}_{\nu,(t,x)}-
j^{bc}_{\mu,(0,0)}j^{ac}_{\nu,(t,x)})$ for $\mu=\nu=1$ and $x=0$. 
The OPE of such a product can be determined from Eq. \reff{anticorrcont}.
Using the results of App. \ref{secB.2}, we have for $t\to 0$,
\be
\sum_c
j^{ac}_{1,(0,0)}j^{bc}_{1,(t,0)}-
j^{bc}_{1,(0,0)}j^{ac}_{1,(t,0)} = 
  {1 \over t} W_{RGI} \left(\overline{g}(\Lambda_\MMS t)\right) 
\left(j_0^{ab}(0) + 
   {t\over2} \partial_0 j_0^{ab}(0)\right),
\label{OPEx=0}
\ee
where, at two loops,
\be
W_{RGI}(\overline{g}) = {N-2\over 2\pi} + 
   {N-2\over 4\pi^2} \overline{g},
\label{Wilson_x=0}
\ee
and $\overline{g}(\Lambda_\MMS t)$ is defined in Eq. \reff{gdix}.
We also consider the angular average of the product of the currents 
for $\mu=0$ and $\nu = 1$. 
Using the results of App. \ref{secB.1}, we have for $r\to 0$
\begin{eqnarray}
&& \hskip -1truecm
{\cal I}^{ab}(r) \equiv \overline{j^{ac}_{0,(0,0)}j^{bc}_{1,(t,x)}-
                      j^{bc}_{0,(0,0)}j^{ac}_{1,(t,x)}} \approx
\nonumber \\
&& \hskip -1truecm \qquad
 -\,{3(N-2)\over 16\pi}\,
[\partial_0 j_1^{ab} (0) + \partial_1 j_0^{ab} (0)] + 
{1\over 2 \overline{g}}\left(1 + {N-6\over 4\pi} \overline{g}\right) 
\, [\partial_0 j_1^{ab} (0) - \partial_1 j_0^{ab} (0)]\, .
\label{sec4.5:SOPE}
\end{eqnarray}
Again, we have tested the validity of the OPE by considering matrix elements 
between one-particle states.
The matrix elements of the product of the currents are obtained from 
\be
\sum_{abc}
\<\pb ,a|j^{(ac)}_{(0,0),\mu}j^{(bc)}_{(t,x),\nu}-
j^{(bc)}_{(0,0),\mu}j^{(ac)}_{(t,x),\nu}|\qb ,b\> = \,
N \sqrt{4\omega(\pb )\omega(\qb )} \lim_{t_s\to\infty }
\widehat{G}^{(a)}_{\mu\nu}(t,x;\pb ,\qb ;2t_s).
\label{AsymptoticAntiSymmetric}
\ee
In Fig. \ref{AntisymmetricCurrentsLeadingData} we show a plot 
of ${\rm Re}\, \widehat{G}^{(a)}_{11}(t,x;\pb ,-\pb ;2t_s)$
obtained on 
lattice \ref{Lattice128x256} for $\pb = 2\pi/L$ and $t_s = 10$, and  
in Fig. \ref{AntisymmetricCurrentsNextToLeadingData} we show 
the angular average of 
${\rm Im}\, \widehat{G}^{(a)}_{01}(t,x;\pb ,0 ;2t_s)$
on the same lattice and again for 
$t_s = 10$. Comparing these graphs with those for the scalar product
of the currents, one sees that the matrix elements show here a larger 
variation with distance, and thus this should provide a 
stronger test of the validity of the OPE.

\begin{figure}
\begin{tabular}{cc}
\hspace{0.0cm}\vspace{-1.5cm}\\
\hspace{-2.5cm}
\epsfig{figure=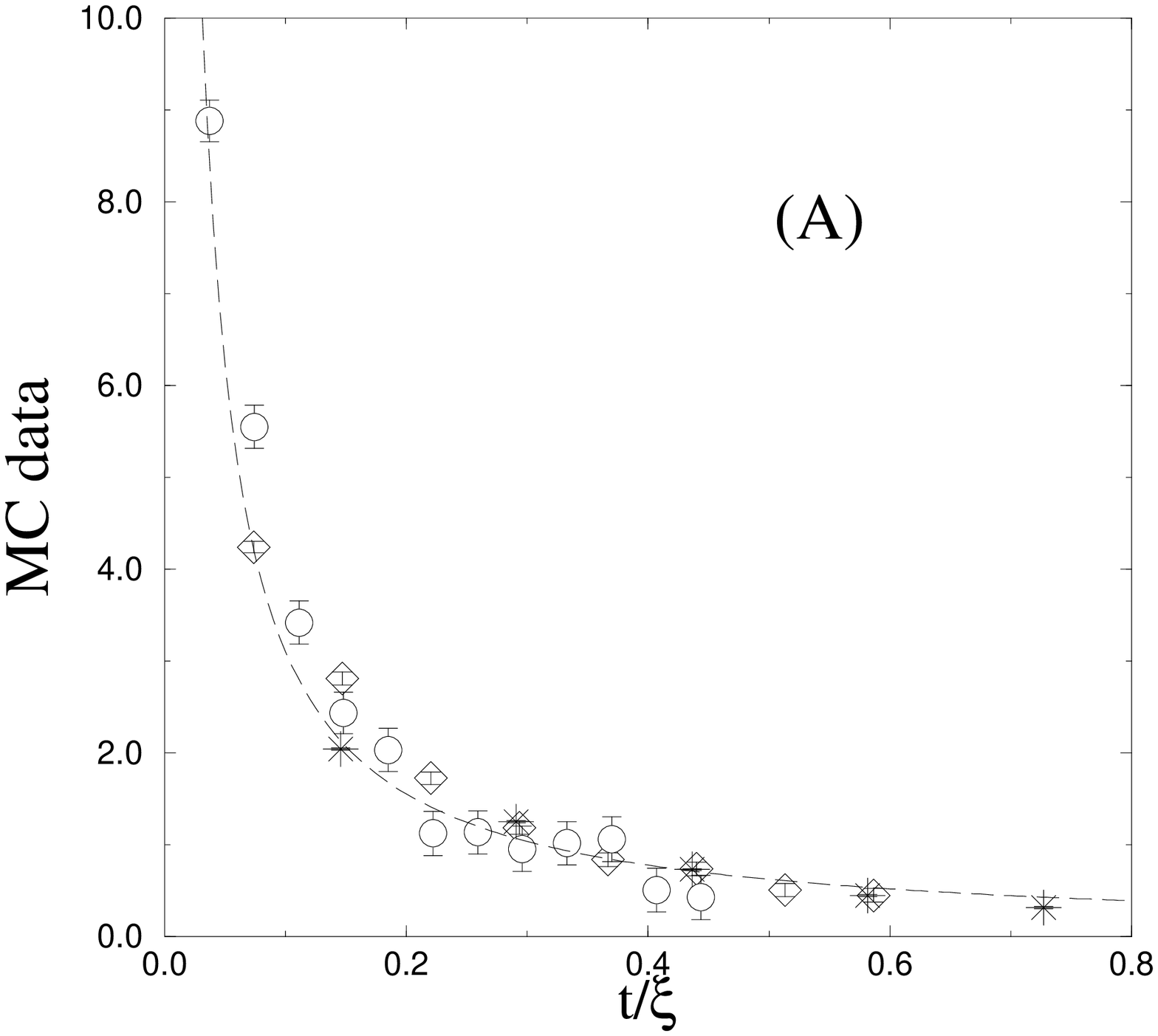,angle=-90,
width=0.6\linewidth}&\hspace{-0.5cm}
\epsfig{figure=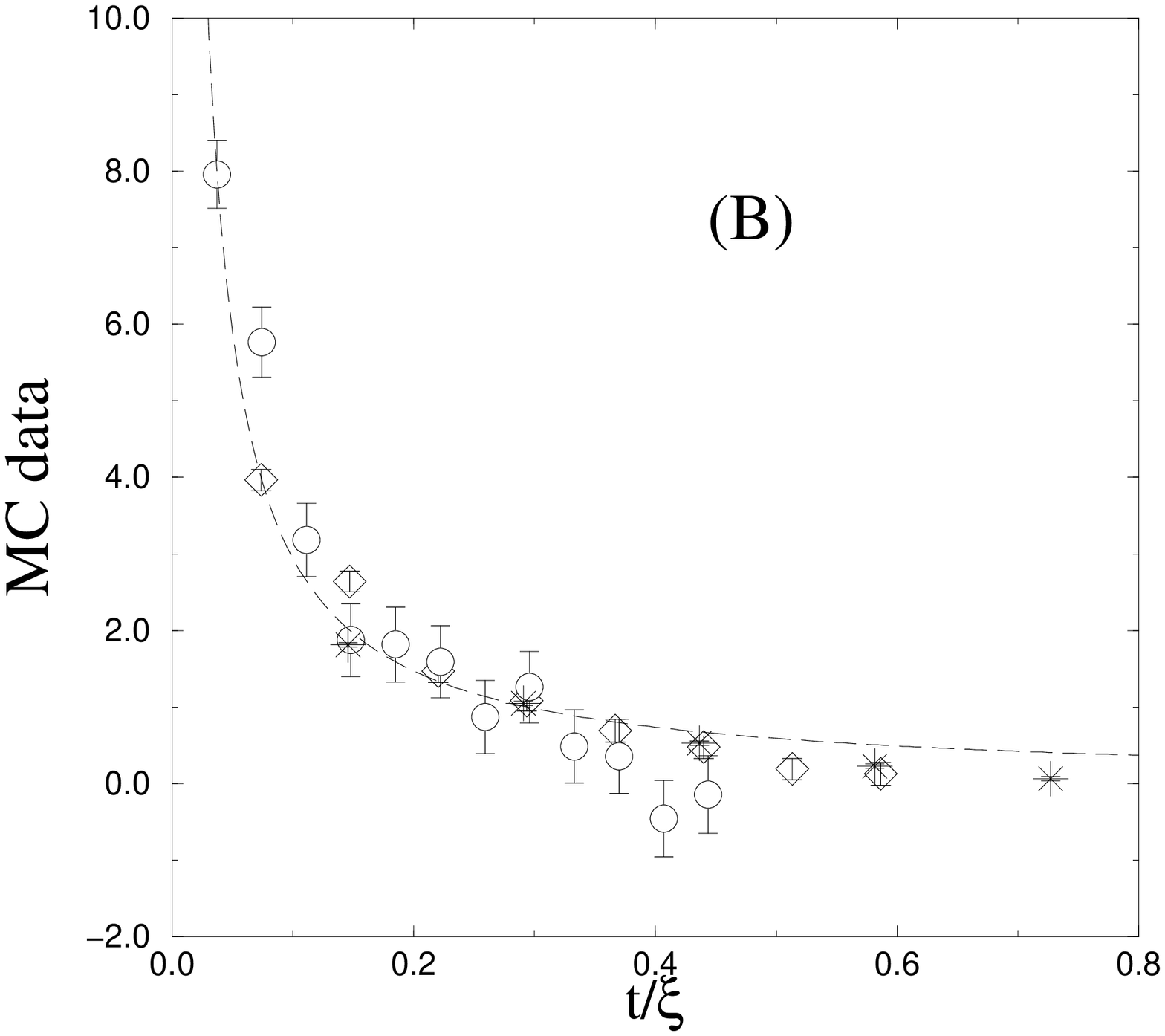,angle=-90,
width=0.6\linewidth}\\
\hspace{0.0cm}\vspace{-1.5cm}\\
\hspace{-3.5cm}
\epsfig{figure=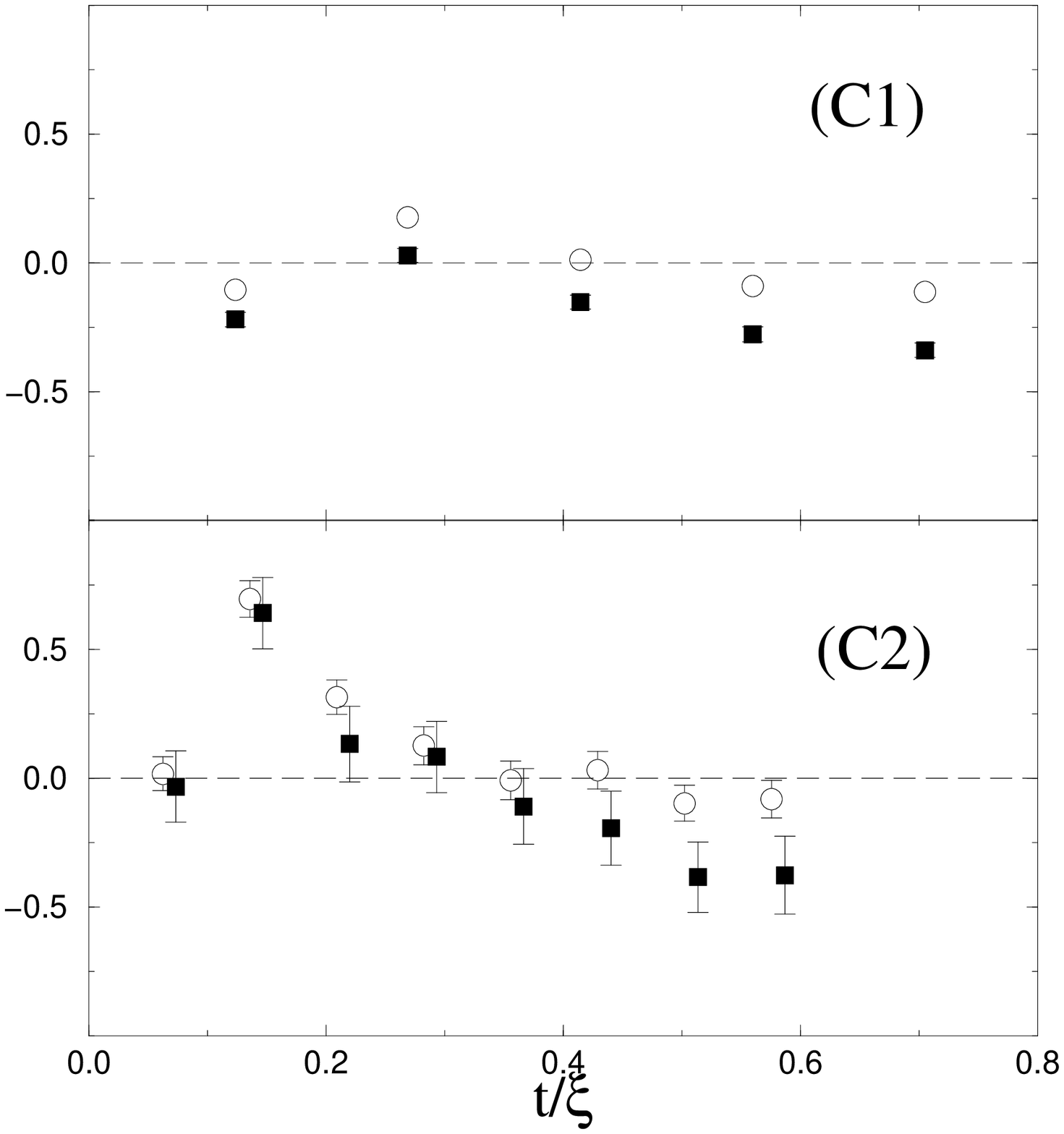,angle=-90,
width=0.6\linewidth}&\hspace{-1.5cm}
\epsfig{figure=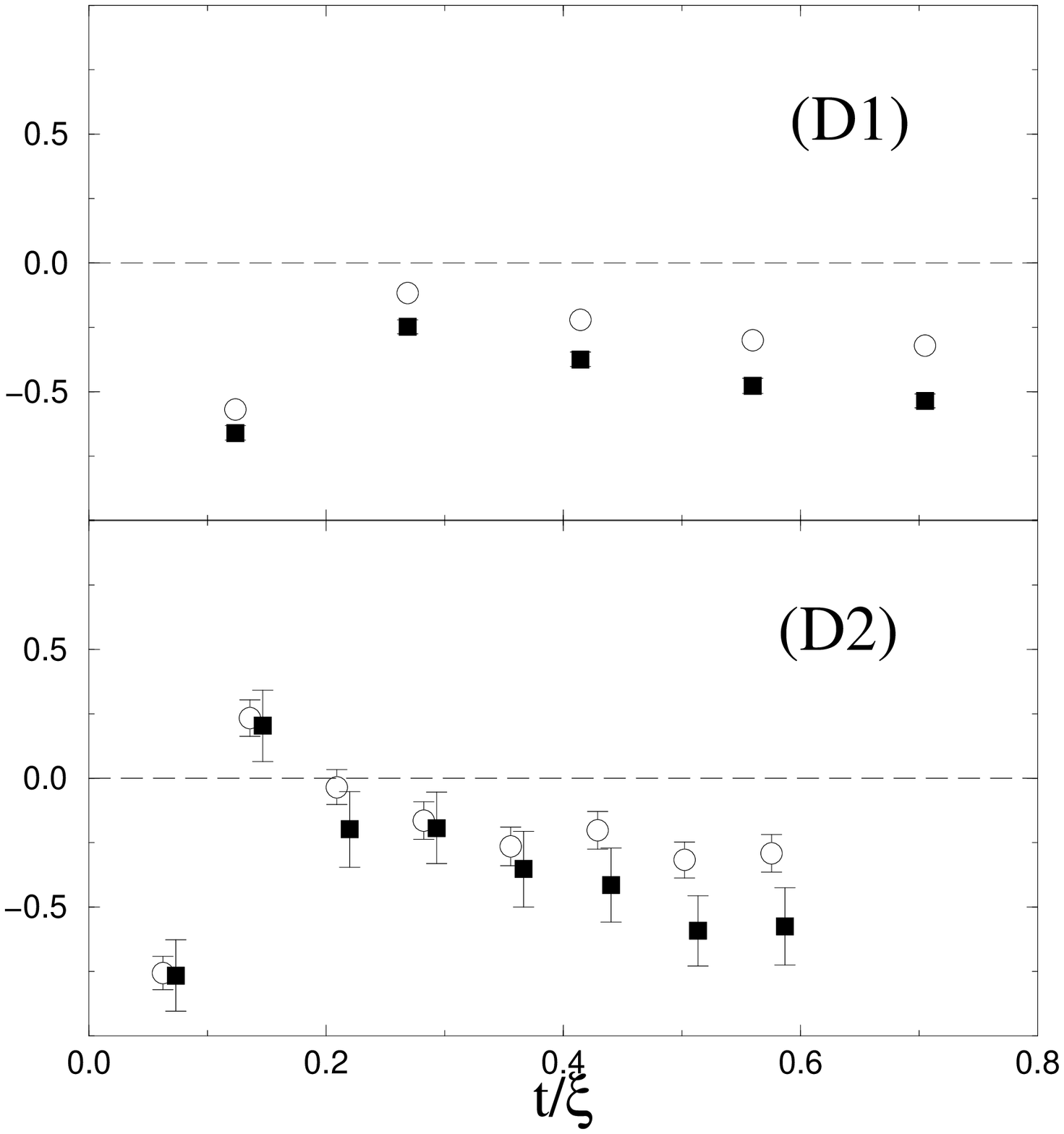,angle=-90,
width=0.6\linewidth}
\end{tabular}
\caption{Antisymmetric
product of two Noether currents for $x=0$ and $t\neq 0$: in graphs (A) and (B)
we report estimates  of $V(t)$, cf. Eq. \reff{ScalingCombination}, 
and of the OPE prediction (dashed line)
$V^{\rm OPE}(t)$, cf. Eq. \reff{Vope}. The numerical data correspond to
$\pb = 2\pi/L$ (graph (A)) and $\pb = 4\pi/L$ (graph (B)). 
Stars, diamonds, and circles refer to lattices \ref{Lattice64x128},
\ref{Lattice128x256}, \ref{Lattice256x512} respectively.
In graphs (C) and (D) we show $V(t) - V^{\rm OPE}(t)$.
Empty circles and filled squares refer to $\pb = 2\pi/L$ and 
$\pb = 4\pi/L$ respectively. (C1) and (D1) refer to lattice \ref{Lattice64x128},
(C2) and (D2) to lattice \ref{Lattice128x256}.}
\label{AntisymmetricCurrentsLeadingDatavsTheory}
\end{figure}

In Fig. \ref{AntisymmetricCurrentsLeadingDatavsTheory}
we compare the results for 
${\rm Re}\, \widehat{G}^{(a)}_{11}(t,x;\pb ,-\pb ;2t_s)$
with the OPE perturbative predictions. 
Here, as always in this Section, we use the RG-improved perturbative method 
to compute $\overline{g}$, using the four-loop expression for the $\beta$
function. As we explained at length in the previous Section, no significant
difference is observed if one uses the finite-size scaling method, 
or improved lattice perturbation theory. 

In graphs (A) and (B)  we show the combination 
\begin{equation}
V(t) \equiv \frac{1}{2}\, \xi^{\rm exp}\, \left[
\widehat{G}^{(a)}_{11}(t,0;\pb ,-\pb ;2t_s)-
\widehat{G}^{(a)}_{11}(-t,0;\pb ,-\pb ;2t_s)
\right]
\, ,
\label{ScalingCombination}
\end{equation}
for two different values of $\pb$.
In the scaling limit $V(t)$ is a function of $\pb\xi^{\rm exp}$ and
of $t/\xi^{\rm exp}$. As it can be seen from the graphs our results 
show a very nice scaling: The data corresponding to the three different 
lattices clearly fall on a single curve. In the same graphs we also 
report the OPE prediction \reff{OPEx=0}, i.e.
\be
V^{\rm OPE}(t) \equiv 2 N \omega(\pb) {\xi^{\rm exp}\over t} 
   W_{RGI}(\overline{g}(\Lambda_\MMS t))
  \<\pb,a|j^{ab}_0(0)|-\pb,b\>.
\label{Vope}
\ee
Note that $\<\pb,a|\partial_0j^{ab}_0|-\pb,b\> = 0$, so that 
the corrections due to higher-order terms in the OPE expansion are of order $t$.
In graph (A) and (B) we use only the one-loop Wilson coefficient 
for $W_{RGI}(\overline{g})$.
There is a good agreement between the OPE prediction and the numerical 
data: quite surprisingly the agreement extends up to  2 
lattice spacings. 

To better understand the discrepancies, in 
graphs (C) and (D) we report $V(t)-V^{\rm OPE}(t)$.
In graphs (C1) and (C2) we use the one-loop Wilson coefficient, and
in (D1) and (D2) the two-loop Wilson coefficient given in Eq. 
\reff{Wilson_x=0}. The numerical data
refer to lattice \ref{Lattice64x128} for (C1) and (D1) and to
lattice \ref{Lattice128x256} for (C2) and (D2).
There is clearly agreement, although here 
deviations are quantitatively somewhat large, since $V(t)$ is strongly varying. 
Let us consider for instance the data obtained on lattice \ref{Lattice128x256}
for $\pb = 2\pi/L$. If we evaluate the matrix element of the
Noether current using the numerical estimate
of $V(t)$ with $t = 3,4,5,6$ we obtain the result with a systematic error
of $2\%, 14\%, 32\%, 27\%$ respectively.

\begin{figure}
\hspace{2cm}
\epsfig{figure=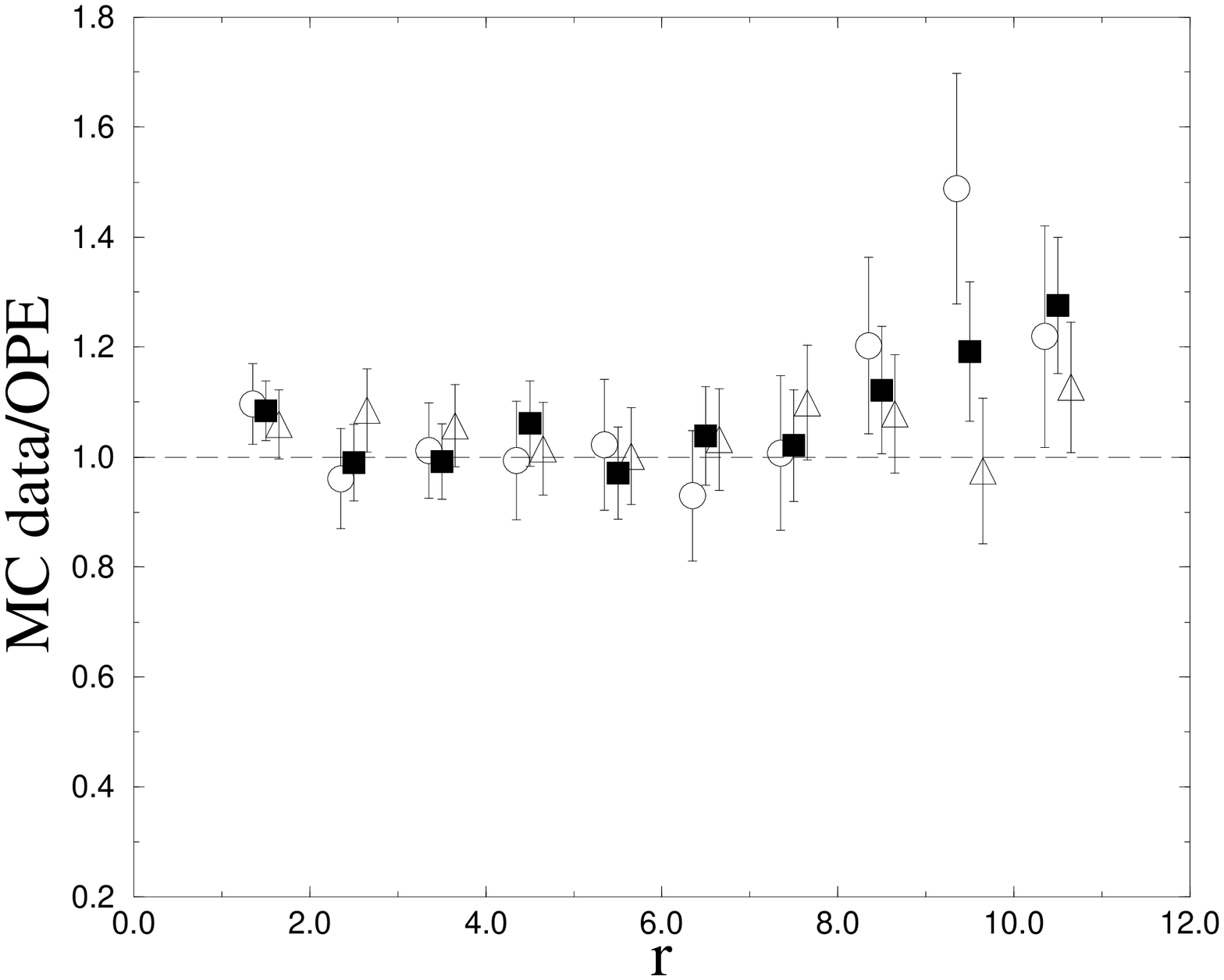,angle=-90,
width=0.6\linewidth}
\caption{The antisymmetric product of two Noether currents compared with 
the OPE prediction: graphs of $Y(r)$, cf. Eq. \protect\reff{defRatioY}, 
obtained using \MS RG-improved perturbation theory. Circles,
filled squares, and triangles correspond to $\pb=2\pi/L$, $4\pi/L$,
and $6\pi/L$ respectively. The data are for lattice 
\ref{Lattice128x256}, $\xi^{\rm exp} = 13.636(10)$.}
\label{AntisymmetricCurrentsNextToLeadingOPE}
\end{figure}

In Fig. \ref{AntisymmetricCurrentsNextToLeadingOPE} we compare the angular 
average of ${\rm Im}\, \widehat{G}^{(a)}_{01}(t,x;\pb ,0 ;2t_s)$
(cf. Fig. \ref{AntisymmetricCurrentsNextToLeadingData})
with the OPE prediction, by defining 
\be
Y(r) = {\<\pb,c|{\cal I}^{ab}(r)|0,c\>\over 
        \<\pb,c|{\cal I}^{OPE,ab}(r)|0,c\>},
\label{defRatioY}
\ee
where ${\cal I}^{OPE,ab}(r)$ is the OPE one-loop prediction
\reff{sec4.5:SOPE}.
Here we use the form-factor prediction \reff{form-factor} for the
matrix elements of the currents, but no significant difference 
would have been observed if the matrix elements of the currents 
had been determined numerically. Use of the form-factor prediction allows 
only a reduction of the statistical errors and thus gives the 
opportunity for a stronger check of the OPE. Again we observe 
a nice agreement and a very large window in which the data are well 
described by perturbative OPE.
The systematic error is below the statistical one (which is approximately 
$10\%$) as soon as $r>2$.

\section{Conclusions} \label{sec5}

In this paper we have studied the OPE of the Noether currents 
in the two-dimensional $\sigma$-model. We have investigated in detail 
the possible problems that may arise in numerical applications. 
We can summarize the main results of this paper as follows:
\begin{enumerate}
\item[1)] The relevant correlation (four-point) functions show a good 
scaling behaviour and can be determined on relatively coarse
lattices (for instance on lattice \ref{Lattice64x128}). Lattice corrections
of order $a^2$ are negligible compared to statistical and 
systematic errors. Notice that we have used the standard action, so that 
we could have easily improved the scaling behaviour by using
an improved action \cite{improvement} and improved operators.
\item[2)] The running coupling constant $\overline{g}(\Lambda x)$ can be 
determined quite precisely. For instance, using  the finite-size scaling method
$\overline{g}(\Lambda x)$ can be computed with an error of a few percent 
for the values of $\Lambda x$ of interest. This error is practically
negligible in our applications.
\item[3)] The main source of error is in our case the perturbative 
truncation of the Wilson coefficients. If one uses one-loop
RG-improved Wilson coefficients---this also requires the two-loop
anomalous dimensions of the operators---one can compute the 
matrix elements with a precision of 5--10\%. In practice, the precision of the 
results is limited by the statistical error, which is quite large:
using $O(10^5)$ independent configurations, we obtain results with a 
statistical uncertainty of 10\%, so that improving the expressions 
for the OPE coefficients at two loops appears of little interest.
\end{enumerate}
In conclusion the method works nicely for the cases we consider.
In all cases, we observe a large window in which  the numerical 
data are well described by one-loop OPE. 

\section*{Acknowledgements}

We thank Peter Weisz for useful correspondence, Guido Martinelli,
Roberto Petronzio, Giancarlo Rossi, and Massimo Testa for suggestions and 
discussions.
\appendix
%
%
\section{Renormalization of composite operators} \label{secA}

\subsection{Continuum ${\overline{MS}}$ scheme}

In this Appendix we consider the $O(N)$ nonlinear $\sigma$-model 
in $d=2+\epsilon$ dimensions. In terms of bare fields, 
the action is written as
\begin{eqnarray}
S(\sg)&=&\frac{1}{g_B}\int d^dx \left[\frac{1}{2}
(\partial \sg_B(x))^2-h_B\sigma_B^N\right].
\label{ContinuumAction}
\end{eqnarray}
In order to perform a perturbative expansion, the $N$-vector field
$\sg_B$ is parametrized as
\be
\sg_B(x)\equiv (\pg_B(x),\sigma_B(x)), \qquad\qquad
\sigma_B(x)\equiv\sqrt{1-\pg_B^2(x)}.
\ee
The theory is renormalized by introducing the renormalized quantities
\begin{eqnarray}
g_B&\equiv&\mu^{2-d} N_d^{-1} Z_g g,\\
\pg_B(x)&\equiv & Z^{1/2}\pg(x),\\
\sigma_B(x)&\equiv& Z^{1/2}\sigma(x) = \sqrt{1-Z\pg^2(x)},\\
h_B & = & \frac{Z_g}{Z^{1/2}}h,
\end{eqnarray}
where $N_d = (4 \pi)^{-\epsilon/2}/\Gamma(1+\epsilon/2)$. 
Such a factor 
is introduced to implement naturally the \MS prescription.
The renormalization constants defined above are known to four-loop order
in perturbation theory 
\cite{Hikami:1981hi,Bernreuther:1986js,Wegner:1989ss}. Here, 
we only need their one-loop expression:
\be
Z_g  =  1+\frac{(N-2)}{2\pi\epsilon}g+ O(g^2), \qquad\qquad
Z  =  1+\frac{(N-1)}{2\pi\epsilon}g+ O(g^2).
\ee
The $\beta$-function and the anomalous dimension $\gamma(g)$ 
of the field are obtained from the renormalization constants:
\be
\beta(g) =\, \epsilon g\left[1 + {g\over Z_g} {dZ_g\over dg}\right]^{-1},
\qquad\qquad
\gamma(g) =\, {\beta(g)\over Z}\, {dZ\over dg}.
\label{RGfunctions-MS}
\ee
We also consider the renormalization of composite operators. In general,
a renormalized composite operator can be written as
\be
\opl A \opr (x)\equiv \sum_B Z_{AB} B(x),
\label{ContinuumRen}
\ee
where the $B$'s are unrenormalized composite operators, that is products of
the {\em renormalized} fields $\pg$'s, $\sigma$'s and of their derivatives. 
Which operators $B$ appear in the r.h.s. of Eq. (\ref{ContinuumRen})?
The naive answer is: all operators that 
transform under $O(N)$ as $A$ and that satisfy ${\rm dim}[B]\le {\rm dim}[A]$. 
This answer is wrong because of the magnetic term in 
Eq. (\ref{ContinuumAction})
that breaks explicitly the $O(N)$ symmetry.\footnote{
A similar problem is encountered in nonabelian gauge theories.
In order to renormalize gauge-invariant operators,
gauge noninvariant quantities must be considered.}
The correct answer has been given in \cite{Brezin:1976ap,Brezin:1976an}.
If we define the $O(N)$ noninvariant operator
\be
\alpha(x)\equiv {1\over \sigma_B} (\partial^2 \sigma_B + h_B) = 
    {1\over \sigma}\left(\partial^2 \sigma + Z_g Z^{-1} h\right),
\label{defalpha}
\ee
then we must also include all products of the type
$\alpha^k\cdot B'$ where $B'$
has the correct $O(N)$ transformation properties and satisfies
${\rm dim}[B']+2k\le {\rm dim}[A]$. 
Note that $\alpha(x)$ can be rewritten as
\be
\alpha(x) = h_B \sigma_B - (\partial\sg_B)^2 + 
   g_B \bpi_B\cdot{\delta S\over \delta \bpi_B},
\label{alphaeqbare}
\ee
so that, on-shell and in the limit $h\to 0$, we recover an 
$O(N)$ covariant mixing matrix. In other words, matrix elements 
between physical states have the correct $O(N)$ transformation properties. 

Finally, we note that, if the \MS scheme is adopted,
the renormalization constants are dimensionless so that
the operators $B$ appearing in the r.h.s.
of Eq. (\ref{ContinuumRen}) satisfy ${\rm dim}[B]={\rm dim}[A]$.

In the following we report the renormalization constants for a few cases of 
interest. 

\subsubsection{$O(N)$ invariant operators of dimension 2}

We consider $O(N)$ invariant operators of dimension 2. There are two such 
operators, $\partial_\mu \sg\cdot \partial_{\nu}\sg$ and 
           $(\partial \sg)^2$. 
For the renormalization we must also consider the 
operator $\alpha(x)$ defined in Eq. \reff{defalpha}.
They renormalize as follows:\footnote{We adopt the 
following convention in reporting the renormalization constants.
Given a basis $\{Q_1,\dots,Q_l\}$ of operators of
dimension $d$ and isospin $s$, their renormalization constants are
defined by $\opl Q_i\opr \equiv \sum_j Z^{(d,s)}_{ij}Q_j$.}
\begin{eqnarray}
\opl \partial_\mu   \sg\cdot   \partial_{\nu}\sg \opr & = &
Z^{(2,0)}_{11}\partial_\mu   \sg\cdot   \partial_{\nu}\sg+
Z^{(2,0)}_{12}(\partial\sg)^2\delta_{\mu   \nu} +
Z^{(2,0)}_{13}\alpha\delta_{\mu   \nu},
\label{renormMS}\\
\opl (\partial\sg)^2\opr & = & Z^{(2,0)}_{22}(\partial\sg)^2+
Z^{(2,0)}_{23}\alpha,
\label{renormMS2}\\
\opl \alpha\opr & =  &Z^{(2,0)}_{32}(\partial\sg)^2+
Z^{(2,0)}_{33}\alpha.
\label{renormMS3}
\end{eqnarray}
A particular combination of these operators is the energy-momentum tensor
\begin{eqnarray}
\opl T_{\mu   \nu}\opr 
\equiv\frac{1}{g}\opl \partial_\mu \sg\cdot \partial_{\nu}\sg-
\delta_{\mu   \nu}\frac{1}{2}(\partial\sg)^2\opr 
= Z_{T,T}\frac{1}{g} 
\left(\partial_\mu   \sg\cdot   \partial_{\nu}\sg 
-\frac{1}{2}\delta_{\mu   \nu}(\partial\sg)^2\right).
\label{EnergyMomentum}
\end{eqnarray}
For this particular set of operators, the renormalization constants 
can be directly related to $Z$ and $Z_g$.
Indeed, one first observes that $T_{\mu \nu}$ is conserved. 
Therefore, it is finite
when it is expressed in terms of bare fields. It follows:
\begin{eqnarray}
Z_{TT} = Z^{(2,0)}_{11} = \frac{Z}{Z_g}.
\label{EMRen}
\end{eqnarray}
Then, one notices that differentiation of the action
(\ref{ContinuumAction}) with respect to $g$ (and keeping
$d$, $h$ and $\pg$ constant) gives a finite operator, which is 
a combination of $\alpha$ and $(\partial\sg)^2$. For 
$g=0$ this combination reduces to $(\partial\sg)^2$. Therefore, it can be 
identified with $\opl (\partial\sg)^2\opr$. The renormalization constants 
immediately follow:
\begin{eqnarray}
Z^{(2,0)}_{22} & = & \frac{Z}{Z_g}-g\frac{\partial}{\partial g}
\left(\frac{Z}{Z_g}\right)= 1+O(g^2),\\
Z^{(2,0)}_{23} & = & -\frac{1}{Z_g}g\frac{\partial}{\partial g}\log Z=
-\frac{N-1}{2\pi\epsilon}g+O(g^2).
\end{eqnarray}
A second relation is obtained by using the fact that the equations of motion 
do not need any renormalization. It follows that the combination
\be
{Z\over Z_g} (\partial\sg)^2 + {1\over Z_g}\alpha
\ee
is finite. Combining this result with what we obtained above, we conclude that 
\begin{eqnarray}
Z^{(2,0)}_{32} & = & {Z\over Z_g} - Z_{22}^{(2.0)} = 
  g\frac{\partial}{\partial g}\left(
\frac{Z}{Z_g}\right)=\frac{1}{2\pi\epsilon}g + O(g^2), \\
Z^{(2,0)}_{33} & = & {1\over Z_g} - Z_{23}^{(2.0)} = 
\frac{1}{Z_g}\left(1+ g\frac{\partial}{\partial g}
\log Z\right)=1+\frac{1}{2\pi\epsilon}g + O(g^2).
\label{AlphaRen}
\end{eqnarray}
The remaining renormalization constants can be obtained by using 
Eqs. (\ref{renormMS}-\ref{EnergyMomentum}):
\begin{eqnarray}
Z^{(2,0)}_{12} & = & {1\over2}\left(Z^{(2,0)}_{22} - Z_{TT}\right) = 
 -\frac{1}{2}g\frac{\partial}{\partial g}
\left(\frac{Z}{Z_g}\right) = -\frac{1}{4\pi\epsilon}g + O(g^2),\\
Z^{(2,0)}_{13} & = & {1\over2} Z^{(2,0)}_{23} = 
-\frac{1}{2Z_g}g\frac{\partial}{\partial g}
\log Z = -\frac{N-1}{4\pi\epsilon}g + O(g^2).
\end{eqnarray}
Note that Eq. \reff{alphaeqbare} holds also for the renormalized operators. 
Indeed, using the expressions of the renormalization constants reported above,
we can show that 
\begin{eqnarray}
\opl \alpha\opr (x) = h\sigma(x)-\opl (\partial\sg)^2\opr (x)+
g \bpi\cdot {\delta S\over \delta \bpi(x)}.
\label{AlphaOnShell}
\end{eqnarray}
In the limit $h\to 0$ and on-shell,
Eq. (\ref{AlphaOnShell}) allows us to express
matrix elements of the $O(N)$ noninvariant operator $\opl \alpha\opr$
in terms of those of the $O(N)$ invariant operator $\opl(\partial\sg)^2\opr$.

Finally, it is possible to compute the trace of the renormalized
energy-momentum tensor that gives the conformal anomaly. Using 
Eq. \reff{EnergyMomentum} we have
\be
T \equiv \delta_{\mu\nu} \opl T_{\mu\nu}\opr =
         \opl T_{11} + T_{22}\opr = - {\epsilon\over2} Z_{TT} 
   {1\over g} (\partial\sg)^2.
\ee
Then, using Eqs. \reff{renormMS2} and \reff{renormMS3}, we can rewrite 
\be
(\partial\sg)^2 =\, 
 {Z_g\over Z\epsilon} \left\{ \gamma(g) \opl \alpha\opr + 
   \left({\beta(g)\over g} + \gamma(g)\right)
\opl(\partial\sg)^2\opr\right\},
\ee
where we have expressed $dZ/dg$ and $dZ_g/dg$ in terms of 
$\beta(g)$ and $\gamma(g)$ using Eqs. \reff{RGfunctions-MS}. 
We finally obtain
\be
T = - {1\over 2g^2}[\beta(g) + g \gamma(g)] \opl(\partial\sg)^2\opr - 
      {\gamma(g)\over 2 g} \opl \alpha\opr.
\label{tracciaemt-continuo}
\ee
 
\subsubsection{Antisymmetric operators in two $O(N)$ indices}

We shall now consider operators that are antisymmetric $O(N)$ tensors 
with two indices. 
Obviously, there is no such operator of 
dimension zero. The only antisymmetric operator of dimension $1$ is the 
Noether current corresponding to the $O(N)$ symmetry:
\begin{eqnarray}
j^{ab}_\mu   \equiv\frac{1}{g}\opl\sigma^{a}\partial_\mu   \sigma^{b}-
\sigma^{b}\partial_\mu   \sigma^{a}\opr & = &
Z^{(1,1)}\frac{1}{g}\left(\sigma^{a}\partial_\mu   \sigma^{b}-
\sigma^{b}\partial_\mu   \sigma^{a}\right).
\end{eqnarray}
Conservation of $j^{ab}_\mu   $ implies
\begin{eqnarray}
Z^{(1,1)} = \frac{Z}{Z_g}.
\end{eqnarray}
There are three operators of dimension $2$ with the same symmetry. 
All of them can be expressed in terms of $\partial_\mu j_\nu^{ab}$. 
Therefore, they renormalize as the current itself.
%
%
\subsection{Lattice} \label{secA.2}

We now consider the lattice action (\ref{action-lattice-conmassa}).
If one neglects lattice artifacts of order $a^2\log^p a$,
lattice correlation functions of $\pg$ and $\sigma$ fields differ 
by a finite renormalization from those computed in any other scheme.
Here we want to relate lattice correlations to the corresponding 
continuum ones computed in the \MS scheme. The finite 
renormalization is given by
\begin{eqnarray}
g_L & = & Z_g^L g,\\
\pg_x & = & (Z^L)^{1/2}\pg(x),\\
\sigma_x & = & (Z^L)^{1/2}\sigma(x),
\end{eqnarray}
where the lattice fields appear in the left-hand side,
the renormalized continuum fields in the right-hand side.
The renormalization constants $Z^L$ and $Z^L_g$ 
have been computed to three-loop order in
perturbation theory \cite{Caracciolo:1995bc,Alles:1999fh}. We shall only need
the one-loop result:
\be
Z_g^L  =  1+\frac{N-2}{4\pi}g_L\log (\frac{\mbar^2 a^2}{32})-\frac{1}{4}g_L 
+ O(g_L^2), \qquad\qquad
Z^L  =  1+\frac{N-1}{4\pi}g_L\log (\frac{\mbar^2 a^2}{32}) + O(g_L^2).
\ee
Analogously, we can relate \MS-renormalized composite operators to the 
lattice corresponding ones. As in the continuum, besides $O(N)$ invariant
operators, we must also consider operators involving 
$\alpha^L$, cf. Eq. \reff{defalphalatt}. However, also on the lattice, 
$O(N)$ invariance is recovered when considering matrix elements 
between on-shell states. A second complication in the lattice approach
is the loss of Lorentz invariance: as a consequence, additional 
operators that are only cubic invariant must be considered. Below we report
the renormalization constants for a few cases of interest.

\subsubsection{$O(N)$-invariant operators of dimension 2} \label{secA.2.1}

On the lattice we have:\footnote{In the following equations we have not reported
an additional contribution proportional to $\mbox{\bf 1}/a^2$, where 
$\mbox{\bf 1}$ is the identity operator. This term does not 
contribute to connected correlation functions, and for this reason,
we have not written it explicitly.}
\begin{eqnarray}
\opl \partial_\mu   \sg\cdot \partial_{\nu}\sg\opr (x)
&=& Z^{L(2,0)}_{11}
   (\overline{\partial}_\mu  \sg\cdot  \overline{\partial}_{\nu}\sg)_x
+Z^{L(2,0)}_{12}\delta_{\mu   \nu}(\overline{\partial}_\mu   \sg)_x^2+
\nonumber\\
&&+Z^{L(2,0)}_{13}\delta_{\mu   \nu}
(\overline{\partial}\sg)_x^2
+Z^{L(2,0)}_{14}\delta_{\mu   \nu}\alpha^L_x, \label{Ren20.1}\\
\delta_{\mu   \nu}\opl(\partial_\mu   \sg)^2\opr (x)& = & 
Z^{L(2,0)}_{22}\delta_{\mu   \nu}(\overline{\partial}_\mu   \sg)_x^2
+Z^{L(2,0)}_{23}\delta_{\mu   \nu}
(\overline{\partial}\sg)_x^2+\nonumber\\
&&+Z^{L(2,0)}_{24}\delta_{\mu   \nu}\alpha^L_x, \\
\opl(\partial \sg)^2\opr (x) & = & Z^{L(2,0)}_{33}
(\overline{\partial}\sg)_x^2+Z^{L(2,0)}_{34}\alpha^L_x,\\
\opl \alpha \opr (x) & = & Z^{L(2,0)}_{43}
(\overline{\partial}\sg)_x^2+Z^{L(2,0)}_{44}\alpha^L_x. \label{Ren20.4}
\end{eqnarray}
Note the presence of an additional operator, which is cubic but not 
$O(N)$ invariant. At one loop  we obtain:\footnote{
Some of the renormalization constants given below 
have been computed in Ref. \cite{Buonanno:1995us}. We report them here 
to correct a few misprints.}
\begin{eqnarray}
Z^{L(2,0)}_{11} & = & 1-\frac{N-2}{4\pi}g_L\log (\frac{\mbar^2 a^2}{32})+
\frac{1}{\pi}g_L + O(g_L^2), \\
Z^{L(2,0)}_{12} & = & \left(\frac{1}{2}-\frac{3}{2\pi}\right)g_L 
+ O(g_L^2),\\
Z^{L(2,0)}_{13} = 
Z^{L(2,0)}_{23} & = & -\frac{1}{8\pi}g_L\log (\frac{\mbar^2 a^2}{32})
-\frac{1}{2\pi}g_L+ O(g_L^2),\\
Z^{L(2,0)}_{14} = 
Z^{L(2,0)}_{24} & = & -\frac{N-1}{8\pi}g_L\log (\frac{\mbar^2 a^2}{32})-
\frac{N-1}{8}g_L + O(g_L^2),\\
Z^{L(2,0)}_{33} & = & 1-\frac{N-1}{4\pi}g_L\log (\frac{\mbar^2 a^2}{32})+
\left(\frac{1}{2}-\frac{5}{4\pi}\right)g_L + O(g_L^2),\\
Z^{L(2,0)}_{34} & = & -\frac{N-1}{4\pi}g_L\log (\frac{\mbar^2 a^2}{32})+
(N-1)\left(\frac{1}{4\pi}-\frac{1}{4}\right)g_L + O(g_L^2),\\
Z^{L(2,0)}_{43} & = & \frac{1}{4\pi}g_L\log (\frac{\mbar^2 a^2}{32})+
\frac{1}{4}g_L + O(g_L^2),\\
Z^{L(2,0)}_{44} & = & 1+\frac{1}{4\pi}g_L\log (\frac{\mbar^2 a^2}{32})+
\frac{1}{4}g_L + O(g_L^2),
\end{eqnarray}
and $Z^{L(2,0)}_{22} = Z^{L(2,0)}_{11} + Z^{L(2,0)}_{12}$.
Finally, for the energy-momentum tensor, we have
\begin{eqnarray}
\opl T_{\mu   \rho}\opr (x)&=& Z^{L(2,0)}_{TT}T^{L}_{\mu   \rho,x}+
Z^{L(2,0)}_{T2}
\frac{1}{g_L}\delta_{\mu   \rho}(\overline{\partial}_\mu   \sg)_x^2+
\nonumber \\
&&+Z^{L(2,0)}_{T3}\frac{1}{g_L}\delta_{\mu   \rho}
(\overline{\partial}\sg)_x^2
+Z^{L(2,0)}_{T4}\frac{1}{g_L}\delta_{\mu   \rho}\alpha^L_x,
\label{EnergyMomentumRenormalization}
\end{eqnarray}
with
\begin{eqnarray}
Z^{L(2,0)}_{TT} & = & 1+\left(\frac{1}{\pi}-\frac{1}{4}\right)g_L + O(g_L^2), 
\label{EMZLatticePT}\\
Z^{L(2,0)}_{T2} & = & \left(\frac{1}{2}-\frac{3}{2\pi}\right)g_L + O(g_L^2),\\
Z^{L(2,0)}_{T3} & = & \left(\frac{5}{8\pi}-\frac{1}{4}\right)g_L + O(g_L^2),\\
Z^{L(2,0)}_{T4} & = & -\frac{N-1}{8\pi}g_L + O(g_L^2).
\end{eqnarray}
Note that, since $\opl T_{\mu\rho}\opr$ is conserved, 
\begin{eqnarray}
\left.\frac{\partial}{\partial a}\right|_{g_L}Z^{L(2,0)}_{Tj}=0.
\end{eqnarray}
Finally, notice that we can eliminate $\alpha_x^L$ using the 
lattice equations of motion 
\be
\alpha_x^L = h \sigma_x^N  +\sg_x\cdot\partial^2 \sg_x + 
   {1\over a^2} 
   g_L \bpi_x\cdot {\delta S^{\rm latt}_{\rm TOT}\over \delta \bpi_x}-
   {g_L\over a^2} \, .
\ee
However, $-\sg_x\cdot\partial^2 \sg_x$ is different from 
$(\overline{\partial} \bsigma)^2_x$ appearing in all previous formulae. 
The two operators are related by a finite renormalization:
\be
-\sg_x\cdot\partial^2 \sg_x = \zeta_1 (\overline{\partial} \bsigma)^2_x + 
                     \zeta_2 \alpha_x^L.
\ee
At one loop
\be
\zeta_1 = 1 + {g_L\over 2} \left(1 - {5\over 2\pi}\right), \qquad\qquad
\zeta_2 = {N-1\over4} \left({1\over \pi} - 1\right) g_L.
\ee
Then, considering only connected correlation functions, 
on-shell and for $h\to 0$, we have
\be
\alpha_x^L =\, - {\zeta_1\over 1 + \zeta_2} (\overline{\partial} \bsigma)^2_x.
\ee
Using this relation, we can compute $\widetilde{Z}_{T3}^{L,(2,0)}$ appearing 
in Eq. \reff{Tmunulattice}. Explicitly
\be
\widetilde{Z}_{T3}^{L,(2,0)} = Z^{L(2,0)}_{T3} - 
    {\zeta_1\over 1 + \zeta_2} Z^{L(2,0)}_{T4}.
\ee
\subsubsection{Antisymmetric operators in two $O(N)$ indices} \label{secA.2.2}

Let us consider operators which are $O(N)$
antisymmetric tensor with two indices. As we discussed in the continuum
case, there is only one relevant operator of dimension 1, the $O(N)$
current. Since the $O(N)$ symmetry is preserved on the lattice, there is 
a lattice discretization, given
in Eq. \reff{jlattice}, that is exactly conserved.  It follows that 
\begin{eqnarray}
j^{ab}_\mu   (x) = j^{L,ab}_{\mu,x},
\end{eqnarray}
i.e. $Z^{L(1,1)} = 1$. 

As in the continuum there are three operators of dimension 2. If on the lattice 
we express them in terms of $\partial^- j^{L,ab}_{\mu,x}$, then also these
operators do not need any renormalization. 

%
\section{OPE for $O(N)$ Noether currents} \label{secB}
In this Appendix we report the explicit one-loop expressions of the 
OPE coefficients reported in Sec. \ref{sec.OPE}.

\subsection{Continuum} \label{secB.1}
At one-loop order the Wilson coefficients appearing in Eq. \reff{scalcorrcont}
are the following:
\begin{eqnarray}
W_{0,\mu   \nu}(x)&=& \delta_{\mu   \nu}\frac{N-1}{8\pi}\left[
1-\frac{N-2}{2\pi}g(\gamma+\log(\mbar x))\right]-\nonumber\\
&&-x_\mu   x_{\nu}\frac{N-1}{4\pi x^2}\left[1-\frac{N-2}{2\pi}g\left(\gamma+
\log(\mbar x)+\frac{1}{2}\right)\right] + O(g^2),
\\
W_1(x) & = & -\frac{N-2}{4\pi}g + O(g^2),\\
W_2(x) & = & \frac{N-2}{2\pi}g + O(g^2),\\
W_3(x) & = & \frac{N-2}{2\pi}g + O(g^2),\\
W_4(x) & = & 1-\frac{N-2}{2\pi}g(\gamma+\log(\mbar x)) + O(g^2),\\
W_5(x) & = & \frac{3N-5}{4\pi}g + O(g^2),\\
W_6(x) & = & \frac{1}{2}\left[ 1-\frac{N-2}{4\pi}g-\frac{N-3}{2\pi}g
\left(\gamma+\log(\mbar x)\right)\right]  + O(g^2),\\
W_7(x) & = & \frac{N-1}{4\pi}g + O(g^2),\\
W_8(x) & = & \frac{N-1}{4\pi}g\left(\gamma+\log(\mbar x)\right) + O(g^2).
\end{eqnarray}
The coefficients appearing in Eq. \reff{anticorrcont} are given by:
\begin{eqnarray}
U_{00}(x) & = & \frac{N-2}{2\pi}g + O(g^2),  \\
U_{01}(x) & = & - \frac{N-2}{4\pi}g + O(g^2), \\
U_{02}(x) & = &  \frac{N-2}{4\pi}g + O(g^2), \\
U_1(x) & = &  1-\frac{N-2}{2\pi}g(\gamma+ \log(\mbar x)) + \frac{N-6}{4\pi}g 
     + O(g^2),\\
U_2(x) & = & -\frac{N-2}{4\pi}g + O(g^2).
\end{eqnarray}
\subsection{Constraints on the OPE coefficients} \label{secB.2}

In this Section we want to derive the constraints due to the current 
conservation on the coefficients of the OPE. First, we need 
the Ward identity related to the $O(N)$ invariance in the presence of a magnetic
term $h$. A simple calculation gives:
\be
\< \partial_\mu j_\mu^{ab}(x) {\cal O}\> = \,
  {h\over g} \< (\delta^{Na} \sigma^b(x) - \delta^{Nb} \sigma^a(x)) {\cal O}\>
  + \left \< {\delta {\cal O}\over \delta \sigma^a(x)} \sigma^b(x) -
       {\delta {\cal O}\over \delta \sigma^b(x)} \sigma^a(x) \right\>.
\label{Wardid-conmassa}
\ee
Then we need the OPE of $j_\mu^{ab}(x)\sigma^c(0)$. 
The leading term for $x\to 0$ has the form 
\be
j_\mu^{ab}(x)\sigma^c(0) = {x_\mu\over x^2} f(\mu x; g) 
   (\delta^{ac} \sigma^b(0) - \delta^{bc}\sigma^a (0) ).
\ee
Using Eq.
\reff{Wardid-conmassa} and noticing that $\sigma^a(x)\sigma^b(0)\sim O(1)$
for $x\to0$, we have $\partial f(\mu x; g)/\partial x^2 = 0$. Thus, 
$f(\mu x; g)$ is a function 
of $g$ only. But the Wilson coefficient satisfies the RG equation
\be
\left( \mu{\partial\over\partial\mu} + \beta(g){\partial\over\partial g}\right)
   f = 0.
\ee
Thus, if it is independent of $x$, and therefore of $\mu$, it is also 
independent of $g$. A simple calculation at tree level gives then
\be
j_\mu^{ab}(x)\sigma^c(0) = {1\over 2\pi} {x_\mu\over x^2}
   (\delta^{ac} \sigma^b(0) - \delta^{bc}\sigma^a (0) ).
\label{OPE-j-sigma}
\ee
The same result has been obtained in \cite{Luescher:1978} using the 
canonical formalism. 

We now consider the OPE of the scalar product of currents. 
Using the Ward identity \reff{Wardid-conmassa} and Eq. \reff{OPE-j-sigma}
we have for $x\to 0$
\be
{1\over2} \sum_{ab} \partial_\mu j^{ab}_\mu(x) j^{ab}_\nu(0) = 
   {h\over g} \sigma^b(x) j_\nu^{Nb} (0) =
   {h\over g} {N-1\over 2\pi} {x_\nu\over x^2} \sigma^N(0),
\ee
where we have discarded contact terms. Then, using Eq. \reff{AlphaOnShell} 
and discarding again contact terms, we obtain for $x\to 0$
\be
{1\over2} \sum_{ab} \partial_\mu j^{ab}_\mu(x) j^{ab}_\nu(0) =
 {N-1\over 2\pi g} {x_\nu\over x^2} \left\{
 \opl\alpha\opr(0) + \opl (\partial\sg)^2\opr (0) \right\}.
\ee
This equation implies the following relations on the Wilson coefficients:
\begin{eqnarray}
&& x^2 \partial_\mu W_{0,\mu\rho} (x) = 2x_{\mu}W_{0,\mu\rho} (x) ,
\\
&& 2 x^2 {\partial\over \partial x^2}\left[
   W_1(x) + W_2(x) + W_3(x) \right] = 2 W_1(x) - W_2(x) + 2 W_3(x),
\\
&& x^2 {\partial\over \partial x^2}\left[
   W_3(x) + W_4(x) \right] = - W_1(x) - W_3(x),
\\
&& 2 x^2 {\partial\over \partial x^2}\left[
   W_5(x) + W_6(x) \right] = 
   {N-1\over 2\pi} g + {1\over 2g} \left[\beta(g) + g\gamma(g)\right] W_3(g) 
  - W_5(x),
\\
&& 2 x^2 {\partial\over \partial x^2}\left[
   W_7(x) + W_8(x) \right] = 
   {N-1\over 2\pi} g + {1\over 2} \gamma(g) W_3(g) 
  - W_7(x).
\end{eqnarray}
In the derivation we used Eq. \reff{tracciaemt-continuo}
for the trace of the energy-momentum tensor. 

Now let us consider the antisymmetric case. Using the Ward identity
\reff{Wardid-conmassa} and Eq. \reff{OPE-j-sigma}, we obtain for $x\to 0$
\be
\sum_c \partial_\mu j_\mu^{ac}(x) j_\nu^{bc}(0) - 
\partial_\mu j_\mu^{bc}(x) j_\nu^{ac}(0) =\,
 {N-2\over 2\pi} {x_\nu\over x^2} {h\over g} 
 \left(\delta^{Na}\sigma^b(0) - \delta^{Nb}\sigma^a(0)\right) = 
 {N-2\over 2\pi} {x_\nu\over x^2} \partial_\mu j_\mu^{ab}(0).
\ee
Again, contact terms have been discarded in the derivation.
Using this relation, we obtain the following constraints 
on the Wilson coefficients\footnote{
Equations \reff{anti-constr-eq1}--\reff{anti-constr-eq5} have been
derived in Ref. \cite{Luescher:1978}. Eq. \reff{anti-constr-eq6} 
is due to the matching of the terms proportional to 
$\partial_\mu j_\mu^{ab}(0)$. It is also a simple consequence 
of Eqs. \reff{anti-constr-eq2} and \reff{anti-constr-eq4} and of the 
RG equations.}
\begin{eqnarray}
&& x^2 {\partial\over \partial x^2}
  \left[ U_{00}(x) + U_{01}(x) + U_{02}(x) \right] = 
  U_{01}(x) + U_{02}(x),
\label{anti-constr-eq1} \\
&& 2 x^2 {\partial\over \partial x^2} U_{02}(x) = 
  - U_{01}(x)  - U_{02}(x),
\label{anti-constr-eq2} \\
&& x^2 {\partial\over \partial x^2}
  \left[ U_1(x) + U_{02}(x)\right] = - U_{02}(x),
\label{anti-constr-eq3} \\
&& x^2 {\partial\over \partial x^2}
  \left[ U_2(x) + U_{02}(x)\right] = - U_{01}(x)  - U_{02}(x),
\label{anti-constr-eq4} \\
&& 2 x^2 {\partial\over \partial x^2}
  \left[ - U_2(x) + U_{00}(x) + U_{01}(x) + U_{02}(x) \right] =
\nonumber \\
   && \qquad\qquad - 2 U_2(x) - U_{00}(x) + 2 U_{01}(x) + 2 U_{02}(x),
\label{anti-constr-eq5} \\
&& U_{02}(x) - U_2(x) = {N-2\over2\pi} g.
\label{anti-constr-eq6} 
\end{eqnarray}
Using Eqs. \reff{anti-constr-eq1} and \reff{anti-constr-eq2} 
we obtain immediately
\be
{\partial\over \partial x^2}
   \left[ U_{00}(x) + U_{01}(x) + 3 U_{02}(x)\right] =\, 0,
\ee
which implies that this combination is $x$ and $\mu$ independent. 
By making use of the RG equations \reff{RG-Wilson-antisymmetric}
one proves that this combination 
is determined uniquely by its one-loop value. Then, using the 
results of the previous Section, we obtain:
\be
U_{00}(x) + U_{01}(x) + 3 U_{02}(x) = {N-2\over \pi} g.
\label{vincolo1}
\ee
Thus, using \reff{vincolo1} and \reff{anti-constr-eq2}, 
$U_{00}(x)$ and $U_{01}(x)$ are uniquely determined by $U_{02}(x)$. Moreover,
using Eqs. \reff{vincolo1} and \reff{anti-constr-eq6} one immediately verifies 
that Eqs. \reff{anti-constr-eq4} and \reff{anti-constr-eq5} are 
equivalent to Eq. \reff{anti-constr-eq2}. 

Finally, consider \reff{anti-constr-eq3}. We will now show that this equation
provides the two-loop estimate of $U_{02}(x)$. Indeed, since 
\be
\left(\mu {\partial\over \partial\mu} + 
      \beta(g) {\partial\over \partial g} - {\beta(g)\over g}\right)
   [U_1(x) + U_{02}(x)] = 0
\ee
and $U_1(x) + U_{02}(x) = 1 + O(g)$, cf. previous Section, we have
\be
U_1(x) + U_{02}(x) = 1 - (\beta_0 \log \mu x + a_0) g -
                      (\beta_1 \log \mu x + a_1) g^2 + O(g^3),
\ee
where $\beta(g) = - g^2 \sum_{k=0}\beta_k g^k$, and $a_0$, $a_1$ 
are constants that are not fixed by the RG equation. Plugging 
this expression into \reff{anti-constr-eq3}, we obtain
\be
U_{02}(x) = {\beta_0\over2} g + {\beta_1\over2} g^2 + O(g^3) = \, 
   {N-2\over 4\pi} g + {N-2\over 8 \pi^2} g^2 + O(g^3).
\ee
Of course, the result at order $g$ agrees with the expression reported
in Sec. \ref{secB.1}. Correspondingly we obtain
\begin{eqnarray}
U_{00}(x) &=&   {N-2\over 2\pi} g - {N-2\over 4 \pi^2} g^2 + O(g^3), \\
U_{01}(x) &=& - {N-2\over 4\pi} g - {N-2\over 8 \pi^2} g^2 + O(g^3), \\
U_{2}(x)  &=& - {N-2\over 4\pi} g + {N-2\over 8 \pi^2} g^2 + O(g^3).
\end{eqnarray}
Let us finally mention that in Ref. \cite{Luescher:1978} it was argued 
that the functions $U_{0i}(x)$ are one-loop exact, in the sense that 
there are no corrections of order $g^k$, $k\ge 2$. As we have shown above
and it has also been recognized by the 
author,\footnote{In Ref. \cite{Luescher:1978Erratum}
it was also shown that, even though the expressions for the Wilson
coefficients were incorrect, one could still modify the argument
so that the main result (existence of a conserved charge) of 
Ref. \cite{Luescher:1978} remains true.}
this is inconsistent with the RG equations.

%
\subsection{Lattice} \label{secB.3}
At one-loop order the Wilson coefficients appearing in Eq. \reff{scalcorrlatt}
are the following:
\begin{eqnarray}
W^L_{0,\mu   \rho}(x)&=& \delta_{\mu   \rho}\frac{N-1}{8\pi }\left[
1-\frac{N-2}{4\pi}g_L(2\gamma+\log(32 x^2))-\frac{1}{4}g_L\right]-\nonumber\\
&&-x_\mu   x_{\rho}\frac{N-1}{4\pi x^2}\left[1-\frac{N-2}{4\pi}g\left(2\gamma+
\log(32 x^2)+1\right)-\frac{1}{4}g_L\right] + O(g^2_L), 
\\
W^L_1(x) & = & -\frac{N-2}{4\pi}g_L + O(g^2_L),\\
W^L_2(x) & = & \frac{N-2}{2\pi}g_L + O(g^2_L),\\
W^L_3(x) & = & \frac{N-2}{2\pi}g_L + O(g^2_L),\\
W^L_4(x) & = & 1-\frac{N-2}{4\pi}g_L(2\gamma+\log(32 x^2))+\left(\frac{1}{\pi}
-\frac{1}{2}\right)g_L + O(g^2_L),\\
\widehat{W^L_1}(x) & = & O(g^2_L),\\
\widehat{W^L_2}(x) & = & O(g^2_L),\\
\widehat{W^L_3}(x) & = & O(g^2_L),\\
\widehat{W^L_4}(x) & = & \left(\frac{1}{2}-\frac{3}{2\pi}\right) g_L + 
              O(g^2_L), \\
W^L_5(x) & = & \frac{3N-5}{4\pi}g_L + O(g^2_L),\\
W^L_6(x) & = & \frac{1}{2}\left\{1-\frac{N-2}{4\pi}g_L-\frac{N-3}{4\pi}g_L
\left(2\gamma+\log(32 x^2)\right)-\frac{1}{4}g_L\right\} + O(g^2_L),\\
W^L_7(x) & = & \frac{N-1}{4\pi}g_L + O(g^2_L),\\
W^L_8(x) & = & \frac{N-1}{8\pi}g_L\left(2\gamma+\log(32 x^2)-\pi\right)
+ O(g^2_L).
\end{eqnarray}
At one-loop order the Wilson coefficients appearing in Eq. \reff{anticorrlatt}
are the following:
\begin{eqnarray}
U^L_{00}(x) & = & \frac{N-2}{2\pi}g_L + O(g_L^2), \\
U^L_{01}(x) & = & - \frac{N-2}{4\pi}g_L + O(g_L^2), \\
U^L_{02}(x) & = &  \frac{N-2}{4\pi}g_L + O(g_L^2), \\
U^L_1(x) & = &  1-\frac{N-2}{4\pi}g_L(2\gamma+
\log(32 x^2))-\frac{1}{4}g_L + \frac{N-6}{4\pi}g_L + O(g_L^2),\\
U^L_2(x) & = & -\frac{N-2}{4\pi}g_L + O(g_L^2).
\end{eqnarray}
\section{Evaluation of the running coupling constant}
\label{RunningCoupling}
The determination of the running coupling constant is a key ingredient in 
the application of RG-improved perturbation theory to any
asymptotically free theory. 
If we use the lattice OPE, the coupling is the bare $g_L$,
one of the input parameters of our numerical calculations. However,
perturbation theory in $g_L$ is poorly behaved, so that one expects 
a poor agreement with the numerical data. It is known that it is much
more convenient to use perturbative expansions in the \MS scheme. 
Perturbative coefficients are smaller, so that truncations in the number 
of loops give smaller systematic errors. For these reasons, it is important
to relate the \MS coupling $g_{\overline{MS}}(\mu)$ to the bare coupling $g_L$.

Here, we shall outline several different procedures---all of them
are exact in the continuum limit---and we shall compare 
their efficiency. We consider the following methods:
\begin{enumerate}
\item \label{NaivePerturbative}
The {\em naive perturbative} method. 

In this scheme, one computes 
the continuum coupling as a function of the lattice coupling
by matching the continuum and the lattice perturbative expansion of
some physical quantity, e.g., of the two-point correlation function.
At $l$-loops one obtains a truncated series of the form
$g_{\overline{MS}} (\mu a ) = g_L + \sum_{k=2}^{l+1} c_k(\mu a) g_L^k$.
We shall call $g_{\overline{MS}}^{{\rm np},l}(\mu a )$ 
the value obtained in this way. The relevant perturbative expansions 
are known to three loops \cite{Caracciolo:1995bc,Alles:1999fh}.
\item \label{RGImprovedPerturbative}
The {\em RG improved perturbative} method. 

The idea of this method---and also of those that will be presented below---is
to compute $g_{\overline{MS}}(\mu)$ starting from some quantity that can 
be computed
numerically at the given value of the bare lattice coupling constant.

In this case, we consider the RG prediction for the mass gap
$m$ in the \MS scheme:
\begin{eqnarray}
m & = & \widehat{C}_N\Lambda_{\overline{MS}}(\mu ,g_{\overline{MS}}(\mu )),
\label{MassGap}
\end{eqnarray}
where 
\begin{eqnarray}
\Lambda_{\overline{MS}}\equiv \mu   
\; e^{-\frac{1}{\beta_0 g}}(\beta_0 g)^{-\beta_1/\beta_0^2}
\exp\left\{-\int_0^g \left(\frac{1}{\beta(z)}+
\frac{1}{\beta_0z^2}-\frac{\beta_1}{\beta_0^2z}\right)dz\right\},
\label{LambdaDefinitionMS}
\end{eqnarray}
and $\beta(g) = - g^2 \sum \beta_n g^n$ is the \MS $\beta$-function. 
The constant $\widehat{C}_N$ is not known for a general theory. For the 
two-dimensional $\sigma$-model it has been computed 
\cite{Hasenfratz:1990ab,Hasenfratz:1990zz} using the 
thermodynamic Bethe ansatz. The result, in the \MS scheme,
reads:
\begin{eqnarray}
\widehat{C}_N= \left(\frac{8}{e}\right)^{\frac{1}{N-2}}
\frac{1}{\Gamma(1+(N-2)^{-1})}\; .
\label{BetheAnsatzPrediction}
\end{eqnarray}
The method works as follows: For a given value of the lattice coupling 
$g_L$, compute numerically (for instance, by means of a Monte Carlo simulation)
the mass gap $ma$. 
Then, fix $\mu a$ and solve numerically
Eq. \reff{MassGap}, obtaining $g_{\overline{MS}}(\mu/m)$. Note that,
since the $\beta$-function is known only to a finite
order in perturbation theory, we have to substitute the function 
$\Lambda_{\overline{MS}}(\mu ,g)$ with its truncated perturbative expansion.
There is some arbitrariness in this truncation. We shall make the
simplest choice
\begin{eqnarray}
\Lambda_{\overline{MS}}^{(l)}(\mu ,g)\equiv \mu   \; e^{-\frac{1}{\beta_0 g}}
(\beta_0 g)^{-\beta_1/\beta_0^2}\left[1+\sum_{k=1}^{l-2}\lambda_k g^k\right],
\label{LambdaTruncatedDefinition}
\end{eqnarray}
where the coefficients $\lambda_k$ are obtained by expanding perturbatively
Eq. (\ref{LambdaDefinitionMS}). Equation
(\ref{LambdaTruncatedDefinition}) gives the $l$-loop
approximation of the $\Lambda$-parameter. The solution of the
corresponding Eq. (\ref{MassGap}) will be denoted as
$g^{{\rm rgp},l}_{\overline{MS}}(\mu/m )$.
The perturbative expansion of the \MS $\beta$-function is known
to four loops \cite{Bernreuther:1986js}.
\item \label{FiniteSizeNonPerturbative}
The {\em finite-size nonperturbative} method. 

This method,
due to L\"uscher \cite{Luscher:1982aa}, was initially 
tested in the two-dimensional $O(3)$ $\sigma$-model
\cite{Luscher:1991wu}.
Recently, it has been successfully
employed in the computation of the $\Lambda$-parameter in quenched QCD
\cite{Luscher:1994gh}. The idea is to consider the  
theory in a finite box and to define a ``finite-size scheme''
in which the renormalization scale is the size of the box. 
For the $\sigma$-model, Ref. \cite{Luscher:1991wu} introduces a 
coupling $g_R(1/L)$ defined as follows:\footnote{This definition 
is by no means unique. For instance, one could also use 
$g_R(1/L) = [m(L)L]^2/(N-1)$, where $m(L)$ is the inverse 
of the second-moment correlation length on a square lattice of size $L/a$. 
The corresponding universal finite-size scaling function---i.e. 
the function that gives the correspondence between 
$g_R(1/L)$ and $mL$---has been determined numerically
in \cite{Caracciolo:1995ud}.}
\be
g_R(1/L) = {2 m(L) L\over N-1},
\ee
where $m(L)$ is the mass gap in a strip of width $L$. Standard finite-size 
scaling theory indicates that $g_R$ is a universal function 
of $mL$, where $m$ is the {\em infinite-volume} mass gap. Such a 
function can be computed nonperturbatively by means of 
Monte Carlo simulations with a good control of the systematic 
errors. If we set $\mu=1/L$, $g_R$ defines a running coupling 
constant that is a function of $\mu/m$. The function $g_R(1/L)$ 
can also be computed in perturbation theory. This provides the 
connection between $g_R$ and any other perturbative scheme. 
This immediately defines
the $l$-loop approximation to the \MS coupling: 
$g^{{\rm fs},l}_{\overline{MS}}(\mu ) = 
 g_R(\mu )+\sum_{k=2}^{l+1} d_k g_R^{k}(\mu )$. The perturbative expansion of 
$g_R(1/L)$ is known to three-loop order \cite{Shin:1996gi}.
\item \label{Hybrid}
The {\em hybrid} method. 

The finite-size scaling method does not provide---at least in the 
implementation of Ref. \cite{Luscher:1991wu}---the coupling 
$g_{\overline{MS}}(\mu )$ for any $\mu $, but only on a properly chosen 
mesh of values, say $\{\mu _i\}_{i=1,\dots}$. 
The next step consists in interpolating between these
values using RG equations. Since the mass gap is a
RG-invariant quantity, at order $l$, we may require
\begin{eqnarray}
\Lambda_{\overline{MS}}^{(l)}(\mu_i,g_{\overline{MS}}(\mu_i)) =
\Lambda_{\overline{MS}}^{(l)}(\mu ,g_{\overline{MS}}(\mu )) .
\label{HybridDefinition}
\end{eqnarray} 
Using $g_{\overline{MS}}(\mu_i)$, one can then obtain 
$g_{\overline{MS}}(\mu)$ for any given $\mu$. 
We shall call $g_{\overline{MS}}^{\rm hybr}(\mu)$
the running coupling obtained by this procedure.
\item \label{DressedCoupling}
The {\em improved-coupling} method. 

Method 
\ref{NaivePerturbative} does not work well because lattice
perturbation theory is not ``well behaved'': Perturbative 
coefficients are large, giving rise to large truncation errors.
Parisi \cite{Parisi:1980aa,Martinelli:1981tb} noticed that much smaller
coefficients are obtained if one expands in terms of 
``improved" (or ``boosted") couplings 
defined using ``short-distance" observables. In the $\sigma$-model one can 
define a new coupling in terms of the energy density
\begin{eqnarray}
g_E \equiv \frac{4}{N-1}(1-\<\sg_x\cdot \sg_{x+\mu }\>),
\label{DressedCouplingEnergy}
\end{eqnarray}
which is then related to $g_{\overline{MS}}(\mu)$ perturbatively. At order $l$,
we can write
$g_{\overline{MS}}^{{\rm dc},l}(a\mu )=g_E+\sum_{k=2}^{l+1} c^E_k(\mu a) 
g_E^k$. 
In practice the method works as follows: Given $g_L$,
one computes numerically $g_E$; then, one uses the previous perturbative 
expansion to determine the \MS coupling constant. This method is expected to 
be better than the naive one. Indeed, one expects 
$|c^E_k(\mu a)|\ll |c_k(\mu a)|$, so that truncation errors should be 
less important. The perturbative coefficients $c_k^E$ can be computed 
up to $l=3$ using the results of 
\cite{Caracciolo:1995bc,Alles:1999fh,Alles:1997rr}.
\end{enumerate}
Notice that the list above is by no means exhaustive. For instance, an 
alternative nonperturbative coupling may be defined 
using off-shell correlation functions:
\begin{eqnarray}
g_{R}(\mu = l^{-1}) = \xi ^d
\frac{\<(\sg_{x+l/a}-\sg_x)^2\>}{\sum_x \sg_0\cdot \sg_x}\; .
\label{RunningOffShell}
\end{eqnarray}
Something similar has been proposed in Refs. 
\cite{Becirevic:1999uc,Becirevic:1999sc,Becirevic:1999hj}, with the purpose of
computing the QCD $\Lambda$-parameter. This approach opens the Pandora
box of possible definitions of the running coupling 
in substitution of Eq. (\ref{RunningOffShell}). A scheme that
has been intensively studied in the context of QCD employs the
three-gluon vertex 
(see Refs. \cite{Becirevic:1999rv,Becirevic:1999sc,Henty:1995gv,
Parrinello:1994fu,Parrinello:1994wd,Parrinello:1997wm,Alles:1997fa}).
\begin{table}
\hspace*{-0.2truecm}
\begin{tabular}{|c|c|c|c|c|c|c|}
\hline
$g_R(\mu)$ & $\frac{m}{\mu}$ & $\frac{1}{\mbar a}$ &
$g^{{\rm fs},3}_{\overline{MS}}(\mu)$ & $g^{{\rm rgp},4}_{\overline{MS}}(\mu)$ &
$g^{{\rm np},3}_{\overline{MS}}(\mu)$ & $g^{{\rm dc},3}_{\overline{MS}}(\mu)$\\
\hline
\hline
0.5372&0.00071(11) & 0.0097(15) &0.5870[-3]&  
 0.5892[-1] & 0.574[-5]  & 0.5902[-28]\\
0.5747&0.00143(11) & 0.0195(15) &0.6321[-4]&
 0.6351[-1] & 0.617[-6]  & 0.6354[-31]\\
0.6060&0.00237(15) & 0.0323(20) &0.6703[-5]&
 0.6736[-2] & 0.652[-6]  & 0.6737[-16]\\
0.6553&0.00478(15) & 0.0652(20) &0.7312[-7]&  
 0.7362[-2] & 0.708[-8]  & 0.7364[+5]\\
0.6970&0.00794(19) & 0.1083(25) &0.7835[-9]&  
 0.7900[-3] & 0.755[-12] & 0.7903[+6]\\
0.7383&0.01231(15) & 0.1678(20) &0.8361[-11]&  
 0.8437[-4] & 0.800[-16] & 0.8436[-13]\\
0.7646&0.01589(15) & 0.2166(20) &0.8701[-13]&  
 0.8788[-5] & 0.828[-20] & 0.8777[-35]\\
0.8166&0.02481(22) & 0.3382(30) &0.9382[-16]&  
 0.9486[-7] & 0.881[-28] & 0.9434[-99]\\
0.9176&0.04958(38) &0.6759(51)  &1.0742[-26]& 
 1.0852[-13] &  0.975[-45] & 1.0623[-275]\\
1.0595&0.1033(6) &1.4082(77) &1.2743[-47]& 
 1.2895[-27] & 1.090[-73] & 1.2135[-593]\\
1.2680&0.2092(5) &2.8519(67) &1.5886[-96]& 
 1.59626[-680] & 1.217[-108]& 1.3863[-1056]\\
\hline
\end{tabular}
\caption{The \MS running coupling constant. We use here 
several different methods as explained in the text.
The errors
on the second and third columns are statistical.}
\label{RunningComparison1}
\end{table} 

Let us compare the different methods. In
Tab. \ref{RunningComparison1} we compare the procedures 
\ref{NaivePerturbative}, \ref{RGImprovedPerturbative},
\ref{FiniteSizeNonPerturbative}, and \ref{DressedCoupling}.
In the first column we report a collection of values of $g_R$.
A subset of the values given in the table have been considered for the first 
time in Ref. \cite{Luscher:1991wu}. Later, the mesh was enlarged by
Hasenbusch  \cite{Hasenbusch-unpublished}.
For these values of $g_R$, Hasenbusch 
computed the corresponding value of $mL=m/\mu$ which is reported in the second 
column. Note a peculiarity of the finite-size approach: 
usually, one fixes $\mu/m$
and then determines the running coupling constant. Here, the 
running coupling constant is fixed at the beginning and the value of the 
scale is determined numerically. In the third column we report
the scale in lattice units for $g_L = 1/1.54$, the value of the lattice 
coupling at which we have done most of our simulations. The results
are obtained by using $(ma)^{-1} = 13.632(6)$. The error
reported there corresponds to the error on $m/\mu$, the error on $(ma)$
being negligible. In column 4 we report the estimate of 
$g_{\overline{MS}}^{\rm fs,3} (\mu)$ obtained by using $g_R$ and three-loop 
perturbation theory \cite{Shin:1996gi}. In brackets we report the difference 
$g_{\overline{MS}}^{\rm fs,2} (\mu) - g_{\overline{MS}}^{\rm fs,3} (\mu)$. 
In the next column we report the four-loop coupling 
$g_{\overline{MS}}^{\rm rgp,4}(\mu)$ 
obtained by using the value of $m/\mu$ given in the second column.
Again, in brackets we report 
$g_{\overline{MS}}^{\rm rgp,3}(\mu) - g_{\overline{MS}}^{\rm rgp,4}(\mu)$. 
In the last two column 
we report the results obtained by using three-loop lattice perturbation theory
\cite{Caracciolo:1995bc,Alles:1999fh,Alles:1997rr}. In the fifth
column we use $g_L=1/1.54$ as the expansion parameter.
In the sixth column the
improved coupling defined by Eq. (\ref{DressedCouplingEnergy}) is used.
The connection with the bare coupling is obtained by using the perturbative 
expressions given in Ref. \cite{Alles:1997rr}.
The relevant expectation value has been evaluated in a Monte Carlo
simulation at $g_L=1/1.54$ on a lattice $128\times 256$ with
statistics $N_{\rm stat} =10000$, yielding 
$g_E = 0.768133(49)$. 

\begin{table}
\begin{center}
\begin{tabular}{|c|c|c|l|}
\hline
$\frac{1}{\mbar a}$ & $g^{{\rm rgp},4}_{\overline{MS}}(\mu)$ &
$g^{{\rm hybr},A}_{\overline{MS}}(\mu)$ & $g^{{\rm hybr},B}_{\overline{MS}}(\mu)$ \\
\hline
\hline
1 & 1.18422 & 1.171(2)  & 1.1804(6) \\
2 & 1.42282 & 1.403(3)  & 1.417(1)  \\
3 & 1.62532 & 1.598(4)  & 1.617(1)  \\
4 & 1.81923 & 1.783(5)  & 1.809(2)  \\
5 & 2.01713 & 1.970(7)  & 2.003(2)  \\
6 & 2.22905 & 2.167(9)  & 2.211(3)  \\
7 & 2.46722 & 2.385(12) & 2.443(4)  \\
8 & 2.75208 & 2.637(16) & 2.717(6)  \\
\hline
\end{tabular}
\end{center}
\caption{The \MS running coupling constant.
In the first column we report the RG coupling 
obtained using $(ma)^{-1} = 13.632(6)$ at $1/g_L=1.54$.
The coupling reported in the third and fourth column 
are obtain using the hybrid scheme:
the columns differ in the choice of
the ``integration constant'', see text. 
The reported error is due to the error on $\mu_i$ appearing in the l.h.s. of
Eq. (\ref{HybridDefinition}), see Tab. \ref{RunningComparison1}, 
second column.}
\label{RunningComparison2}
\end{table}

In Tab. \ref{RunningComparison2} we compare, on a broad range of scales, 
the outcome of RG-improved perturbation theory and the 
interpolation procedure which we called ``hybrid.'' In both cases four-loop 
perturbation theory is used.
The two hybrid couplings referred to as $A$ and $B$ differ in the ``boundary
condition'' for the RG interpolation, that
is in the value used in the left hand side of Eq. (\ref{HybridDefinition}).
We used respectively, $g_{\overline{MS}} = 1.074254$ for 
$m/\mu = 0.04958$ and $g_{\overline{MS}} = 1.588556$ for $m/\mu = 0.2092$,
obtained using the finite-size scaling method, cf.
Tab. \ref{RunningComparison1}.
In all cases we fixed $(ma)^{-1} = 13.632(6)$. 

What do we learn from this comparison? First of all, lattice (naive)
perturbation theory (sixth column of Tab. \ref{RunningComparison1})
is a very bad tool. Even at energies as high as $50$ times the mass gap
$g_{\overline{MS}}(\mu)$
is affected by a $\sim 5\%$ systematic error. 
However, it is reassuring that the expansion tells us its own unreliability.
Indeed, the observed discrepancy is of the order (at most twice as large) 
of the difference between the two-loop and the three-loop result.
The perturbative expansion in terms of the improved coupling
is much better. The results are quite precise up to $\mu\approx 10m$. 
For smaller values of $\mu$ the discrepancy increases, but it is nice
that it is again of the order of the difference between 
the two- and the three-loop result. Perturbative RG
supplemented with the prediction (\ref{BetheAnsatzPrediction})  gives 
results which are in agreement with the nonperturbative ones obtained 
using the finite-size scaling method within a few
percent for all the energy scales given in Tab. \ref{RunningComparison1}.
The accuracy remains good (if the comparison is made with the ``hybrid'' 
procedure) also for scales of the order of the mass gap. 

Up to now we have discussed the \MS scheme and how to obtain the value 
of the \MS coupling. However, as we already mentioned above,
reasonably good results can also be obtained if we use 
the coupling $g_E$. In this scheme we introduce the $\Lambda_E$ parameter
\be
\Lambda_{E} (a,g_E) = {1\over a}\
   \left( \beta_0 g_E\right)^{-\beta_1/\beta_0^2}
       \exp\left[ - {1\over \beta_0 g_E} - \int^{g_E}_0 dt
       \left( {1\over \beta_E(t)} + {1\over \beta_0 t^2} -
              {\beta_1\over \beta_0^2 t}\right)\right] ,
\label{LambdaEdef}
\ee
where $\beta_E(g_E) = - g_E^2 \sum_{k=0} \beta_k^E g^k_E$. If 
$g_L = f(g_E)$, then $\beta_E(g_E) = \beta^{\rm latt}(f(g_E))/f'(g_E)$. 
The mass gap is invariant and thus $m = C_{N,E} \Lambda_{E} (a,g_E)$,
where 
\be
C_{N,E} = \left( {8\over e} \right)^{1/(N-2)}
{1\over \Gamma\left(1+{1\over N-2}\right)} 2^{5/2}
  \exp\left[{\pi \over 4(N-2)}\right].
\label{costanteHasenfratz_schemaE}
\ee


\clearpage
%
%
%
%
%
%
%
%
%
%
%
%
%
%
%
%
%
%
%
%
%
%

\end{document}